\RequirePackage{lineno}
\documentclass[rmp,twocolumn,eqsecnum,floatfix]{revtex4-1}

\usepackage{cancel}
\usepackage{graphicx}
\usepackage{epsfig}
\usepackage{amsmath,amsfonts,amssymb}
\usepackage{t1enc}
\usepackage{url}

\newcommand{\oh}{\frac{1}{2}}
\newcommand{\ohs}{{\textstyle\frac{1}{2}}}

\newcommand{\afb}{A_{FB}}
\newcommand{\ac}{A_C}
\newcommand{\acll}{A_{C}^{\ell \ell}}
\newcommand{\acA}{A_\text{C}^{t \bar t \gamma}}
\newcommand{\acW}{A_\text{C}^{t \bar t W}}
\newcommand{\afbl}{A_{FB}^\ell}
\newcommand{\afbll}{A_{FB}^{\ell \ell}}
\newcommand{\au}{A_u}
\newcommand{\ad}{A_d}
\newcommand{\ttb}{t \bar t}

\newcommand{\uub}{u \bar u}
\newcommand{\ddb}{d \bar d}
\newcommand{\qqb}{q \bar q}
\newcommand{\ppb}{p \bar p}
\newcommand{\ttA}{t \bar t \gamma}
\newcommand{\mttb}{m_{t \bar t}}
\newcommand{\pt}{p_T}
\newcommand{\pTl}{p_T^\ell}
\newcommand{\pTttb}{p_T^{t \bar t}}
\newcommand{\sint}{\sigma_\text{int}}
\newcommand{\squa}{\sigma_\text{quad}}
\newcommand{\fbin}{fb$^{-1}$}
\newcommand{\Pt}{P_{z}}
\newcommand{\Ptb}{P_{z'}}

\def\gev {\ifmmode {\rm GeV} \else GeV$/c$\fi}
\newcommand{\ifb}{ {\rm fb}^{-1} }
\newcommand{\et}{\mbox{${E_T}$}}
\newcommand{\met}{\mbox{$\protect \raisebox{.3ex}{$\not$}\et$}}
\newcommand{\ppbar}{p\bar{p}}

\newcommand{\ttbar}{t\bar{t}}

\newcommand{\dy}{\Delta y}
\newcommand{\yq}{qy_{l}}
\newcommand{\afblep} {A_\mathrm{FB}^{\ell}}

\newcommand{\Ocal}{\mathcal{O}}
\newcommand{\Lcal}{\mathcal{L}}
\newcommand{\gM}{\gamma^\mu}

\newcommand{\yttb}{y_{t\bar{t}}}
\newcommand{\dabsy}{\Delta |y|}

\def\pt{\ensuremath{p_T}} 
\def\et{\ensuremath{E_T}} 

\newcommand{\betattb}{\beta_z^{t \bar t}}

\begin{document}

\title{Asymmetries in top quark pair production at hadron colliders}

\author{J.A. Aguilar-Saavedra}
\affiliation{Departamento de F\'{\i}sica Te\'orica y del Cosmos, Universidad de Granada,
 E-18071 Granada, Spain.}
\affiliation{PH-TH Department, CERN, CH-1211 Geneva 23, Switzerland.}
\author{D. Amidei}
\affiliation{Department of Physics, University of Michigan, Ann Arbor, MI 48109, USA.}
\author{A. Juste}
\affiliation{Instituci\'o Catalana de Recerca i Estudis Avan\c{c}ats (ICREA), E-08010 Barcelona, Spain.}
\affiliation{Institut de F\'{\i}sica d'Altes Energies (IFAE), E-08193 Bellaterra, Barcelona, Spain.}
 \author{M. P\'erez-Victoria}
\affiliation{Departamento de F\'{\i}sica Te\'orica y del Cosmos and CAFPE, Universidad de Granada,
 E-18071 Granada, Spain.}

\begin{abstract}
We review the asymmetries in top quark pair production at the Tevatron and the LHC. We summarize the experimental measurements and the interpretations of a possible excess in terms of new physics. We also review other top quark properties---emphasizing effects related to the $\ttb$ asymmetries---as well as other collider signals.
\end{abstract}

\maketitle
\tableofcontents

\section{Introduction}
Precision tests of the production and decay modes of elementary particles have proved to be of great help to extend our knowledge of the forces that govern their interactions, and search for new physics. The role of $e^+ e^-$ colliders, such as the CERN Large Electron Positron (LEP) collider, has been fundamental in this task~\cite{ALEPH:2005ab}. Hadron colliders, which are usually regarded as discovery machines, offer complementary precision measurements that further test the standard model (SM) of elementary particle interactions. This is indeed the case for the physics of the top quark, which was discovered at the Fermilab Tevatron $\ppb$ collider~\cite{Abe:1995hr,Abachi:1995iq}. The relative ease to identify top quarks in a hadronic collider environment and the large samples produced, not only at the Tevatron but also at the CERN Large Hadron Collider (LHC), have allowed to perform several precision measurements of its properties. In most cases a good agreement has been obtained with the predictions of the SM. However, a notable exception has been found in several forward-backward (FB) asymmetries, often also referred to as charge asymmetries, in top quark pair ($\ttb$) production at the Tevatron. Experimentally, these asymmetries are conveniently defined in terms of the rapidities of the top (anti-)quark and their decay products in the laboratory frame, where the rapidity of a particle is given by
\begin{equation}
y = \frac{1}{2} \log \frac{E+p_z}{E-p_z} \,,
\end{equation}
with $E$ its energy and $p_z$ the component of its three-momentum in the $\hat z$ axis, taken here in the proton direction. The largest deviations were found in the so-called $\ttb$ rest-frame FB asymmetry, 
\begin{equation}
\afb = \frac{N(\Delta y >0) - N(\Delta y <0)}{N(\Delta y >0) + N(\Delta y <0)} \,,
\label{ec:afb1}
\end{equation}
with $\Delta y = y_t-y_{\bar t}$ and $N$ standing for the number of events.\footnote{The denomination for this asymmetry stands for the fact that $\afb$ is the same when the $t$ and $\bar t$ rapidities are taken in the $\ttb$ rest frame, and in this frame $\Delta y = 2 y_t$. Therefore, a ``forward'' event with the top quark following the proton direction in the $\ttb$ rest frame has $\Delta y > 0$ and, conversely, a ``backward'' event has $\Delta y < 0$.} 
The discrepancy between experimental data and the SM predictions surpassed three standard deviations in the 2011 CDF measurement of $\afb$ at high $\ttb$ invariant mass $\mttb \geq 450$ GeV~\cite{Aaltonen:2011kc}, with roughly 5 fb$^{-1}$ of data taken at a centre-of-mass (CM) energy $\sqrt{s} = 1.96$ TeV. Deviations were found in the inclusive asymmetry too, and also by the D0 experiment~\cite{Abazov:2011rq}. This anomaly motivated intense research in model building---invoking new physics in $\ttb$ production as the explanation for the anomaly---as well as the update of the SM calculations. Since then, it has also fostered the theoretical and experimental search of other anomalies in $\ttb$ production, which might appear if the Tevatron asymmetry were indeed a sign of new physics. Among the latter, the charge asymmetry at the LHC has a prominent role. Close to the end of Tevatron operations, in 2011 the LHC began taking data in $pp$ collisions at $\sqrt{s} = 7$ TeV, quickly producing large $\ttb$ samples. However, in $pp$ collisions a FB asymmetry with a fixed $\hat z$ axis, such as the one defined in Eq.~(\ref{ec:afb1}), vanishes due to the symmetry of the initial state. Instead, a ``forward-central'' charge asymmetry can be defined,
\begin{equation}
\ac = \frac{N(\Delta |y| >0) - N(\Delta |y| <0)}{N(\Delta |y| >0) + N(\Delta |y| <0)} \,,
\label{ec:ac}
\end{equation}
with $\Delta |y| = |y_t| - |y_{\bar t}|$, which is a complementary probe of asymmetric $\ttb$ production. (The precise meaning of this statement will be clear in the following section.)
Noticeably, the measurements of this asymmetry performed by the ATLAS and CMS Collaborations at $\sqrt{s} = 7,\,8$ TeV~\cite{Chatrchyan:2012cxa,Aad:2013cea,ATLAS:2012sla,Chatrchyan:2014yta,CMS:2013nfa} are in good agreement with the SM predictions. These results, although not conclusive because the measurements refer to an observable that differs from the Tevatron one, call into question the Tevatron excess. Concurrently, when the full Tevatron dataset of around 10 fb$^{-1}$ has been analyzed, the discrepancies have been reduced with respect to previous results. The CDF Collaboration finds a $1.7\sigma$ excess over the SM predictions~\cite{Aaltonen:2012it} whereas the D0 Collaboration finds agreement within $1\sigma$~\cite{Abazov:2014cca}. Other asymmetries can also be constructed using the momenta of charged leptons $\ell$ produced in the top quark decay $t \to W b \to \ell \nu b$. The Tevatron measurements~\cite{Aaltonen:2013vaf,Aaltonen:2014eva,Abazov:2014oea,Abazov:2013wxa} are above the SM values too, whereas leptonic asymmetries measured at the LHC are consistent with the SM.

This review attempts to provide a self-contained description of the current status of theoretical and experimental research on the subject of the $\ttb$ asymmetries, paying also special attention to other observables that further test the presence of new physics in $\ttb$ production. The $\ttb$ asymmetries, their interrelation and the SM predictions are reviewed in detail in Sec.~\ref{sec:2}. The current experimental status of asymmetry measurements at the Tevatron is presented in Sec.~\ref{sec:3}. Measurements at the LHC are reviewed in Sec.~\ref{sec:4}, where prospects for the next run with 14 TeV are also discussed. In Sec.~\ref{sec:5} we give an overview of the new physics proposals to address  the Tevatron anomaly. We address the correlated effects in $\ttb$ production of these new physics proposals in Sec.~\ref{sec:6}, and other collider effects are briefly discussed in Sec.~\ref{sec:7}.
Finally, conclusions are outlined in Sec.~\ref{sec:8}.


\section{Overview of the asymmetries}
\label{sec:2}

\subsection{The Tevatron $\ttb$ asymmetry}
\label{sec:2a}

\begin{figure}[htb]
\begin{center}
\begin{tabular}{ccc}
\includegraphics[height=2.5cm,clip=]{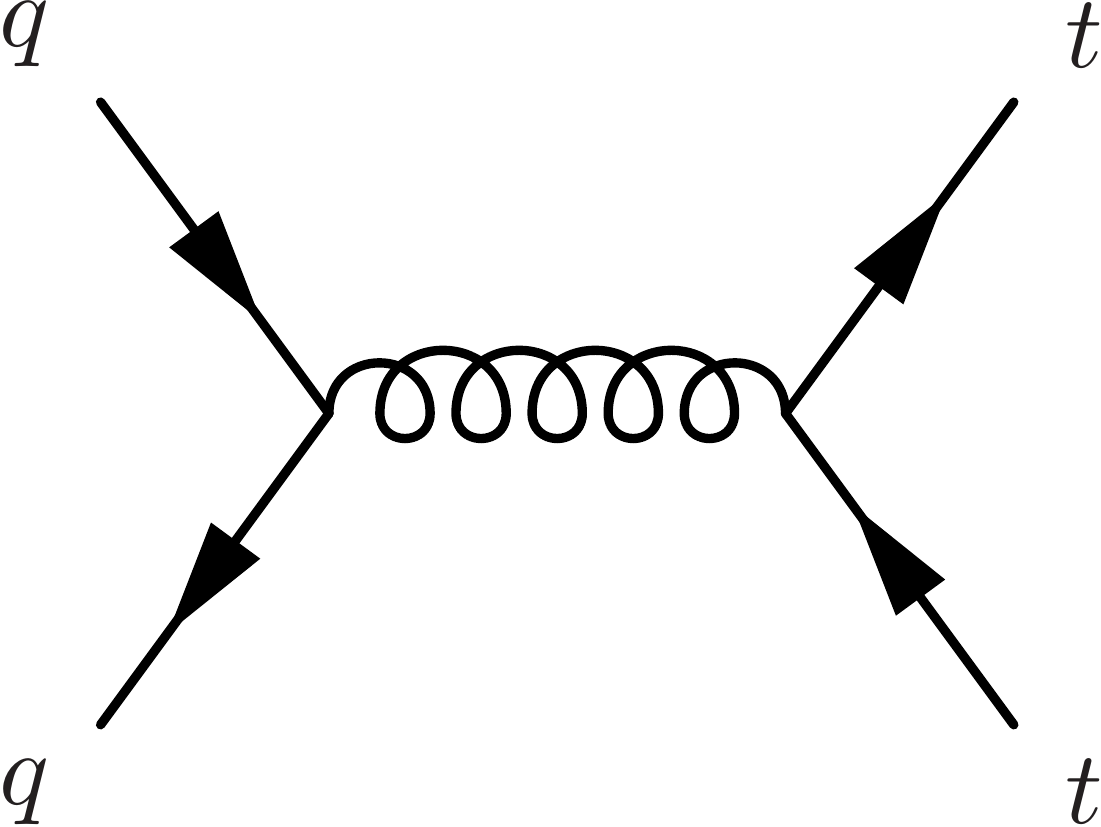} & &
\includegraphics[height=2.5cm,clip=]{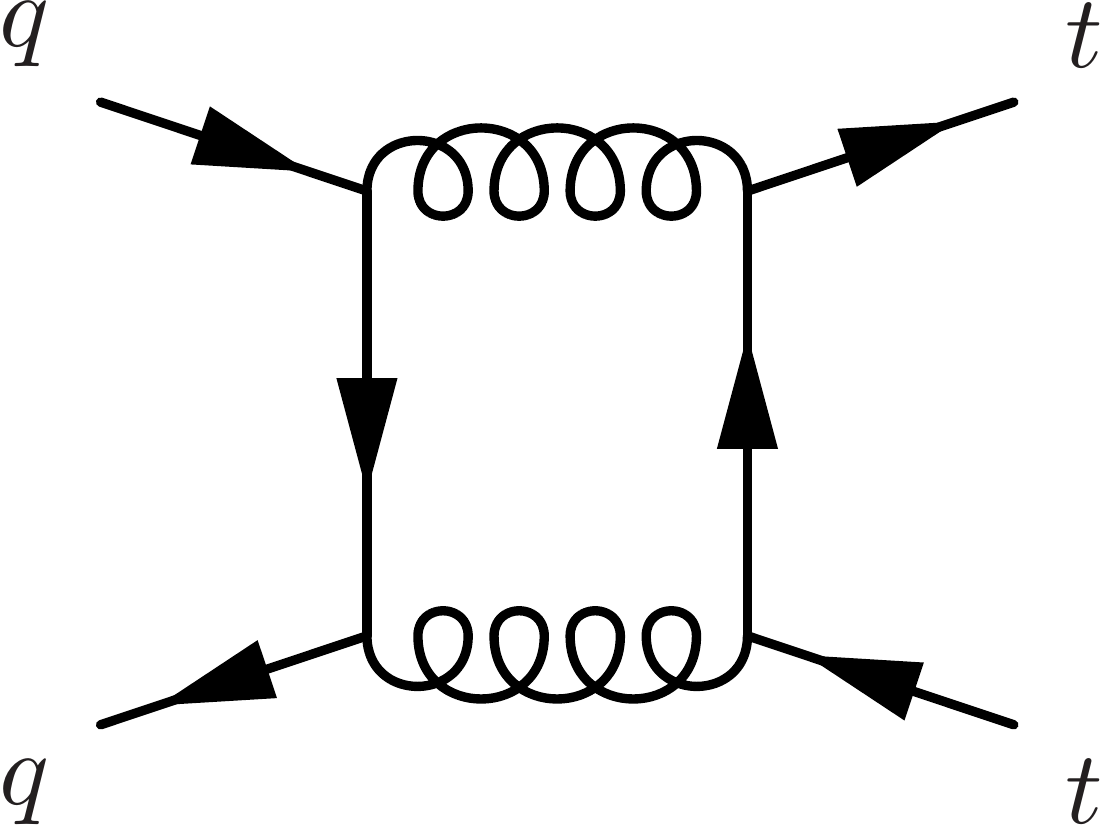} \\
(a) & & (b) \\[3mm]
\includegraphics[height=2.5cm,clip=]{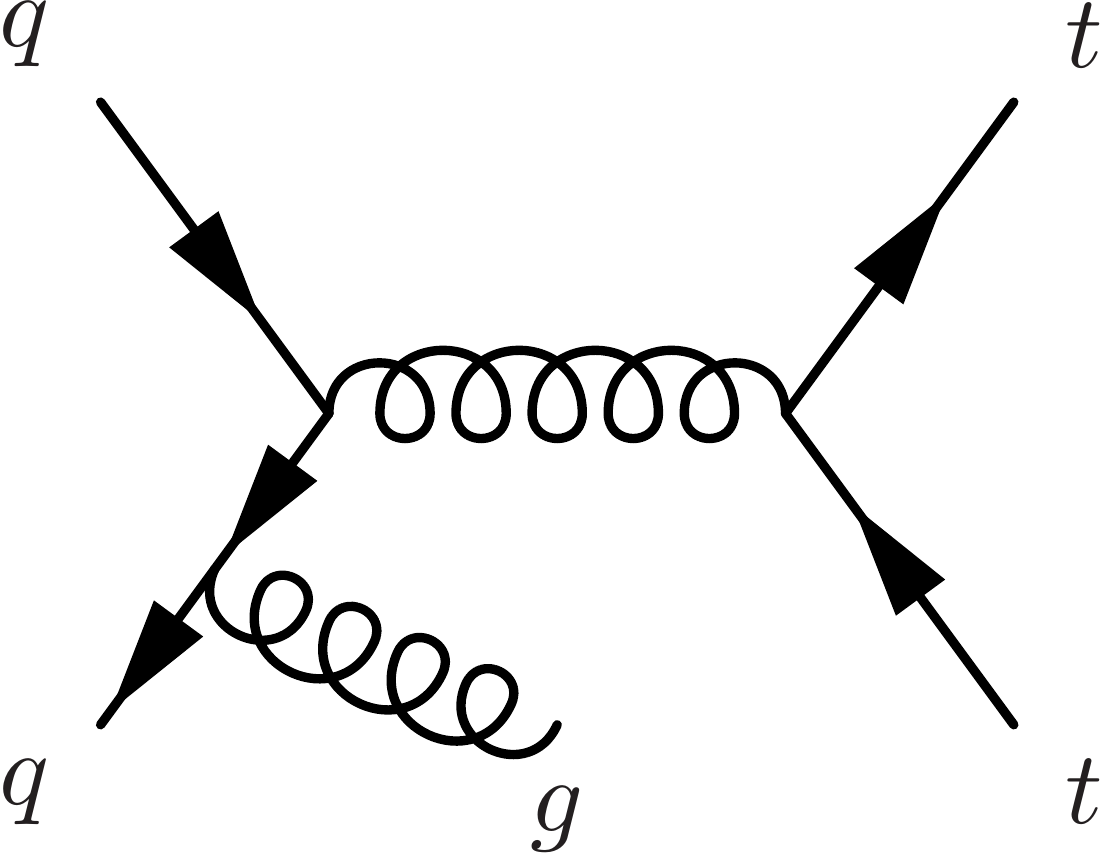} & &
\includegraphics[height=2.5cm,clip=]{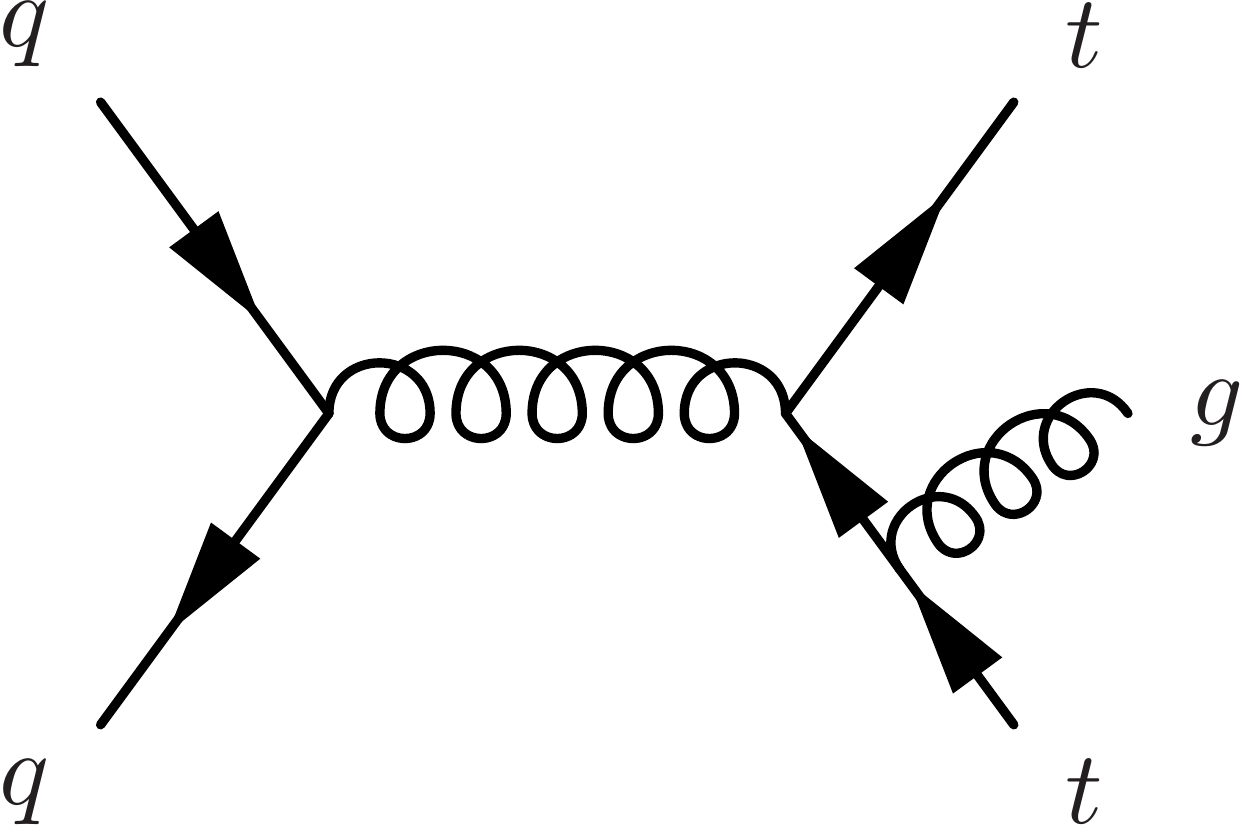} \\
(c) & & (d) \\[3mm]
\end{tabular}
\end{center}
\caption{Sample Feynman diagrams contributing to $q \bar q \to \ttb$ at leading order (LO) (a) and NLO in QCD (b,c,d).}
\label{fig:2A}
\end{figure}

At the Tevatron, top quark pairs are produced mainly in the partonic subprocesses $\qqb ,gg \to \ttb X$, with $q=u,d$ and $X$ denoting possible additional jets. (In this section we explicitly indicate with $X$ the possibility of extra jets, to emphasize the different sources of the inclusive asymmetry; this will be omitted for simplicity in the following sections.) Within the SM, the main contribution to the asymmetry (\ref{ec:afb1}) arises at next-to-leading order (NLO) in QCD due to the interference of order $\alpha_s^3$ terms in the cross section that are odd under the interchange $t \leftrightarrow \bar t$ with the initial quarks fixed---hence the denomination of $\afb$ as ``charge asymmetry'', despite the fact that it does not have any relation with the charge conjugation symmetry $C$. The interference of tree-level and one-loop diagrams for $\qqb \to \ttb$, Fig.~\ref{fig:2A} (a,b), generates a positive asymmetry, while the interference of initial- and final-state radiation in $\qqb \to \ttb g$, for example diagrams (c,d) in Fig.~\ref{fig:2A}, generates a negative asymmetry~\cite{Kuhn:1998jr}. The relative size of these contributions depends on the transverse momentum of the $\ttb$ pair, $\pTttb$, which is zero in $\qqb \to \ttb$ but not in $\qqb \to \ttb g$. It is found that for $\pTttb \lesssim 25$ GeV the asymmetry is positive, while for $\pTttb \gtrsim 25$ GeV it is negative. When integrated over the full $\pTttb$ spectrum, the net contribution to (\ref{ec:afb1}) is positive. An alternative way of explaining the $\pTttb$ dependence involves QCD radiation~\cite{Skands:2012mm}. For forward top quarks the color charge is less accelerated, so they are less likely to emit gluons than backward top quarks. Hence, forward top quarks are associated with smaller $\pTttb$, and vice versa.

The asymmetry generated in $\qqb \to \ttb X$ is diluted by the $gg \to \ttb$ subprocess, which amounts to 10\% of the total cross section and does not contribute to the numerator of $\afb$ because of the symmetry of the initial state. Other quark subprocesses like $s \bar s$, $c \bar c$ do not contribute either because the parton density functions (PDFs) are the same for quarks and antiquarks. They have small cross sections, and do not significantly contribute to the denominator either.
For completeness, let us mention that the Tevatron experiments also measure the so-called laboratory-frame asymmetry,
\begin{equation}
A_{FB}^{p\bar p} = \frac{N(y_t >0) - N(y_t <0)}{N(y_t >0) + N(y_t <0)} \,,
\label{ec:afbalt}
\end{equation}
which is smaller than the $\ttb$ rest frame asymmetry $\afb$ because of kinematics, and was also found above the SM expectation in earlier measurements. Neither the CDF nor the D0 Collaborations measure this asymmetry in their latest full dataset analyses, however, and we will restrict our discussion to $\afb$.

The discrepancy found in the 2011 measurements of the asymmetry~\cite{Aaltonen:2011kc,Abazov:2011rq} motivated the refinement of the SM predictions, including weak, mixed QCD-weak and QCD-QED corrections that increase the asymmetry by 25\% with respect to the NLO QCD value. Because $\afb$ vanishes at the tree level in the SM, a fixed-order expansion at LO in perturbation theory involves the numerator in Eq.~(\ref{ec:afb1}) at NLO, including $\mathcal{O}(\alpha_s^3)$, $\mathcal{O}(\alpha^2)$ and $\mathcal{O}(\alpha_s^2 \alpha)$ terms, and the denominator at LO. Several independent calculations yield similar results, $\afb = 0.089^{+0.008}_{-0.006}$~\cite{Hollik:2011ps}, $\afb = 0.087 \pm 0.010$~\cite{Kuhn:2011ri}, and $\afb = 0.088 \pm 0.006$~\cite{Bernreuther:2012sx}, where the theory uncertainty is due to the variation of the factorization and renormalization scales. The contribution of the different subprocesses can be read in Table~\ref{tab:afb-tev}.
\begin{table}[htb]
\begin{center}
 \caption{The  contributions to the numerator $\Delta N$ of the FB asymmetry~(\ref{ec:afb1})
  at NLO for three different scales. The total LO cross sections and the resulting asymmetries are also given. From~\textcite{Bernreuther:2012sx}.\label{tab:afb-tev}}
\begin{tabular}{ccccc}
\hline
\hline
\multicolumn{1}{c}{$\Delta N$~(pb)} &  & $\mu={m_{t}}/{2}$  & $\mu=m_{t}$  & $\mu=2m_{t}$ \\
\hline 
$O(\alpha_{s}^{3})$ 
     & $u\bar{u}$  & $0.5014$ & $0.3297$  &  $0.2251$ \\
     & $d\bar{d}$  & $0.0899$ & $0.0582$  & $0.0392$ \\
     & $qg$ & $7.6 \times 10^{-5}$ & $3.4 \times 10^{-5}$ & $2.9 \times 10^{-5}$ \\
$O(\alpha^{2})$
     & $u\bar{u}$  & $1.47 \times 10^{-2}$ & $1.29 \times 10^{-2}$  & $1.15 \times 10^{-2}$ \\
     & $d\bar{d}$  & $1.9 \times 10^{-3}$ & $1.6 \times 10^{-3}$  & $1.5 \times 10^{-3}$ \\
$O(\alpha\alpha_{s}^{2})_\text{weak}$
     & $u\bar{u}$  & $10.7 \times 10^{-3}$ & $7.8 \times 10^{-3}$  & $5.8 \times 10^{-3}$ \\
     & $d\bar{d}$  & $-3.4 \times 10^{-3}$ & $-2.4  \times 10^{-3}$  & $-1.8 \times 10^{-3}$ \\
$O(\alpha\alpha_{s}^{2})_\text{QED}$
     & $u\bar{u}$  & $0.1047$ & $0.0761$  & $0.0569$ \\
     & $d\bar{d}$  & $-9.4   \times 10^{-3}$ & $-6.7 \times 10^{-3}$  & $-4.9 \times 10^{-3}$ \\
\multicolumn{1}{c}{total $\Delta N$} &  & $0.7104$ & $0.4772$  & $0.3332$ \\
\multicolumn{1}{c}{$\sigma_\text{QCD}^\text{LO}$} &  & $7.618$ & $5.456$ & $4.030$ \\
\multicolumn{1}{c}{$\afb$~(\%)} &  & $9.33$ & $8.75$ & $8.27$ \\
\hline
\hline
\end{tabular}
\end{center}
\end{table}
On the other hand, it has also been suggested that the Tevatron cross section and FB asymmetry can be reproduced with an unconventional choice of renormalization scale for the strong coupling~\cite{Brodsky:2012ik}, instead of the usual one $\mu_R \sim m_t$. Still, it remains to be shown that the several differential distributions ({\it e.g.} of $\ttb$ invariant mass, particle transverse momenta and pseudo-rapidities, etc.) that are measured with good accuracy in $\ttb$ production are also well reproduced using this proposal.

Higher-order effects beyond NLO were probed with the inclusion of soft-gluon resummation, with different results depending on the method used: while these effects were found to be rather small by \textcite{Almeida:2008ug} and \textcite{Ahrens:2011uf}, they were found larger by \textcite{Kidonakis:2011zn}, and eventually closer to the exact next-to-next-to-leading order (NNLO) calculation. (See~\textcite{Kidonakis:2013zsa} for a comparison between different methods.) Electroweak Sudakov corrections are small, 5\% with respect to the NLO QCD value~\cite{Manohar:2012rs}. Recently, a prediction at NNLO in QCD, with NLO electroweak corrections, has become available~\cite{Czakon:2014xsa}. The full NNLO QCD contribution increases the asymmetry by a factor of $1.13$ with respect to the NLO value. Including NLO electroweak corrections, the prediction is $\afb = 0.095 \pm 0.007$. An approximate next-to-next-to-next-to-leading order  (aN$^3$LO) calculation in QCD based on soft-gluon resummation, including NLO electroweak corrections, is also available~\cite{Kidonakis:2015ona}, yielding $\afb = 0.100 \pm 0.006$. A summary of some of the results obtained at different orders in perturbation theory is shown in Table~\ref{tab:afbsumm}.

\begin{table}[htb]
\begin{center}
 \caption{Summary of selected calculations of the Tevatron $\ttb$ rest-frame asymmetry, at different orders. The label ``EW'' denotes electroweak (weak, QED and mixed) corrections. \label{tab:afbsumm}}
\begin{tabular}{ll}
\hline
\hline
\multicolumn{1}{c}{Order} & \multicolumn{1}{c}{$\afb$} \\ \hline
NLO QCD            & $0.072 \pm 0.009$ \cite{Kuhn:2011ri} \\[1mm]
\begin{tabular}{l}NLO QCD \\ + EW \end{tabular}  & $0.088 \pm 0.006$ \cite{Bernreuther:2012sx} \\[1mm]
NNLO QCD          & $0.083 \pm 0.003$ \cite{Czakon:2014xsa} \\[1mm]
\begin{tabular}{l}NNLO QCD \\ + NLO EW \end{tabular} & $0.095 \pm 0.007$ \cite{Czakon:2014xsa} \\[1mm]
aN$^3$LO QCD          & $0.087 \pm 0.002$ \cite{Kidonakis:2015ona} \\[1mm]
\begin{tabular}{l}aN$^3$LO QCD \\ + NLO EW \end{tabular} & $0.100 \pm 0.006$ \cite{Kidonakis:2015ona} \\
\hline
\hline
\end{tabular}
\end{center}
\end{table}

Finally, let us mention that, alternatively to the NLO predictions with LO denominators discussed above, the asymmetry can be computed using NLO numerator and denominator, and this is the only possibility when using Monte Carlo generators. For example, $\afb = 0.058$ with {\sc mcfm}~\cite{Campbell:1999ah} and $\afb = 0.05$ with {\sc mc@nlo}~\cite{Frixione:2002ik} for $\ttb$ production at NLO in QCD. These values are smaller than the LO denominator predictions partly due to the missing electroweak corrections in the numerator, which amount to an increase by a factor of 1.26, and partly because the total cross section appearing in the denominator is around 25\% larger at NLO. Likewise, the NNLO prediction can be computed with NNLO numerator and denominator, and is $\afb = 0.087$, slightly smaller than the value expanded in powers of $\alpha,\alpha_s$. The corresponding aN$^3$LO prediction is also smaller, $\afb = 0.094$.

\subsection{The polar angle distribution and its asymmetry}
\label{sec:2b}

The asymmetry~(\ref{ec:afb1}) is equivalent to a FB asymmetry in the polar angle $\theta$ between the top quark momentum in the CM frame and the $\hat z$ axis, because $y_t$ and $\cos \theta$ have the same sign.
Also, in $\ppb$ collisions at the Tevatron the initial state quark and antiquark in $\qqb \to \ttb X$ are supplied by the proton and antiproton, respectively, with a small $\lesssim 0.4\%$ probability for the opposite due to PDF suppression. Then, the initial quark direction almost always coincides with the proton direction, and the asymmetry~(\ref{ec:afb1}) nearly equals the FB asymmetry
\begin{equation}
\afb = \frac{N(\cos \theta >0) - N(\cos \theta<0)}{N(\cos \theta>0) + N(\cos \theta<0)}
\label{ec:afb2}
\end{equation}
in the polar angle between the top and the initial quark directions in the CM frame.

In order to view the asymmetry (\ref{ec:afb2}) in a more general context, we consider the $2 \to 2$ process $\qqb \to \ttb$, not necessarily at the tree level, using the helicity formalism~\cite{Jacob:1959at}. For helicities $\lambda_1$, $\lambda_2$, $\lambda_3$, $\lambda_4$ corresponding to the external particles $q$, $\bar q$, $t$, $\bar t$, respectively, angular momentum conservation allows to write the amplitude as
\begin{equation}
A = \sum_J a_{\lambda_1 \lambda_2 \lambda_3 \lambda_4}^J D_{\lambda_i \lambda_f}^{J*}(\phi,\theta,0) \,,
\label{ec:amp}
\end{equation}
where $J$ labels the total angular momentum and $\lambda_i=\lambda_1-\lambda_2$, $\lambda_f=\lambda_3-\lambda_4$; the dependence on the production angles $(\theta,\phi)$ is given by the Wigner functions $D^j_{m'm}(\alpha,\beta,\gamma) \equiv \langle jm' | e^{-i \alpha J_z} e^{-i \beta J_y} e^{-i \gamma J_z} | jm \rangle$ and $a_{\lambda_1 \lambda_2 \lambda_3 \lambda_4}^J$ are constants. The sum runs over all possible values $J=0,1,2,\dots$ (since plane waves ``contain'' all possible orbital angular momenta), but in particular cases it may happen that only a few values of $J$ contribute. For example, in the SM at the tree level the process takes place via a spin-1 $s$-channel gluon, therefore the sum only contains the term $J=1$.

From the general amplitude (\ref{ec:amp}), one obtains with a little algebra the partonic differential cross section,
\begin{eqnarray}
\frac{d\hat \sigma}{d\Omega} & \propto &  \sum_{\lambda_1 \lambda_2 \lambda_3 \lambda_4 J J' l}  a_{\lambda_1 \lambda_2 \lambda_3 \lambda_4}^J a_{\lambda_1 \lambda_2 \lambda_3 \lambda_4}^{J'*} 
\langle J \lambda_i J' -\lambda_i | l 0 \rangle  \notag \\
& & \times  \langle J \lambda_f J' -\lambda_f | l 0 \rangle (-1)^{\lambda_i-\lambda_f} P_l(\cos \theta) \,,
\label{ec:ampsq}
\end{eqnarray}
with $\langle j_1 m_1 j_2 m_2 | j m \rangle$ Clebsch-Gordan coefficients and $P_l$ a Legendre polynomial of degree $l$. The interference between top helicity states can be ignored as long as one is only interested in quantities that are independent of the $W$ boson azimuthal angle in the top quark rest frame (see for example~\onlinecite{AguilarSaavedra:2012xe}), as is the case for all observables involved in the $2 \to 2$ process, in particular $\afb$. (Note, however, that the charged lepton rapidity in the laboratory frame is {\it not} one of such observables.)

One can gain further insight into the asymmetry~(\ref{ec:afb2}) setting $J=J'=1$, as in the SM at the tree level. Then, the Clebsch-Gordan coefficients in Eq.~(\ref{ec:ampsq}) imply that only $l=0,1,2$ contribute, and the corresponding Legendre polynomials are $P_0(x)=1$, $P_1(x)=x$, $P_2(x)=(3x^2-1)/2$. Among them, only $P_1$ produces a FB asymmetry. Moreover, for $l=1$ the symmetry properties of Clebsch-Gordan coefficients imply that only $\lambda_i,\lambda_f\neq 0$ contribute to $\afb$, and
\begin{eqnarray}
\afb & \propto & \left[ |a_{\oh -\oh \oh -\oh}^1|^2 + |a_{-\oh \oh -\oh \oh}^1|^2  -  |a_{\oh -\oh -\oh \oh}^1|^2 \right. \notag \\
& & \left. -  |a_{-\oh \oh \oh -\oh}^1|^2 \right]  \,.
\label{afbamp}
\end{eqnarray}
$C$ (or parity $P$) invariance implies
\begin{align}
& |a_{\oh -\oh \oh -\oh}^1|^2 = |a_{-\oh \oh -\oh \oh}^1|^2 \,, \notag \\
& |a_{\oh -\oh -\oh \oh}^1|^2 = |a_{-\oh \oh \oh -\oh}^1|^2 \,,
\end{align}
so it is clear that the so-called ``charge asymmetry'' $\afb$ does not entail a violation of the charge conjugation symmetry $C$. On the other hand, at the tree-level in QCD the modulus squared amplitudes are invariant under the exchange of the $t$ and $\bar t$ momenta, keeping the initial quarks fixed,
\begin{align}
& |a_{\oh -\oh \oh -\oh}^1|^2 = |a_{\oh -\oh -\oh \oh}^1|^2 \,, \notag \\
& |a_{-\oh \oh -\oh \oh}^1|^2 = |a_{-\oh \oh \oh -\oh}^1|^2 \,,
\end{align}
implying that the right-hand side of Eq.~(\ref{afbamp}) vanishes.

Dropping the $J=1$ restriction, we see from Eq.~(\ref{ec:ampsq}) that in full generality the partonic differential cross section can be expanded in terms of Legendre polynomials,
\begin{equation}
\frac{d\hat \sigma}{d\Omega} = \sum_l a_l P_l (\cos \theta) \,.
\label{ec:Legendre_expansion}
\end{equation}
The coefficients $a_l$ of this expansion are called ``Legendre momenta'' and the first ones have been measured by the CDF Collaboration~\cite{CDF:2013gna}, finding an excess in $a_1$, precisely the coefficient of the lowest-order polynomial generating an asymmetry.

\subsection{Asymmetries at the LHC versus Tevatron}
\label{sec:2c}

In $\ppb$ collisions at the Tevatron, an asymmetry (\ref{ec:afb2}) in the angle between the top and initial quark induces the asymmetry (\ref{ec:afb1}) between the top and proton directions precisely because the proton and quark directions almost always coincide. At the LHC, the two colliding hadrons are protons, and an asymmetry such as (\ref{ec:afb1}) vanishes. Nevertheless, since valence quarks have on average a larger momentum fraction than sea antiquarks, a forward top quark (with respect to the quark direction) in the $\ttb$ CM frame has in average a larger $|y|$ in the laboratory frame---with $y$ of either sign---than the backward antiquark. Hence, the asymmetry (\ref{ec:ac}) is well suited to probe a partonic asymmetry in the direction between the top and initial quark. Other asymmetries~\cite{Hewett:2011wz,Kuhn:2011ri} are numerically different but based on the same idea.

The SM NLO predictions for the LHC asymmetry, including electroweak contributions and taking the denominator at LO are, at 7 TeV and 8 TeV respectively, $\ac = 0.0115 \pm 0.0006$, $\ac = 0.0102 \pm 0.0005$~\cite{Kuhn:2011ri}, and $\ac = 0.0123 \pm 0.0005$, $\ac = 0.0111 \pm 0.0005$~\cite{Bernreuther:2012sx}. (NNLO predictions are not yet available.) The contributions of different subprocesses are given in Table~\ref{tab:afb-lhc}, for a CM energy of 8 TeV. 
\begin{table}[htb]
\begin{center}
 \caption{The  contributions to the numerator $\Delta N$ of the asymmetry~(\ref{ec:ac})
  at NLO for three different scales. The total LO cross sections and the resulting asymmetries are also given. From~\textcite{Bernreuther:2012sx}.\label{tab:afb-lhc}}
\begin{tabular}{ccccc}
\hline
\hline
\multicolumn{1}{c}{$\Delta N$~(pb)} &  & $\mu={m_{t}}/{2}$  & $\mu=m_{t}$  & $\mu=2m_{t}$ \\
\hline 
$O(\alpha_{s}^{3})$ 
     & $q\bar{q}$  & $1.7887$ & $1.2914$  & $0.9567$ \\
     & $qg$        & $0.1162$ & $0.0813$  & $0.0564$ \\
$O(\alpha^{2})$
     & $q\bar{q}$  & $5.13 \times 10^{-2}$ & $4.97 \times 10^{-2}$  & $5.21 \times 10^{-2}$ \\
$O(\alpha\alpha_{s}^{2})_\text{weak}$
     & $q\bar{q}$  & $7.8 \times 10^{-3}$  & $6.3 \times 10^{-3}$   & $5.0 \times 10^{-3}$ \\
     & $qg$        & $2.13 \times 10^{-2}$ & $-1.58 \times 10^{-2}$  & $-4.48 \times 10^{-2}$ \\
$O(\alpha\alpha_{s}^{2})_\text{QED}$
     & $q\bar{q}$  & $0.2020$ & $0.1616$  & $0.1838$ \\
     & $qg$        & $7.2   \times 10^{-3}$ & $5.4 \times 10^{-3}$  & $4.1 \times 10^{-3}$ \\
\multicolumn{1}{c}{total $\Delta N$} &  & $2.1945$ & $1.5799$  & $1.2113$ \\
\multicolumn{1}{c}{$\sigma_\text{QCD}^\text{LO}$~(pb)} &  & $190.77$ & $142.94$ & $113.21$ \\
\multicolumn{1}{c}{$\afb$~(\%)} &  & $1.15$ & $1.11$ & $1.07$ \\
\hline
\hline
\end{tabular}
\end{center}
\end{table}
The SM prediction for $\ac$ is one order of magnitude smaller than for $\afb$ owing to two effects. First, $gg$ fusion is dominant at the LHC, with 80\% of the total cross section at these CM energies, and it does not produce any asymmetry but washes out the one produced in $\qqb$ annihilation. Second, the probability that the antiquark has larger momentum fraction than the quark---in which case a forward top has smaller $|y|$ and contributes negatively to $\ac$---is not negligible, and leads to a further dilution of the generated asymmetry. Note also that at the LHC the $qg$ processes are not suppressed as they are at the Tevatron, but the asymmetry they generate is small.

\begin{figure}[t]
\begin{center}
\begin{tabular}{c}
\includegraphics[height=5.5cm,clip=]{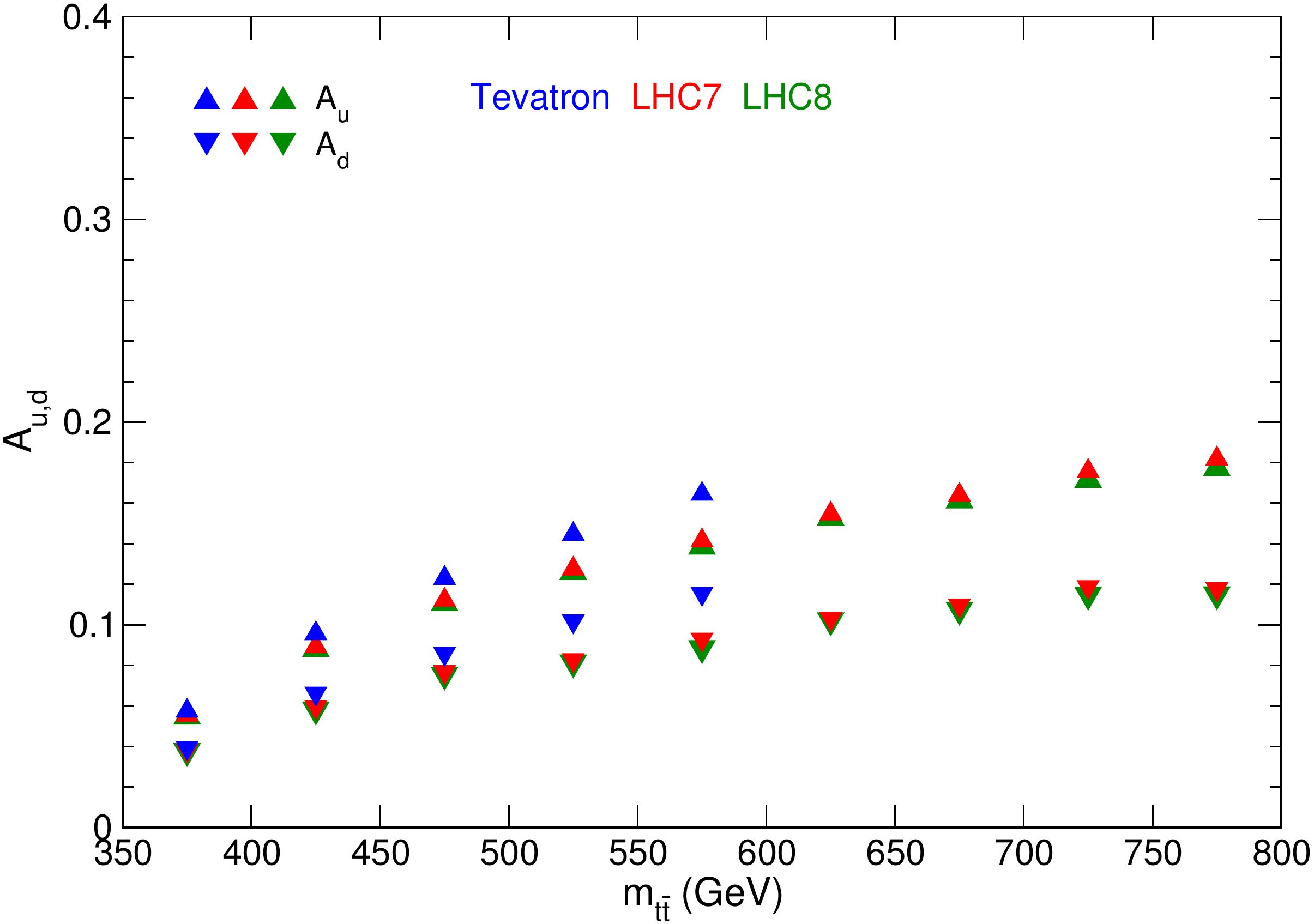} \\[3mm]
\includegraphics[height=5.5cm,clip=]{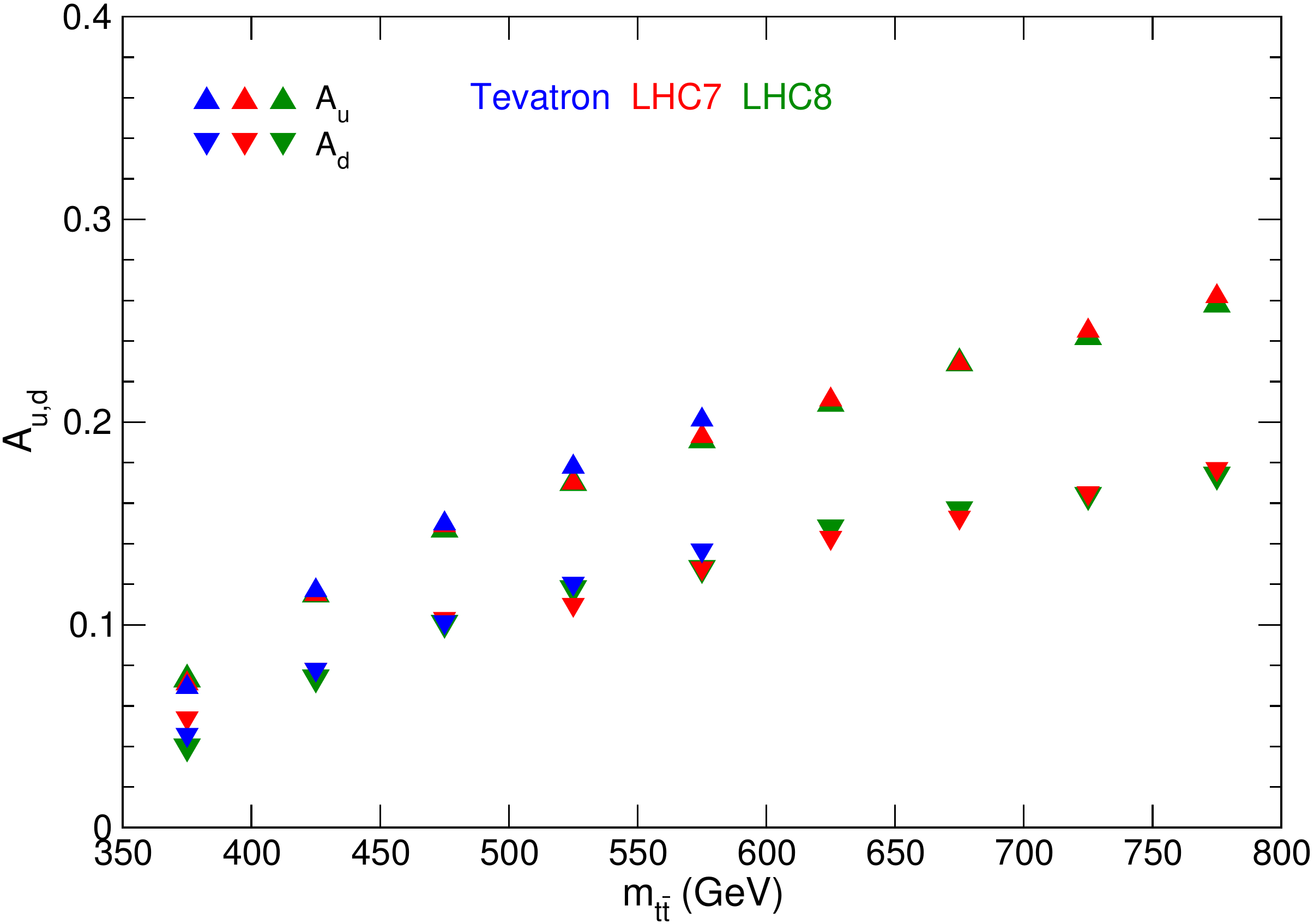}
\end{tabular}
\end{center}
\caption{SM predictions for the collider-independent asymmetries $A_u$, $A_d$ at the Tevatron and the LHC, without $\pTttb$ cut (up) and with $\pTttb < 30$ GeV (down). From~\textcite{AguilarSaavedra:2012rx}.}
\label{fig:2C0}
\end{figure}

It is clear that the asymmetries (\ref{ec:afb1}), (\ref{ec:ac}) are different observables, hence a measured value of the latter consistent with the SM prediction does not preclude an anomaly in the former. In fact, the relation between them is model-dependent~\cite{AguilarSaavedra:2011hz}. Still, $\afb$ and $\ac$ arise from asymmetries in $\qqb$ annihilation that, for a fixed partonic CM energy $\hat s$, are the same at the two colliders. One can write~\cite{AguilarSaavedra:2012va}
\begin{eqnarray}
\afb & = & \au F_u + \ad F_d \,, \notag \\
\ac & = & \au F_u D_u + \ad F_d D_d \,,
\label{ec:AuAd}
\end{eqnarray}
where $A_{u,d}$ are the ``intrinsic'' asymmetries in the partonic processes $\uub \to \ttb X$, $\ddb \to \ttb X$, respectively, $F_{u,d} = \sigma_{\uub, \ddb}/\sigma$ are the $\uub$ and $\ddb$ fractions of the cross section, and $D_{u,d}$, dilution factors arising from the already mentioned fact that sometimes the antiquark has larger momentum fraction than the quark. To a good approximation, $\au$ and $\ad$ are the same at the Tevatron and the LHC, provided one restricts $\mttb$ to a narrow interval, hence their labelling as ``collider-independent''. The SM NLO calculations of $A_{u,d}$ including electroweak corrections are presented in Fig.~\ref{fig:2C0} (upper panel), using $\mttb$ bins of 50 GeV up to 800 GeV. 
The differences beween the Tevatron and LHC values are already smaller than the expected statistical uncertainties, but can be further reduced by applying an upper cut on $\pTttb$ (lower panel).
As yet, there are not any measurements of $A_{u,d}$ neither at the Tevatron nor at the LHC. They could be measured from the two-dimensional distribution of $\afb$ or $\ac$ as a function of $\mttb$ and the velocity of the $\ttb$ pair
\begin{equation}
\betattb = \frac{|p_z^t + p_z^{\bar t}|}{E_t + E_{\bar t}} \,,
\end{equation}
using Eqs.~(\ref{ec:AuAd}) and exploiting the fact that for fixed $\mttb$, $A_{u,d}$ are almost independent of $\betattb$ while $F_{u,d}$ and $D_{u,d}$ are not. This is a demanding but revealing measurement. Being basically the same quantities at the two colliders, the measurement of $A_{u,d}$ at the LHC is a unique direct test of the Tevatron anomaly. Furthermore, measurements of $A_{u,d}$ at the Tevatron and the LHC could be combined for a more precise determination of the two partonic asymmetries.

\begin{figure}[htb]
\begin{center}
\includegraphics[height=7cm,clip=]{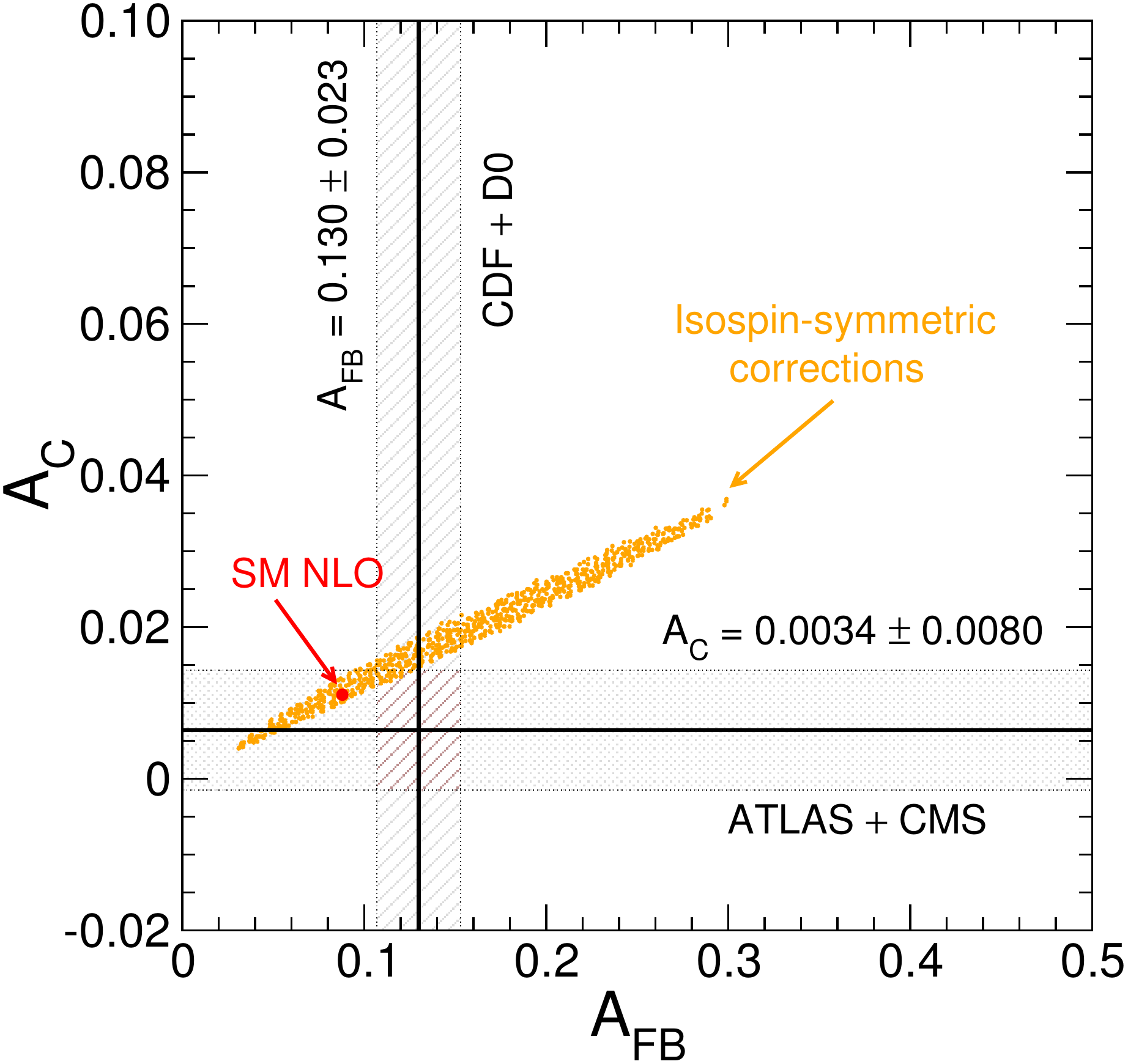}
\end{center}
\caption{Relation between $\ac$ and $\afb$ resulting from random isospin-symmetric variations of $A_{u,d}$ around their SM value, as described in the text. Also shown are the simple error-weighted averages of $\afb$ measurements at the Tevatron (see Sect.~\ref{sec:3d}) 
and of $\ac$ measurements at the LHC at $\sqrt{s}=7$ TeV (see Sect.~\ref{sec:4c}).}
\label{fig:2C1}
\end{figure}

In this context, it is worthwhile pointing out that isospin-symmetric corrections to the SM values of $A_{u,d}$ shift $\afb$ and $\ac$ in the same direction. Figure~\ref{fig:2C1} shows the asymmetries $\afb$ and $\ac$ resulting from random variations of $A_{u,d}$ between $1/4$ and $4$ times their SM NLO values, fixing $F_{u,d}$ and $D_{u,d}$ as in the SM---which is a reasonable approximation given the good agreement of various differential distributions with data. (The random variations are done independently in each bin of $\mttb$.) An increase in the $\afb$ prediction to fit the Tevatron average\footnote{In the absence of official combinations, the quoted Tevatron and LHC averages are weighted averages of the relevant measurements, detailed in Sec.~\ref{sec:3} and~\ref{sec:4}, respectively.} also increases $\ac$ leading to a $1\sigma$ deviation, which reaches almost $2\sigma$ if one wants to reproduce the CDF measurement $\afb = 0.164 \pm 0.045$.

\begin{figure}[htb]
\begin{center}
\includegraphics[width=7.5cm,clip=]{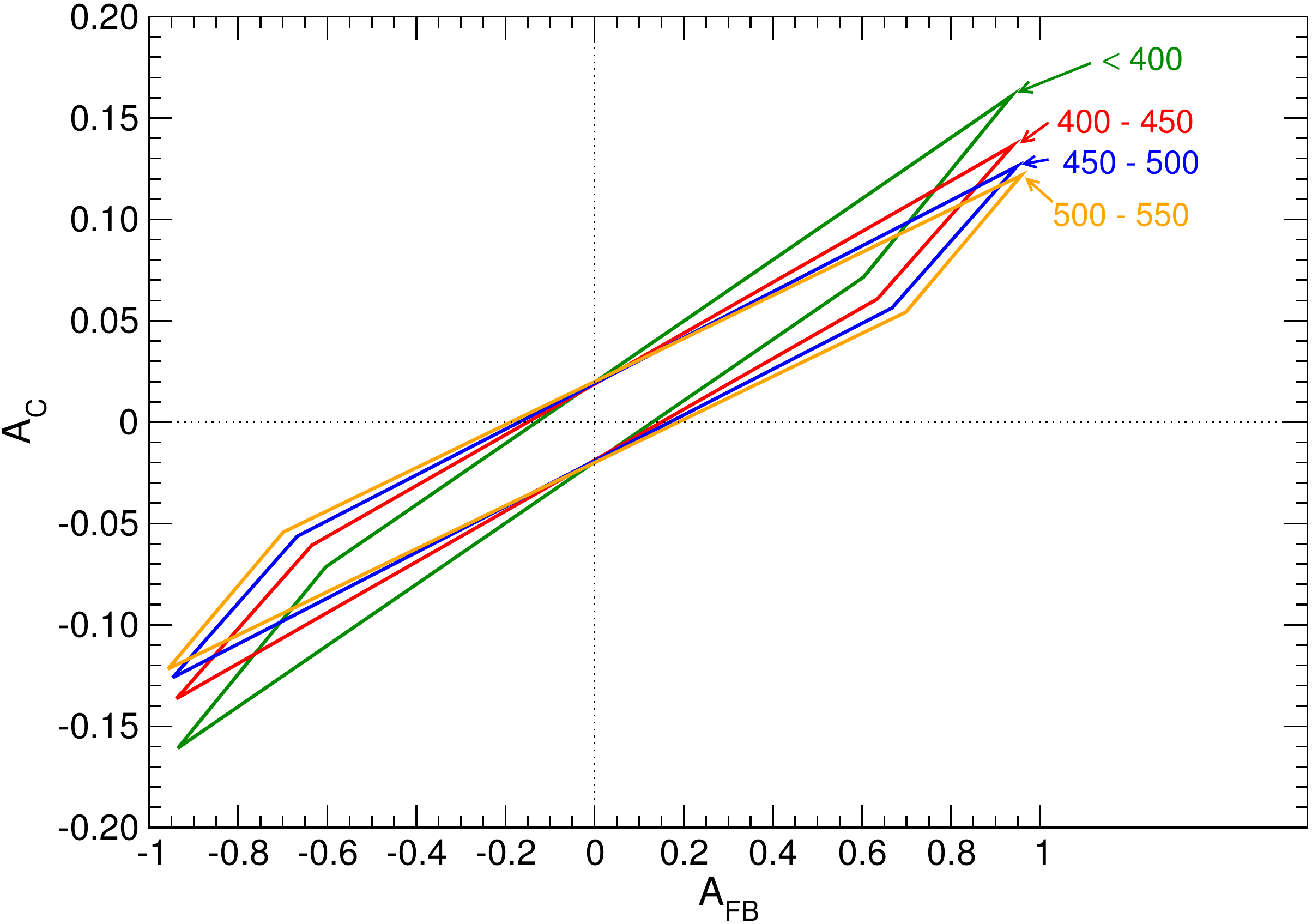}
\end{center}
\caption{Allowed asymmetries $\ac$ and $\afb$ in several $\mttb$ bins (in GeV), which result from arbitrary $A_{u,d}$ within these bins. From~\textcite{AguilarSaavedra:2012va}.}
\label{fig:2C2}
\end{figure}

Conversely, isospin-breaking corrections to $A_{u,d}$ can generate a positive contribution to $\afb$ with small or vanishing contribution to $\ac$. This is illustrated by Fig.~\ref{fig:2C2}, where $A_{u,d}$ are left completely arbitrary within each $\mttb$ bin and $F_{u,d}$, $D_{u,d}$ are fixed to their SM values. A positive asymmetry $\afb > 0$ is compatible with a vanishing or negative $\ac$ provided $A_u$ and $A_d$ have a different sign. The implementation of this condition in actual models is discussed at the end of Sec.~\ref{sec:5}.

\subsection{Leptonic asymmetries}
\label{sec:2d}

In addition to the $\ttb$-based asymmetries, the Tevatron experiments measure asymmetries based on the rapidities of the charged leptons from the top quark decay,
\begin{eqnarray}
\afbl & = & \frac{N(q_\ell y_\ell >0) - N(q_\ell y_\ell <0)}{N(q_\ell y_\ell >0) + N(q_\ell y_\ell <0)} \,, \notag \\
\afbll & = & \frac{N(\Delta y_\ell >0) - N(\Delta y_\ell <0)}{N(\Delta y_\ell >0) + N(\Delta y_\ell <0)} \,,
\end{eqnarray}
with $q_\ell$ the lepton charge and $\Delta y_\ell = y_{\ell^+} - y_{\ell^-}$. The former can be measured in the $\ell+$jets or dilepton decays of the $\ttb$ pair, whereas the latter requires the two charged leptons and is measured only in the dilepton channel. (The $\ell+$jets and dilepton decay modes of the $\ttb$ pair are those in which one or two charged leptons, respectively, are produced from the decay of the two $W$ bosons.)
Within the SM, these asymmetries are generated from the $\ttb$ asymmetry $\afb$, given the fact that the top \mbox{(anti-)quarks} are produced with zero polarization in the production plane, that is, the plane spanned by the top and initial quark momenta in the $\ttb$ CM frame. The SM predictions at NLO are $\afbl = 0.038 \pm 0.003$, $\afbll = 0.048 \pm 0.004$~\cite{Bernreuther:2012sx}, using LO denominators.

In general the leptonic asymmetries and $\afb$ are independent observables in much the same way as $\afb$ and $\ac$ are. This fact is clear when one considers the threshold behavior of $\qqb \to \ttb$ and the possible effect of new physics~\cite{Falkowski:2011zr}. At the threshold, $\ttb$ pairs produced from initial $q_R \bar q_R$ states have their spins aligned in the proton direction, independently of $\theta$. The top decay dynamics makes the positive charge lepton tend to follow the top spin direction, so it is preferentially emitted with $y_{\ell^+} > 0$. The negative charge lepton from the top decay tends to be emitted opposite to the top spin, so $y_{\ell^-} < 0$. For initial $q_L \bar q_L$ states the behavior is the opposite, and the charged leptons preferentially have $y_{\ell^+} < 0$ and $y_{\ell^-} > 0$. For equal $q_R \bar q_R$ and $q_L \bar q_L$ cross sections, as when produced by QCD interactions, the two effects cancel and the asymmetries vanish at threshold. But any excess in $q_R \bar q_R$---or decrease in $q_L \bar q_L$ via interference---caused by new physics will originate positive leptonic asymmetries, independently of $\afb$. Conversely, a decrease in $q_R \bar q_R$ or an excess of $q_L \bar q_L$ will generate negative asymmetries.

At the LHC, a leptonic asymmetry
\begin{equation}
\acll = \frac{N(\Delta |y_\ell| >0) - N(\Delta |y_\ell| <0)}{N(\Delta |y_\ell| >0) + N(\Delta |y_\ell| <0)}
\label{ec:acll}
\end{equation}
has also been measured in the dilepton decay mode of the $\ttb$ pair. The SM predictions at NLO are $\acll = 0.0070 \pm 0.0003$ for 7 TeV and $\acll = 0.0064 \pm 0.0003$ for 8 TeV~\cite{Bernreuther:2012sx}.


\section{Experimental measurements at the Tevatron}
\label{sec:3}

Interest in the $\ttbar$ asymmetry was sparked by  papers from the CDF and D0 Collaborations in 2008~\cite{Abazov:2007ab,Aaltonen:2008hc} where, in small initial samples from Tevatron Run 2, both experiments observed large $\dy$ asymmetries in $\ttbar$ events in the  $\ell$+jets decay mode. Follow up measurements with roughly $5~\ifb$ again showed large asymmetries in both experiments, with a significant dependence on $\mttb$ at CDF~\cite{Abazov:2011rq, Aaltonen:2011kc, cdfdilprelim}.  Subsequent studies  examined the differential behavior of the asymmetry in rapidity and $\ttbar$ invariant mass,  the differential cross-section in the scattering angle, and the asymmetry in the isolated top decay leptons, and also expanded the asymmetry measurements into the dilepton decay mode. All of these measurements are grounded in the techniques used to measure the $\ttbar$ cross section which is the denominator of the asymmetry. The combined Tevatron cross section for $\ttbar$ production in $\ppbar$ collisions at 1.96 TeV is $\sigma_{\ttbar} = 7.60\pm 0.41$~pb for $m_t = 172.5~\gev$~\cite{Aaltonen:2013wca}. 

The Tevatron asymmetry measurements rely on using a well-measured lepton from the decay chain $t\rightarrow Wb\rightarrow \ell\nu b$ to measure the top quark charge. We review here asymmetry results in both $\ell$+jets  and dilepton channels using the full Tevatron Run 2 data set, as reported by both the CDF and D0 experiments. 

\subsection{Inclusive asymmetry in the $\ell$+jets mode}
\label{sec:3a}

When one and only one top quark decays leptonically, the $\ttbar$ final state contains a lepton, missing transverse energy $\met$, and four hadronic jets, two of which are initiated by $b$ quarks. These $\ell+$jets events contain sufficient information to completely reconstruct the $\ttbar$ 4-vectors and the electric charges of the top quarks. The samples are selected requiring a central (here referring to pseudo-rapidity) isolated electron or muon, with  $\pt > 20~\gev$, missing transverse energy $\met > 20~\gev$, and at least 3 (D0) or 4 (CDF) jets with $\pt > 20~\gev$. Decay channels through $\tau$ leptons are not included, although there is a small leakage of $\tau$ events from the $W\to\tau\to e/\mu$ decay chain. The presence of final state $b$-jets is confirmed using information on reconstructed displaced secondary vertices and tracks with significant impact parameter with respect to the hard-scatter primary vertex.  The total efficiency of the selection (including the leptonic branching ratios) is $\sim 3\%$. The non-$\ttbar$ backgrounds in this selection are dominated by $W$+jets events with heavy flavor or incorrectly $b$-tagged jets, plus small contributions from QCD multijets and electroweak processes. These backgrounds are modeled using Monte Carlo generators and detailed detector simulation, except for the pure QCD multijets component, which is derived from data sidebands.  

The most probable 4-vectors for the $\ttbar$ production hypothesis are derived for each event.  Subtraction of the background processes yields the distribution of $\dy$ for the reconstructed top quarks, which has been distorted by acceptance losses and resolution smearing. These distortions can be estimated from study of the simulated NLO signal model and used to construct a regularized linear transformation that deconvolves (or unfolds) the true production-level distribution from the reconstructed one.  The reliability of the unfolding procedure is checked using simulated  $\ttbar$ samples with a variety of $\afb$ models.

\subsubsection{CDF}\label{sec:cdfljets}
\label{sec:3a1}
The CDF Collaboration measured the asymmetry with $9.4~\ifb$~\cite{Aaltonen:2012it}. The selection uses the four leading hadronic jets,  and also requires a transverse energy sum of all objects $H_T> 200$~GeV, giving 2653 candidate events. The sample composition is found using a  detailed accounting of the $b$-tagging rate in all expected processes; the non-$\ttbar$ backgrounds total $530\pm 124$ events.  In the 4-jet sample, the $\ttbar$ 4-vectors are reconstructed using constraints on the $W$ and top masses, varying the jet energies within their expected resolutions, and choosing the jet-parton combination with the lowest $\chi^2$.  The $\dy$ distribution is calculated and background shapes, estimated from simulated samples and data side-bands, are subtracted. The asymmetry at the reconstruction level is found to be $0.087\pm 0.026$. 

\begin{figure}[htb]
\begin{center}
\begin{tabular}{r}
\includegraphics[width=8cm,clip=]{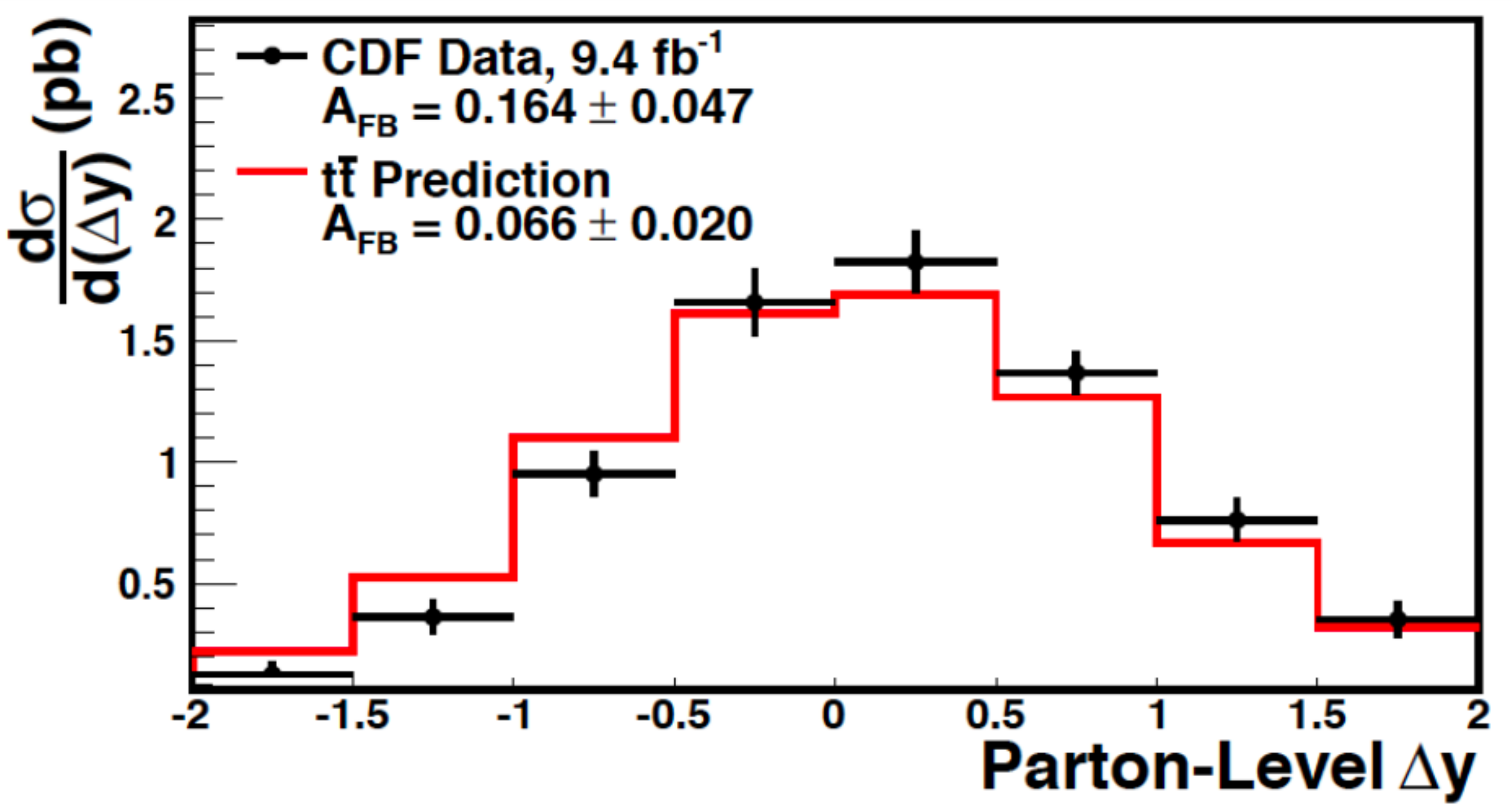} \\[3mm]
\includegraphics[width=8.2cm,clip=]{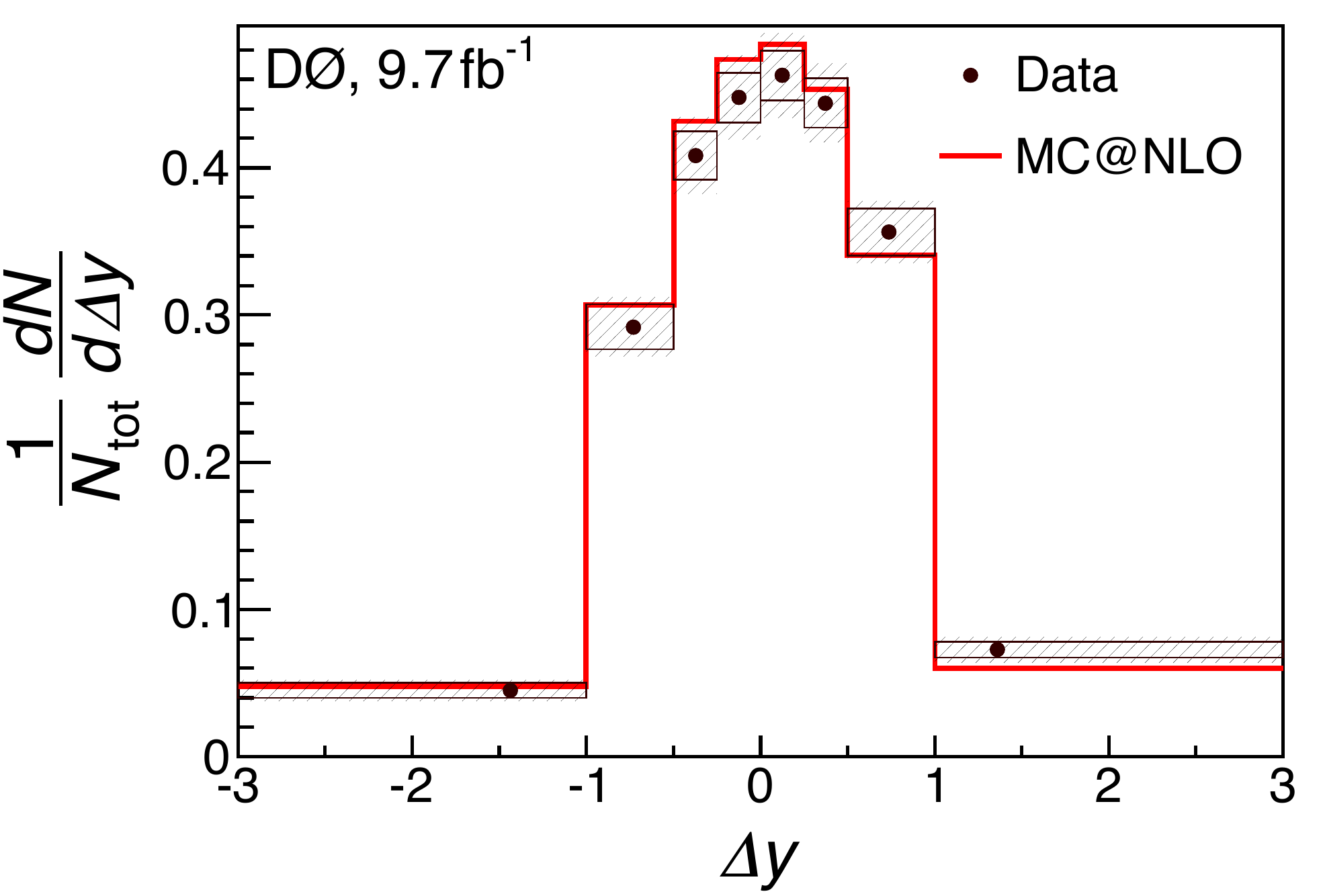}
\end{tabular}
\end{center}
\caption{Production-level $\dy$ distributions. Top: At CDF, where the error bars represent the total statistical and systematic uncertainty in each bin. Bottom: At D0, where the statistical uncertainty is given by the black rectangles and the hashed area represents the total uncertainty including systematic effects. From~\textcite{Aaltonen:2012it,Abazov:2014cca}.} \label{fig:inc_dy}
\end{figure}

The $\ttbar$ signal is modeled using the NLO generator {\sc powheg}~\cite{Frixione:2007nw}. Parton showers are added by  {\sc pythia}~\cite{ Sjostrand:2006za}, and the result is run through a full detector simulation. The electroweak contribution is included by rescaling the asymmetric parts of $\dy$ by an additional factor of 1.26 (see Sec.~\ref{sec:2a}). The one-dimensional $\dy$ distribution is unfolded in eight bins using a response matrix based on {\sc powheg}; the production-level distribution of $\dy$ is given in the top plot in Fig.~\ref{fig:inc_dy}. The inclusive asymmetry is $\afb = 0.164\pm 0.039\,{\rm (stat)}\pm 0.026\,{\rm (syst)}$. The systematic uncertainty of the measurement is dominated by the background modeling and normalization. 

\subsubsection{D0}\label{sec:d0ljets}
\label{sec:3a2} 
The D0 Collaboration has measured the asymmetry in $9.7~\ifb$~\cite{Abazov:2014cca}. Events are selected in both the exclusive 3-jet and inclusive 4-jet mode with the additional requirement that the leading jet has $\pt\geq 40$~GeV.  The $b$-tagging uses a multivariate analysis (MVA)  of the jet fragmentation and track impact parameter information. The $\ttbar$ signal is modeled with {\sc mc@nlo} and showered with {\sc herwig}~\cite{Corcella:2000bw}. Kinematic reconstruction in the 4-jet sample is weighted over jet energy transfer functions and $b$-tagging likelihoods.  Partial reconstruction in the 3-jet sample uses a MVA with kinematic variables to find the most probable reconstruction of the $\ttbar$ system assuming the lost jet is from the hadronically decaying top quark. In both samples the jet-parton combination yielding the highest likelihood value is selected to reconstruct the 4-momenta of the top and antitop quarks. The sample is divided into six channels according to jet and $b$-tag multiplicity, and the composition of each is found using multivariate discriminants. The total sample is 10947 events with an estimated background of $6202\pm 78$. In the 4-jet subsample comparable to the CDF analysis there are 2875 events. The asymmetry in the $W$+jets background sample is re-weighted according to that observed in the background-dominated 3-jet 0-tag sample. After background subtraction the inclusive asymmetry at the reconstruction level is found to be $0.079\pm 0.027$. An unfold in 26/50 bins of true/reconstructed $\dy$, using the {\sc mc@nlo} response model gives the production-level distribution shown in the bottom of Fig.~\ref{fig:inc_dy}. The inclusive asymmetry is $\afb = 0.106\pm 0.030$. The systematic part of the uncertainty is dominated by the background modeling. 

Except for very small effects from common assumptions concerning PDF's, the D0 and CDF results are uncorrelated. In this case it is possible to combine the measurements using a simple error-weighted average, yielding $\afb = 0.124\pm 0.025$.  

\subsection{Kinematic dependence of the asymmetry}
\label{sec:3b}

As the $\ttbar$ cross section is a function of the scattering angle, momentum transfer, and $\ttbar$ transverse momentum, it is interesting to explore the differential behavior of the asymmetry in these variables. 

\subsubsection{Rapidity difference}
\label{sec:3b1}
In the collider environment, the rapidity difference $\dy$ is the natural proxy for the scattering angle. The rapidity dependent asymmetry

\begin{equation}
\afb(\dy) = \frac{N(+\dy) - N(-\dy)}{N(+\dy) + N(-\dy)},
\label{afbvdy_data}
\end{equation}

\noindent  follows directly from the information in Fig.~\ref{fig:inc_dy}. The results from both experiments are compared to the NLO prediction in the top panel of Fig.~\ref{fig:diffafbs}.  In both cases, the NLO model is well fit with a simple linear form. Since the $\dy$ distribution is continuous at $\dy = 0$, the intercept must be consistent with zero. The measurements of both experiments are in good agreement with the linear form, with slopes shown in Table~\ref{tab:slopes}. The D0 measurement is $1.3\sigma$ above the prediction and the CDF measurement is $1.3\sigma$ above the D0 one.

\begin{table}
\begin{center}
\caption{Slope $\alpha$ of $\afb$ as a linear function of $\dy$ and $\mttb$. 
The predicted slopes have been estimated using the {\sc powheg} generator interfaced to {\sc pythia} (see Sect.~\ref{sec:cdfljets} for details).
Predictions and measurements have been extracted from~\textcite{Aaltonen:2012it,Abazov:2014cca}.}
\label{tab:slopes}
\begin{tabular}{c c c c}
\hline
\hline
          & $\alpha(\dy)$    & $\alpha(\mttb)$ \\
predicted & $0.097\pm 0.015$ & $(3.4\pm 1.2)\times 10^{-4}$ \\
CDF       & $0.253\pm 0.062$ & $(15.5\pm 4.8)\times 10^{-4}$ \\
D$0$      & $0.154\pm 0.043$ & $(3.9\pm 4.4)\times 10^{-4}$ \\  
\hline
\hline
\end{tabular}
\end{center}
\end{table}

\begin{figure}[htb]
\begin{center}
\begin{tabular}{c}
\includegraphics[width=8.5cm,clip=]{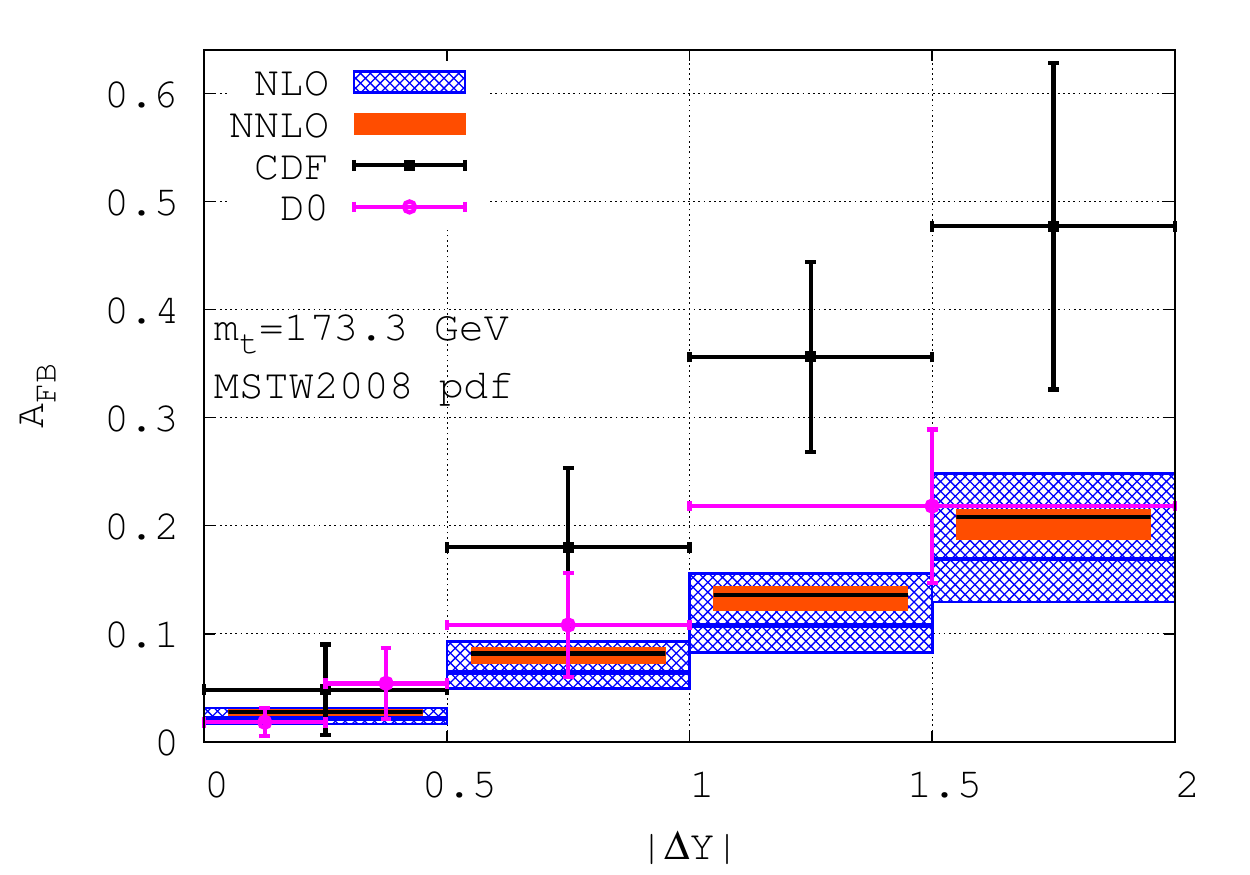} \\
\includegraphics[width=8.5cm,clip=]{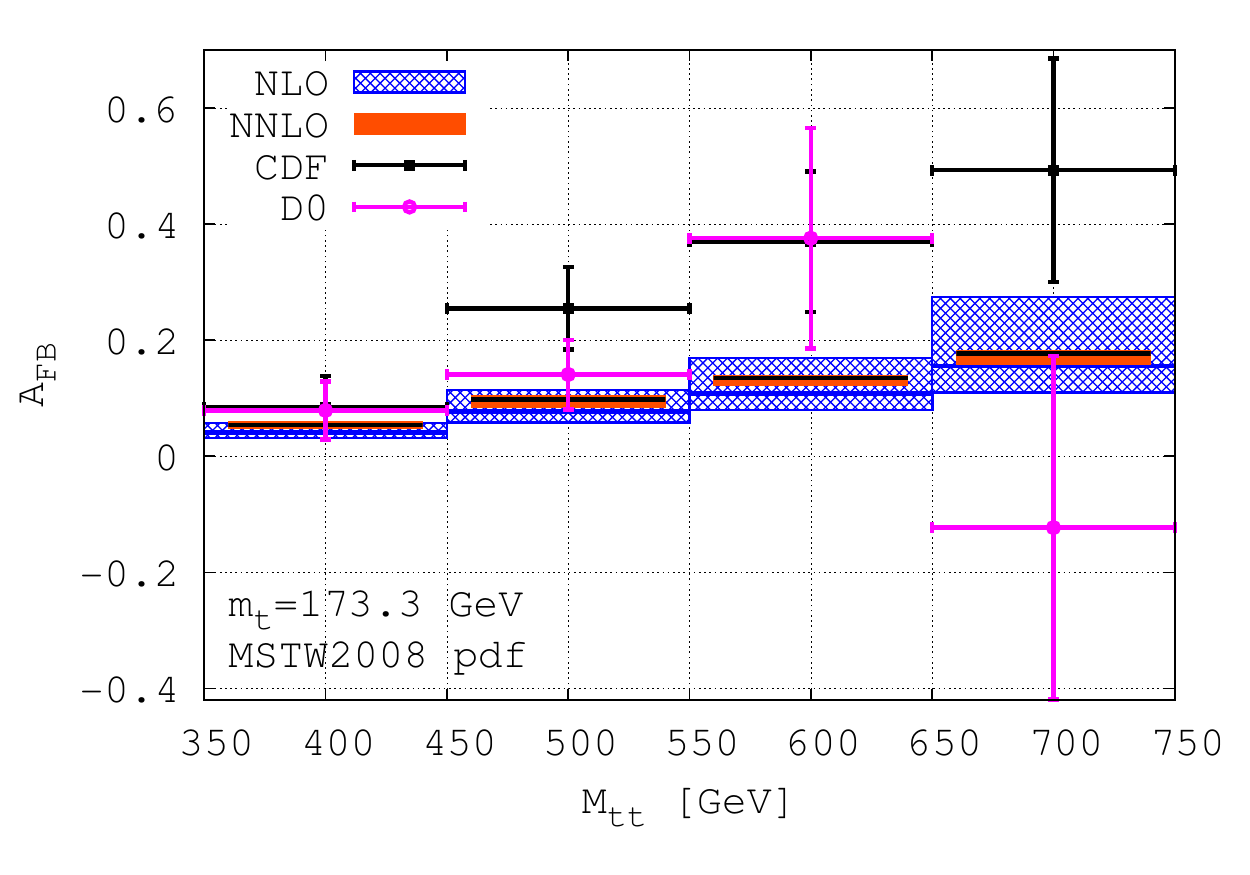}
\end{tabular}
\end{center}
\caption{Production-level dependence of the asymmetry on the rapidity difference (top) and $\ttb$ invariant mass (bottom) for both Tevatron experiments, compared to NLO and NNLO predictions. The horizontal error bars indicate the binning used in each experiment. The CDF measurements~\cite{Aaltonen:2012it} are squares and the D0 measurements~\cite{Abazov:2014cca} are circles. The SM predictions and their scale uncertainties are given by the horizontal lines and associated bands. From~\textcite{Czakon:2014xsa}.}
\label{fig:diffafbs}
\end{figure}

\subsubsection{$\ttb$ invariant mass}
\label{sec:3b2}
The $\hat{s}$ behavior of the asymmetry is measured in the variation of $\dy$ with the $\ttbar$ invariant mass $\mttb$.  A production level measurement requires a two-dimensional unfolding in the space of $\dy$ and $\mttb$. The CDF Collaboration uses  four bins in $\mttb$ and 2 bins in $\dy$; the latter choice confines the need for regularization to $\mttb$ only. The D0 Collaboration uses more granularity, with 26/50 bins in production/reconstruction level $\dy$, 6 bins in $\mttb$, and simultaneous regularization of both variables. The results for the two experiments are shown in Fig.~\ref{fig:diffafbs}, bottom, and Table~\ref{tab:slopes}.  Except at the highest mass, the two results are in modest agreement. The fitted slopes are more discordant, with a difference of $1.8\sigma$.  

\subsubsection{Production angle} \label{sec:cdfcos}

\noindent The reconstruction of the $\ttbar$ four vectors in the $\ell$+ jets mode allows a direct measurement of the differential cross-section in the $\ttbar$ scattering angle $d\hat{\sigma}/d\cos\theta$ (in the $\ttbar$ rest frame), as discussed in Sec.~\ref{sec:2b}.  The top plot of Fig.~\ref{fig:costh} shows the differential  cross section for two SM predictions and two representative new physics models.  The LO QCD prediction shows the characteristic $\sim 1+\cos^2\theta$ behavior, while the NLO curve shows the addition of a small approximately linear correction. A 1.2 TeV $s$-channel color octet with axial couplings shows a large linear correction. Alternatively, a 200 GeV flavor changing $t$-channel $Z^{\prime}$ shows a strong forward scattering component.  

The CDF Collaboration has performed the production angle measurement characterizing the the cross section as an expansion in Legendre polynomials according to Eq~(2.8)~\cite{CDF:2013gna}. The analysis uses the 4-jet sample of Sec.~\ref{sec:cdfljets}, augmented with 3-jet events with an additional soft jet having $\pt > 12$~GeV. A total of 3776 events are used, with estimated background $1026\pm 210$. The 4-jet reconstruction gives $\cos\theta$, and backgrounds are subtracted from the distribution. By discretizing in Legendre moments rather than histogram bins, the transfer matrix to the production level is well-conditioned, and the unfold can be done by simple inversion, avoiding regularization. The Legendre coefficients $a_l$ are shown on the bottom of Fig.~\ref{fig:costh}, normalized to the total cross section $a_0 = 1$. In the predictions (i) LO {\sc pythia} has $a_1=0$ and small $a_4$ from $gg$ initiated $t$-channel scattering; (ii)  NLO QCD has box and and radiative diagrams give corrections to all moments, including the asymmetry producing odd terms; (iii) the $s$-channel octet model adds a non-zero $a_1$ to the LO {\sc pythia} model; (iv) the $t$-channel model, with leading behavior $1/(1-\cos\theta)$, has large higher order Legendre terms.  The data suggests that the asymmetry arises in the linear term in $\cos\theta$, with coefficient $a_1 = 0.40\pm 0.09\,{\rm (stat)}\pm 0.08\,{\rm (syst)}$. The statistical precision is limited, but disfavors the large higher order moments characteristic of  $t$-channel models.

\label{sec:3b3}
\begin{figure}[htb]
\begin{center}
\begin{tabular}{c}
\includegraphics[width=8cm,clip=]{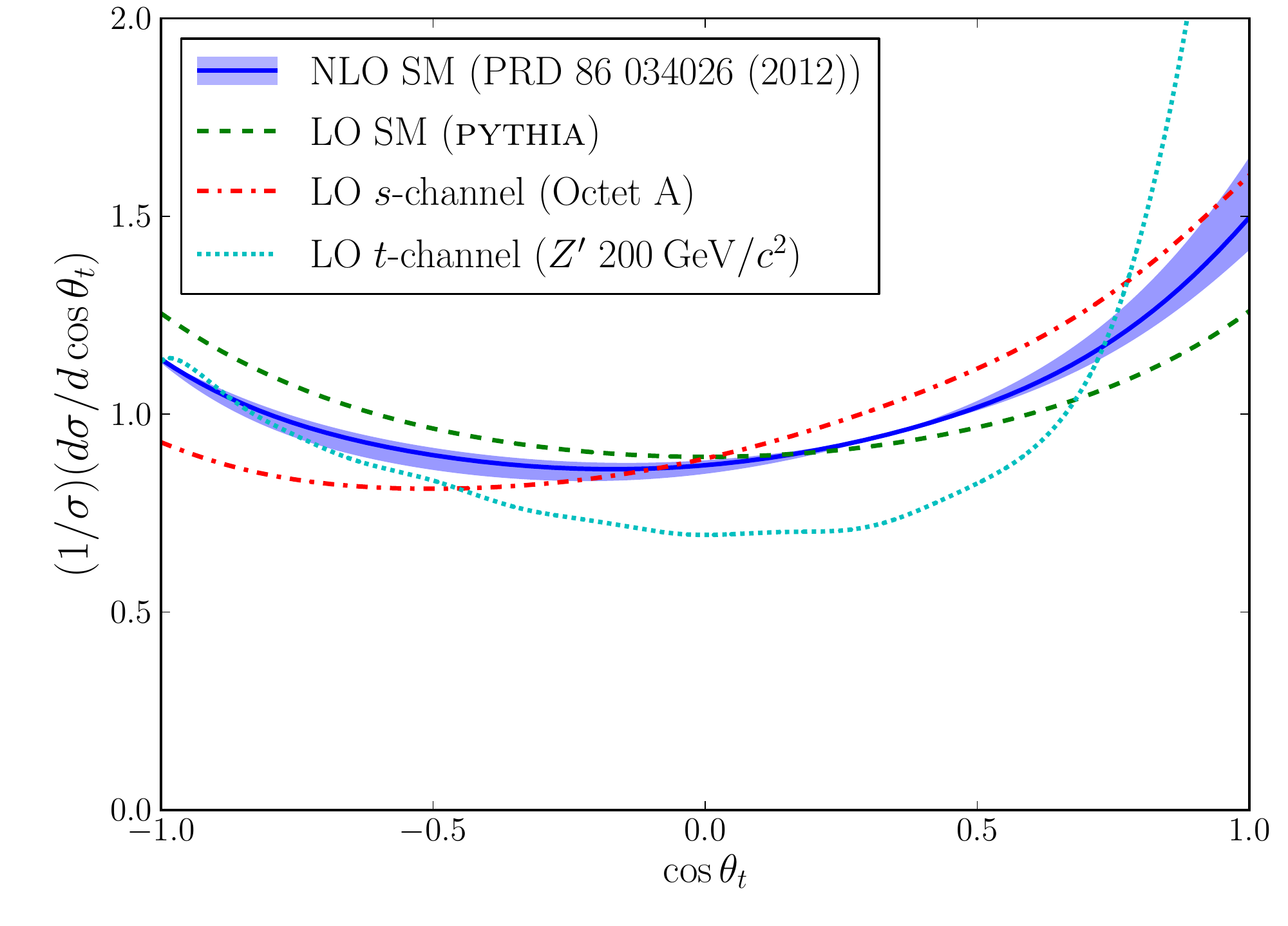} \\
\includegraphics[width=8cm,clip=]{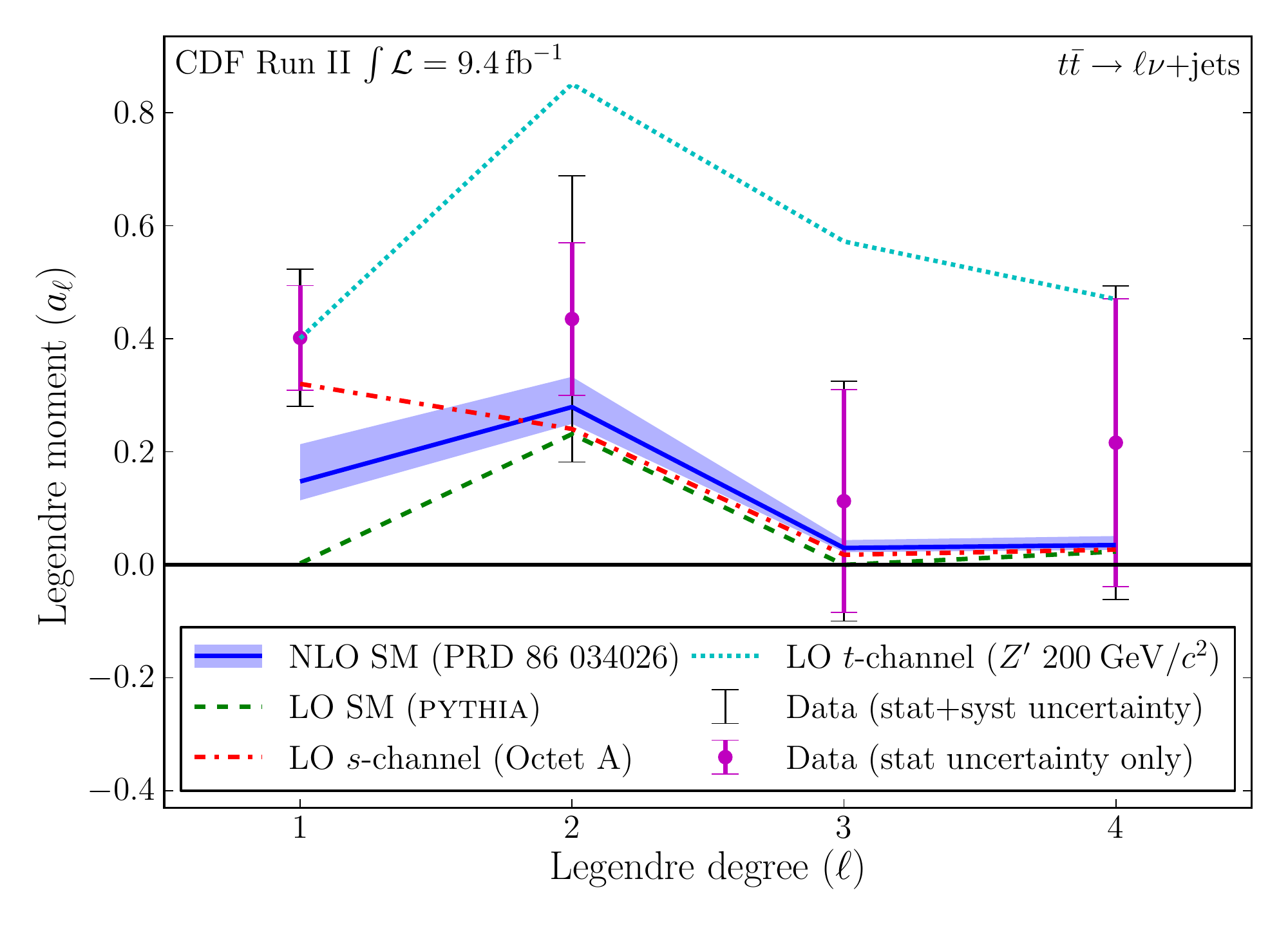}
\end{tabular}
\end{center}
\caption{Top: Angular cross-sections for NLO QCD and new physics models. Bottom: Legendre coefficients for data and models. From~\textcite{CDF:2013gna}.}
\label{fig:costh}\end{figure}

\subsubsection{$\ttb$ transverse momentum}
\label{sec:3b4}

As described in Sec.~\ref{sec:2a}, the NLO QCD asymmetry should have a strong dependence on the transverse momentum of the $\ttbar$ system, $\pt^{\ttbar}$.
\noindent The top plot in Fig.~\ref{fig:Avspt} shows the $\pt^{\ttbar}$-dependent asymmetry 

\begin{equation}
\afb(\pt^{\ttbar}) = \frac{N_F(\pt^{\ttbar}) - N_B(\pt^{\ttbar})}{N_F(\pt^{\ttbar}) + N_B(\pt^{\ttbar})}.
\label{afbvpt_data}
\end{equation}

\noindent for four different SM calculations at the production level. The strong color coherence effect is seen in {\sc pythia}; it is interesting that the average of this dependence over all $\pt^{\ttbar}$ gives the expected LO QCD result $\afb = 0$. The NLO QCD calculation {\sc mcfm} shows the positive Born-box asymmetry at $\pt^{\ttbar}=0$ and the negative initial/final state radiation (ISR/FSR) interference  asymmetry elsewhere. {\sc powheg} has the same NLO matrix elements, with higher order effects approximated by {\sc pythia} showering, smoothing the {\sc mcfm} form.  The curve called ``$\ttbar$+Jet" includes the higher order effects at large $\pt^{\ttbar}$ explicitly as jets\footnote{K.Melnikov, A. Scharf and M. Schulze, private communication.}, and is in good agreement with the showered {\sc powheg}. 

\begin{figure}[htb]
\begin{center}
\begin{tabular}{r}
\includegraphics[width=8cm,clip=]{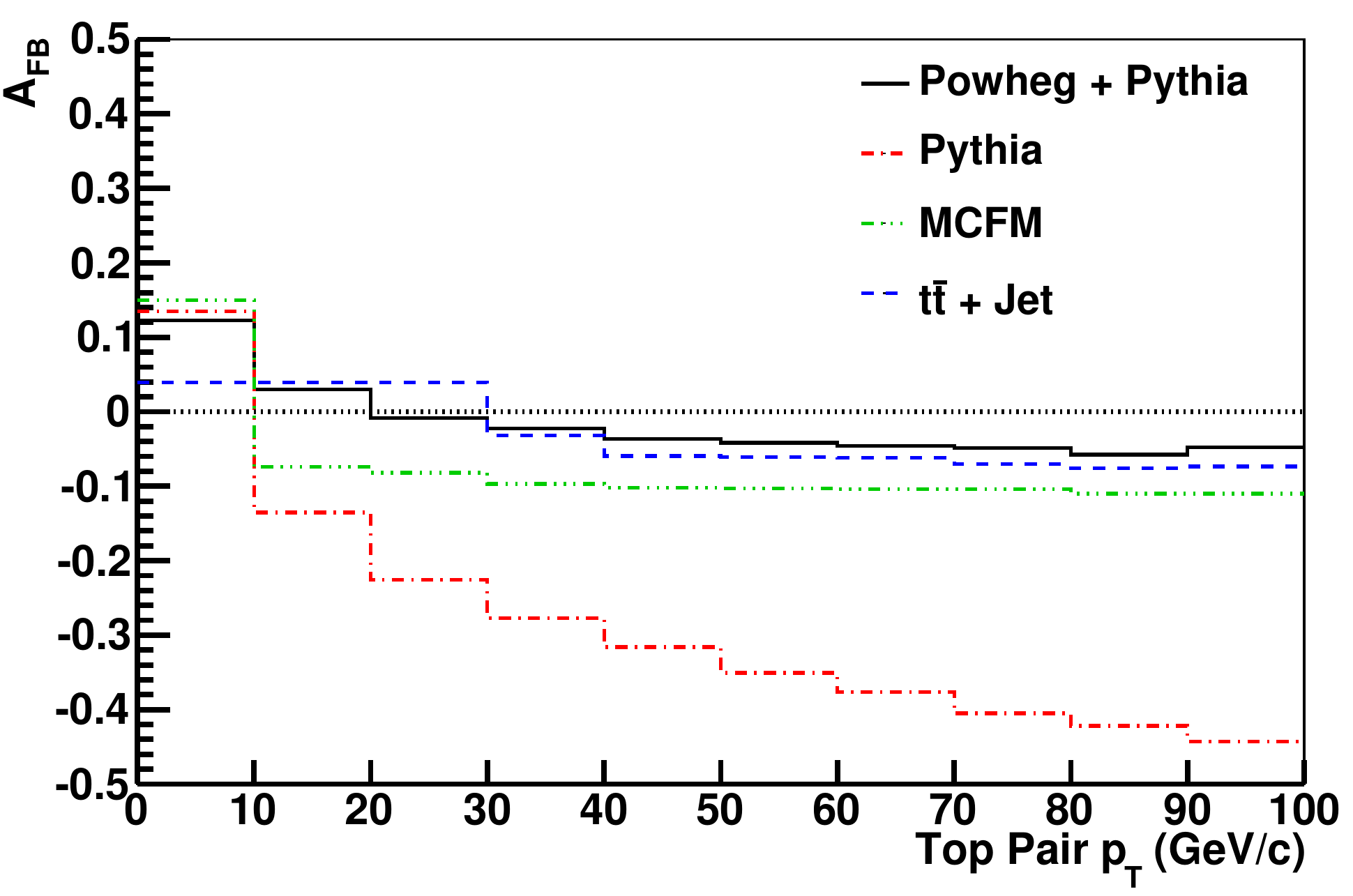} \\[3mm]
\includegraphics[width=8.5cm,clip=]{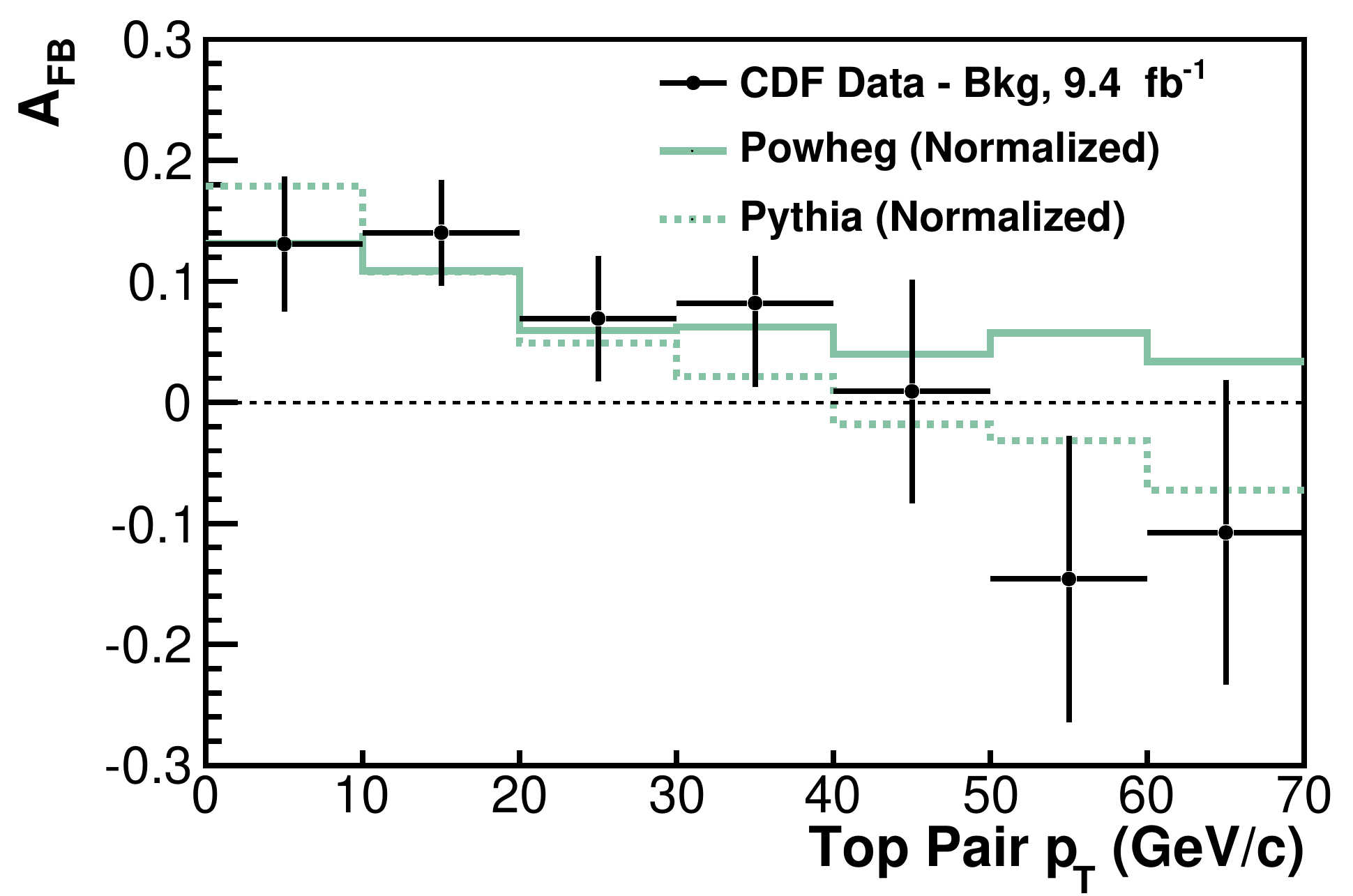}
\end{tabular}
\end{center}
\caption{Top: $\afb$ vs. $\pt^{\ttbar}$ for LO and NLO QCD calculations. Bottom: Comparison of data to predictions including an additional $\pt^{\ttbar}$ independent asymmetry. From~\textcite{Aaltonen:2012it}}.
\label{fig:Avspt}
\end{figure}

The CDF Collaboration measured the $\pt^{\ttbar}$-dependent asymmetry as part of the study described in Sec.~\ref{sec:cdfljets}. The analysis is performed with the reconstructed, background-subtracted data, avoiding issues with unfolding. The bottom plot in Figure~\ref{fig:Avspt} shows $\afb(\pt^{\ttbar})$ for the data after background subtraction.  The asymmetry in data falls with $\pt^{\ttbar}$ as in the top plot, but lies above the predictions there. In order to make a slope comparison they consider a simple normalization ansatz where the excess asymmetry in the data is independent of $\pt^{\ttbar}$, as is approximately the case for the NLO  QCD effect and also some of the new physics models. Since independent asymmetries add, this produces an additive correction in each $\pt^{\ttbar}$ bin equal to the difference of the inclusive $\afb$ in the reconstructed data and the simulated SM. The bottom plot in Fig.~\ref{fig:Avspt} shows the comparison of the data to the {\sc pythia} and {\sc powheg} models normalized in this way. The data are well described by either the {\sc powheg} or {\sc pythia} modeling in conjunction with a $\pt^{\ttbar}$-independent asymmetry according to the inclusive measurement.

\subsection{Inclusive asymmetry in the dilepton mode}
\label{sec:afbdil}

When both top quarks decay leptonically, the $\ttbar$ final state contains two charged leptons, missing transverse energy $\met$ from two overlapping neutrinos, and two hadronic jets initiated by $b$ quarks.  The information loss in the overlapping neutrinos prohibits a direct reconstruction of the $\ttbar$ kinematics, but the top quark asymmetry can be recovered using probabilistic techniques.

The D0 Collaboration has performed a preliminary measurement in the full Tevatron dataset of $9.7~\ifb$~\cite{Abazov:DILinc}. Events are selected requiring two isolated opposite sign electrons or muons having $\pt > 15~\gev$ and two or more jets with $E_T > 20$~GeV. Additional selections based $H_T$, $\met$, and the $\met$ significance are optimized separately for each of the three modes $ee$, $e\mu$, and $\mu\mu$. At least one of the two jets is required to be $b$-tagged by a multivariate discriminant that is also optimized for the flavor of the leptonic mode. The total number of events is 542. Non-$\ttbar$ backgrounds to this selection include $Z$ bosons and electroweak di-bosons in association with jets, and QCD multijets that manage to satisfy the lepton and $\met$ requirements. The electroweak backgrounds are modeled with simulated samples, and the QCD multijets background is modeled with data driven techniques. The estimated background contamination is $62\pm 15$ events.

The top quark $\dy$ distribution is reconstructed using a novel modification of the matrix-element technique used to measure the top-quark mass at D0~\cite{Abazov:2011fc}. In each event a likelihood function for $\dy$ is constructed by comparing the final state kinematic configuration to the LO SM matrix element for $\ttbar$ production. The number of integrations is reduced by assuming that the initial and final states conserve energy and momentum, that the lepton direction, $b$-quark direction, and electron energies are perfectly measured, that both $l\nu$ systems have $m_W=80.4~\gev$, and that both top systems have $m_t = 172.5~\gev$. The muon and jet energy resolution is treated using transfer functions. The final integration over $p_t^{\ttbar}$, $\phi^{\ttbar}$, the energies of the two leading jets, and the energies of the muons (if applicable), produces a likelihood distribution of $\dy$ for each event. The sum of the event likelihoods estimates the distribution of $\dy$ in the sample. The background models are used to derive the $\dy$ distribution for the non-$\ttbar$ components and these are subtracted.

This distribution and its asymmetry includes dilution effects due to the limited detector acceptance, the finite resolution of the detector measurements, and the assumptions in the matrix element integration. The production level asymmetry is recovered from the measured asymmetry using a linear transfer function derived from samples of {\sc mc@nlo} that have been reweighted for various asymmetries according to the scheme of~\textcite{Hong:2014dga}. Tests with simulations of new physics models for the asymmetry show that the technique is unbiased at the level of $ \leq 2\%$ as long as the top-quark decays are SM-like. i.e. have no unexpected polarization. The production level asymmetry is found to be $\afb = 0.180\pm 0.061\,{\rm (stat)}\pm 0.032\,{\rm (syst)}$. The systematic uncertainty is dominated by the hadronization and showering model and the PDF assumptions. 

\subsection{Leptonic asymmetries}
\label{sec:3c}

The leptonic asymmetries defined in Eq. (2.10)  are experimentally attractive because the lepton rapidity is very well-measured and free from the complications of combinatorics and jet resolution present in the $\dy$ reconstruction.  In the SM, the {\sc powheg} generator and NLO calculations~\cite{Bernreuther:2012sx} both suggest a ratio $\afbl/\afb \approx 0.5$.

\subsubsection{Leptonic asymmetries at D0}
\label{sec:3c1}

The D0 Collaboration has measured $\afbl$ in both decay modes. The measurement in $\ell$+jets~\cite{Abazov:2014oea} uses $9.7~\ifb$ and the sample selection of Sec.~\ref{sec:d0ljets}. The D0 lepton acceptance extends to $|\eta|=1.5$ and the analysis and unfold is done with this cut, ignoring leptons of larger rapidity. The significant background from $W$+jets events is calibrated against a control sample derived from $\ell$+3-jet events with no $b$-tag. Similarly to~\textcite{Abazov:2014cca}, a multivariate technique is used to separate signal and background for events with each sign of $\yq$, and for each bin of jet multiplicity ($3, \geq 4$) and $b$-tag multiplicity (1,2). The unfold uses the response model of {\sc mc@nlo}, and the results for each final state category are combined in a weighted average to yield  $\afbl = 0.042\pm0.030$ (stat+syst), to be compared with the {\sc mc@nlo} prediction of $0.02\pm 0.001$ (statistical error only) for lepton pseudo-rapidity $|\eta| < 1.5$. The ratio $\afbl/\afb = 0.44\pm 0.27$ is consistent with the {\sc powheg} prediction. This result differs considerably from the $5~\ifb$ D0 measurement~\cite{Abazov:2011rq}, $\afbl = 0.152\pm 0.040$, that used only the $\ell$+4-jet sample. In the new measurement the large asymmetry in the $\ell$+4-jet sample remains, but smaller asymmetries in the $\ell$+3-jet samples reduce the overall inclusive value. The dependence of $\afbl$ on the lepton $\pt$ in new physics models is discussed in Sec.~\ref{sec:6c}. The dependence in the data is shown on the top panel in Fig.~\ref{fig:afblep}. 
 
\begin{figure}[htb]
\begin{center}
\begin{tabular}{r}
\includegraphics[width=8cm,clip=]{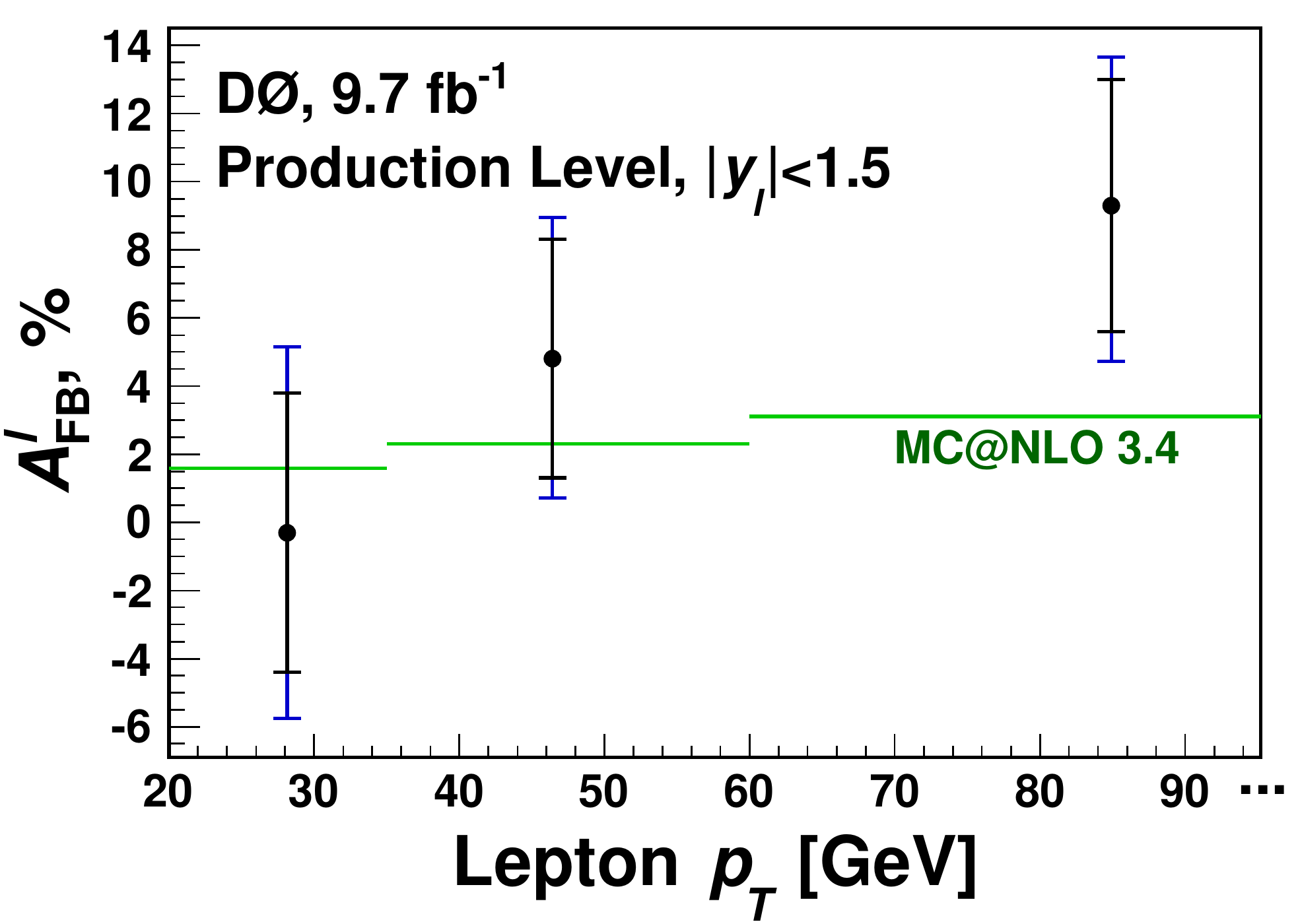} \\[3mm]
\includegraphics[width=8cm,clip=]{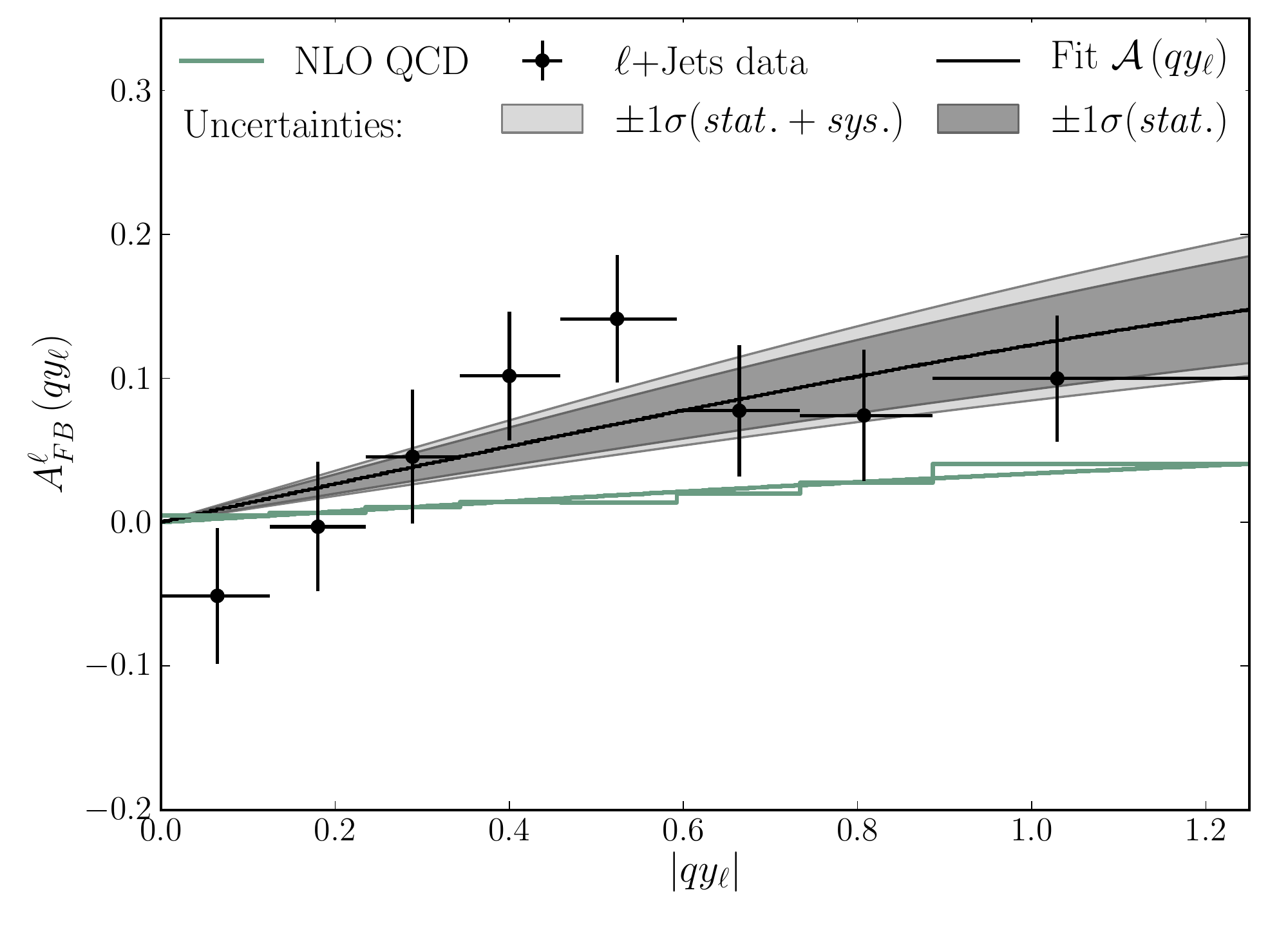}
\end{tabular}
\end{center}
\caption{Top: D0 $\afblep$ versus lepton $\pt$ in $\ell$+jets, showing statistical/statistical+systematic uncertainties in black/blue. Bottom: $A^l_{FB}(\yq)$ in CDF data with statistical uncertainties with best fit. From~\textcite{Abazov:2014oea,Aaltonen:2013vaf}.}
\label{fig:afblep}
\end{figure}

The D0 measurement of $\afbl$ in the dilepton mode uses $9.7~\ifb$~\cite{Abazov:2013wxa}. Reconstructed electrons and muons must be isolated, have $\pt > 15~\gev$, and opposite signs. Events with like-flavor leptons must have two jets with $E_T > 20$~GeV; $e\mu$ events must have at least one such jet. A multivariate technique is used to require that jets are consistent with originating from $b$-quarks. Further specialized cuts select on the $H_T$ and $\met$ significance for each decay mode. The non-$\ttbar$ backgrounds to this selection are modeled using data sidebands, and subtracted from the data. The asymmetries are corrected for the finite lepton acceptance using a scale factor derived from {\sc mc@nlo}. The asymmetries are found to be $\afbl = 0.044\pm0.039$ and $\afbll= 0.123\pm0.056$. The dilepton $\afbl$ can be combined with the $\ell$+jets result using scale factors obtained from Monte Carlo simulations. The combined asymmetry is $\afbl = 0.047\pm0.027$~\cite{Abazov:2014oea}.

\subsubsection{Leptonic asymmetries at CDF}
\label{sec:3c2}
The CDF Collaboration has also measured the leptonic asymmetry in both the $\ell$+jets and dilepton decay modes. The $\ell$+jets measurement ~\cite{Aaltonen:2013vaf} uses the same sample as the $\cos\theta$ analysis of Sec.~\ref{sec:cdfcos}, with a total sample of $3864$ events and an expected non-$\ttbar$ background of $1026\pm 210$.  The limited central lepton acceptance of the CDF detector $|\eta|\leq 1.2$ makes an unfolding correction for the full rapidity range impossible. Instead, the measurement relies on the observation that the asymmetric part of the asymmetry
\begin{equation}
 A^l_{FB}(\yq) = \frac{N\left(\yq\right) - N\left(-\yq\right)}
                {N\left(\yq\right) + N\left(-\yq\right)}
\end{equation}
is described by a simple phenomenological function $F(\yq) = a\tanh({\yq}/{2})$ for all models tested~\cite{Hong:2014dga}, while the symmetric part is model independent. $A^l_{FB}(\yq)$ in the measured region can be corrected for backgrounds and acceptance, and used to find the best fit to $F(\yq)$ in the data. The function $F(\yq)$ can then be extrapolated to the full rapidity range, and integrated with the model-independent symmetric part (using any generator), allowing a measurement of the production level asymmetry.  The $A^l_{FB}(\yq)$ distribution and fit, shown on the bottom in Fig. ~\ref{fig:afblep}, gives a production level asymmetry of $\afbl = 0.094^{+0.032}_{-0.029}$.  
 
The dilepton measurement is done in a sample of $9.1~\ifb$~\cite{Aaltonen:2014eva}. The selection requires exactly two opposite sign leptons with $\pt > 20~\gev$  and combined invariant mass $m_{\ell \ell} > 10$~\gev,  $\met > 25$~GeV, two or more jets with $E_T > 15$~GeV and $|\eta| < 2.5$, and $H_T > 200$~GeV. A total of 569 events are found. The shape and normalization of the non-$\ttbar$ component is estimated with a combination of Monte Carlo and data driven techniques, giving an expected background of $160\pm 21$ events.  The measurement of $\afbl$ uses both leptons in each event, doubling the statistics. The results are $\afbl = 0.072\pm0.060$ and $\afbll = 0.076\pm0.081$.  The single lepton result here can be combined with the $\ell$+jets channel to give an overall $\afbl = 0.090^{+0.028}_{-0.026}$.  The ratio $\afbl/\afb = 0.55\pm 0.24$ is consistent with the NLO prediction. 

Appealing, again,  to the near independence of the CDF and D0 measurements, a simple error weighted average of the two gives a combined Tevatron $\afbl = 0.069\pm 0.019$ compared to the NLO prediction of $0.038\pm 0.003$~\cite{Bernreuther:2012sx}.

\subsection{Tevatron summary}
\label{sec:3d}

A compendium of the Tevatron measurements is shown in Fig.~\ref{fig:afbsum}.  With the final results from Run 2 at the Tevatron, the significance of the top $\afb$ and $\afbl$ discrepancies are around $1.5\sigma$, with a spread between the two experiments of roughly $1\sigma$. One of the most interesting experimental issues is the evolution of the D0 measurements toward smaller $\afb$,  reducing the tension with the SM suggested by the earlier results. An important part of that evolution was the addition of the 3-jet decay mode, which adds a statistically-independent sample, but also mixes in a different $\pt^{\ttbar}$ spectrum, raising the issue of the $\pt^{\ttbar}$ modeling for both experiments. The time development of the D0 measurements is discussed in further detail in \textcite{Abazov:2014cca, Abazov:2014oea}. 

\begin{figure}[htb]
\begin{center}
\includegraphics[width=7.5 cm,clip=]{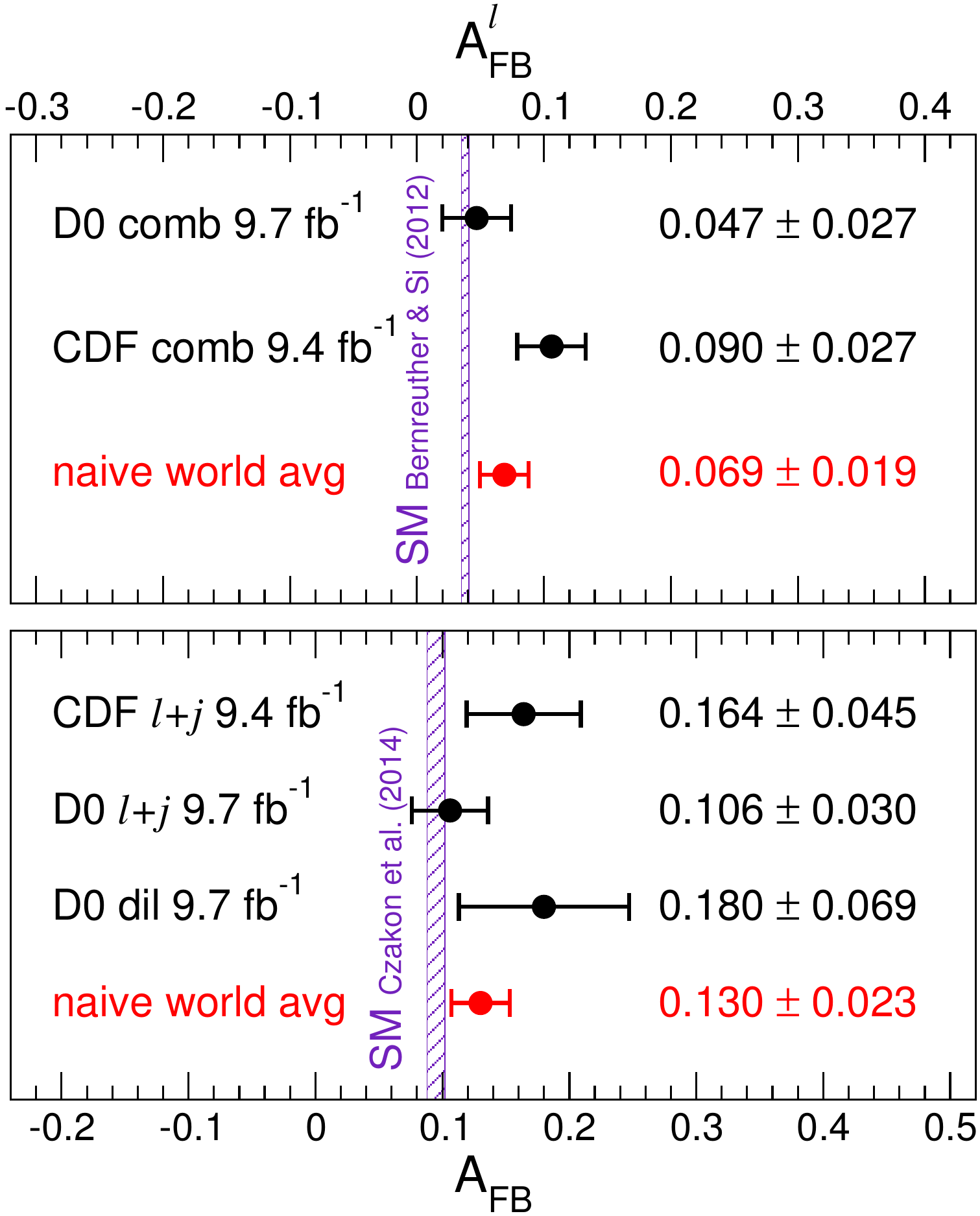}
\end{center}
\caption{Summary of $\afb$ and $\afbl$ measurements at the Tevatron}
\label{fig:afbsum}
\end{figure}


\section{Experimental measurements at the LHC}
\label{sec:4}

Following the end of operations of the Tevatron collider, measurements at the LHC are being carried out in an attempt
to further clarify the experimental picture, which would otherwise remain inconclusive based on the measurements
by the CDF and D0 Collaborations using the full dataset, discussed in the previous section. Despite the small
$\ttb$ asymmetries expected in $pp$ collisions, the very high statistics $\ttb$ samples available at the LHC
can be exploited in the context of selections that are optimized to increase the
fraction of $q\bar{q}$ events. This fact, together with the higher kinematic reach at the LHC, makes differential
measurements of the charge asymmetry particularly interesting. Indeed, beyond confirming or ruling out the
Tevatron anomaly, a new kinematic regime is being explored at the LHC that may unveil signs of new physics
the Tevatron could not be sensitive to. Here we review the most recent results from the LHC Run 1 (2011-2012) and give some prospects for run 2.

\subsection{Charge asymmetry measurements}
\label{sec:4a}

Measurements of the charge asymmetry have been performed by the ATLAS and CMS Collaborations
using the full datasets collected at CM energies of 7~TeV and 8~TeV. These measurements
have been carried out in the $\ell$+jets channel at 7~TeV and 8~TeV, and in the dilepton channel, so far only at 7~TeV.

The measurement of the charge asymmetry involves the reconstruction of the event kinematics under
the $\ttb$ hypothesis in order to determine the rapidities of the top quark and antiquark, c.f. Eq.~(\ref{ec:ac}). This is possible, not only in the $\ell$+jets channel, where the presence of a single
neutrino still leaves sufficient measurements for kinematic reconstruction, but also in the dilepton channel,
where the a-priori under-constrained kinematics from the presence of two neutrinos can be overcome
through the application of additional assumptions. The ability to reconstruct the event kinematics is
exploited to measure the charge asymmetry differentially, as a function of $\mttb$, $\pTttb$ and the rapidity of the $\ttb$ system, $\yttb$,
in addition to inclusively. The reconstructed distributions used for these measurements, $\dabsy$, as well as
the above kinematic variables of the $\ttb$ system, are distorted by effects related to selection efficiencies, 
detector resolution effects as well as ambiguities in the kinematic reconstruction. Unfolding techniques
are used in order to correct these measurements to the parton level and thus be able to compare them
with theoretical predictions.

\subsubsection{Inclusive asymmetry in the $\ell$+jets channel}
\label{sec:4a1}

The first measurement of the inclusive charge asymmetry at the LHC was performed by the CMS Collaboration
in the $\ell$+jets channel using a total integrated luminosity of $1.1~\ifb$ at 7~TeV, about one fifth of the total 
dataset eventually cumulated at this CM energy. Since then, improved measurements using the 
full datasets at 7~TeV and 8~TeV have become available. Here we report only on those most
precise measurements.

The ATLAS Collaboration has performed a measurement of the inclusive charge asymmetry in
the $\ell$+jets channel using the full dataset collected in 2011 at $\sqrt{s}=7$~TeV, corresponding to 
an integrated luminosity of $4.7~\ifb$~\cite{Aad:2013cea}. Events were collected using single-lepton
triggers. The offline selection requires exactly one isolated electron or muon with $\pt>25$~GeV and
$\pt>20$~GeV, respectively. In order to suppress background from QCD multijets production, requirements
are placed on $\met$ and the transverse mass reconstructed from the lepton and $\met$.
In addition, it is required that the event has at least four jets reconstructed with the anti-$k_t$ algorithm~\cite{Cacciari:2008gp} 
with a radius parameter $R=0.4$ and satisfying $\pt>25$~GeV and $|\eta|<2.5$. In order to suppress
background from $W$+jets production, at least one jet is required to be $b$-tagged. The typical per-jet 
$b$-tagging efficiency and light-jet mistag rate are 70\% and 0.7\%, respectively. 
This results in a selected sample of approximately $60000$ events, with an estimated $\ttb$ purity
of $80\%$. Simulated $\ttb$ events are modeled using the LO multi-parton matrix-element Monte Carlo
generator {\sc alpgen}~\cite{Mangano:2002ea} interfaced with {\sc herwig} for the simulation
of showering and fragmentation.
After selection, the dominant background is $W$+jets production, whose normalization
is estimated using data, while the shape of the distributions is estimated using the simulation.
Smaller backgrounds originate from QCD multijets, single top, $Z$+jets and diboson production.
With the exception of QCD multijets, which is entirely estimated from data, the remaining backgrounds
are estimated with the simulation. A likelihood-based kinematic fit is used to reconstruct the
4-momenta of the top and antitop quarks. This likelihood calculation takes as input the measured
kinematic quantities of the lepton, $\met$ and the leading four jets, and employs transfer functions associating the
measured variables to the parton-level ones. The ambiguities resulting from the reconstruction of the 
leptonically-decaying $W$ boson and the jet-parton assignments are solved by choosing the reconstruction
hypothesis leading to the highest likelihood value. Figure~\ref{fig:dabsy}~(top) displays the reconstructed $\dabsy$ distribution.
\begin{figure}[htb]
\begin{center}
\begin{tabular}{r}
\includegraphics[width=8cm, clip]{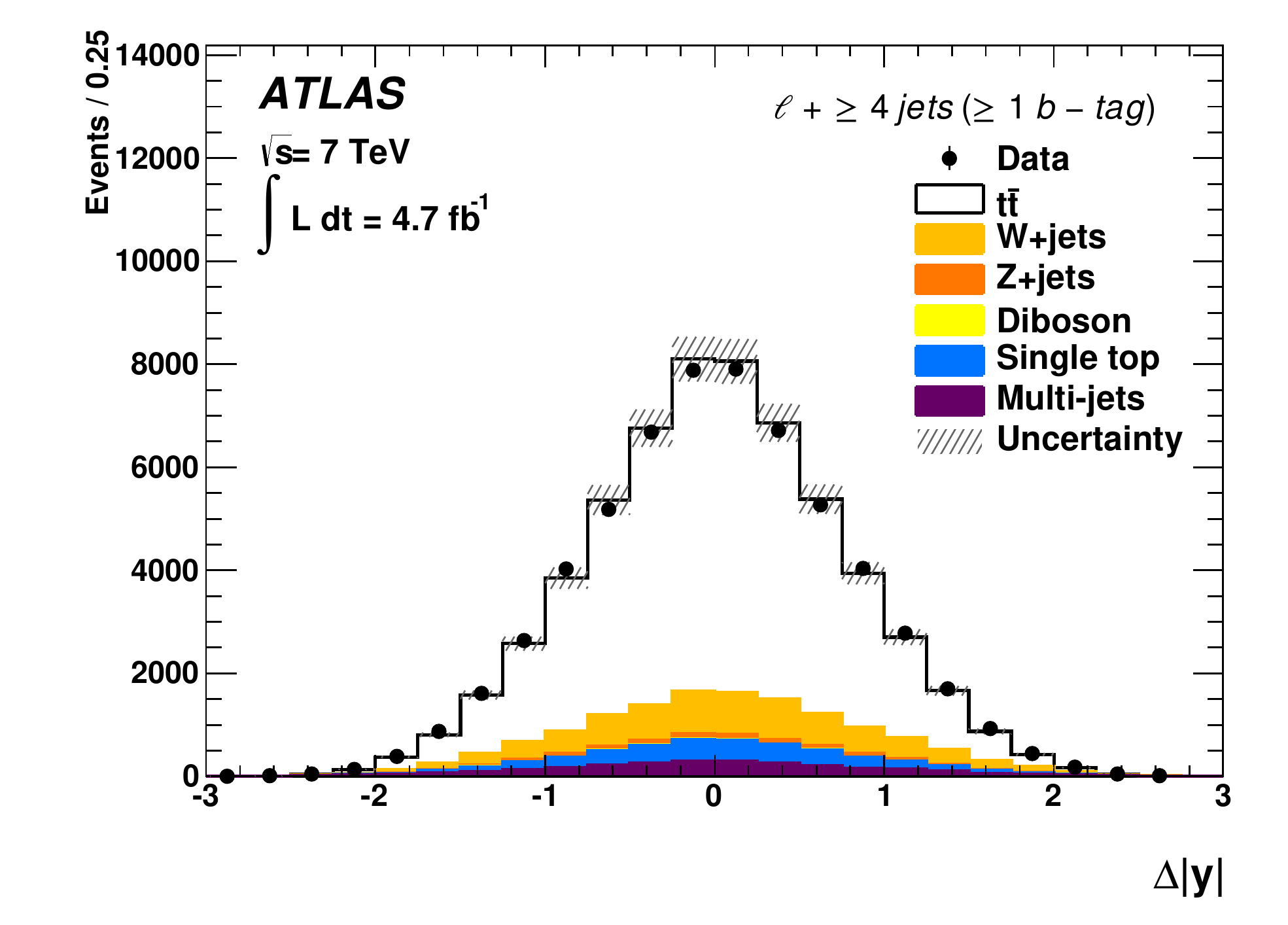} \\
\includegraphics[width=8cm, clip]{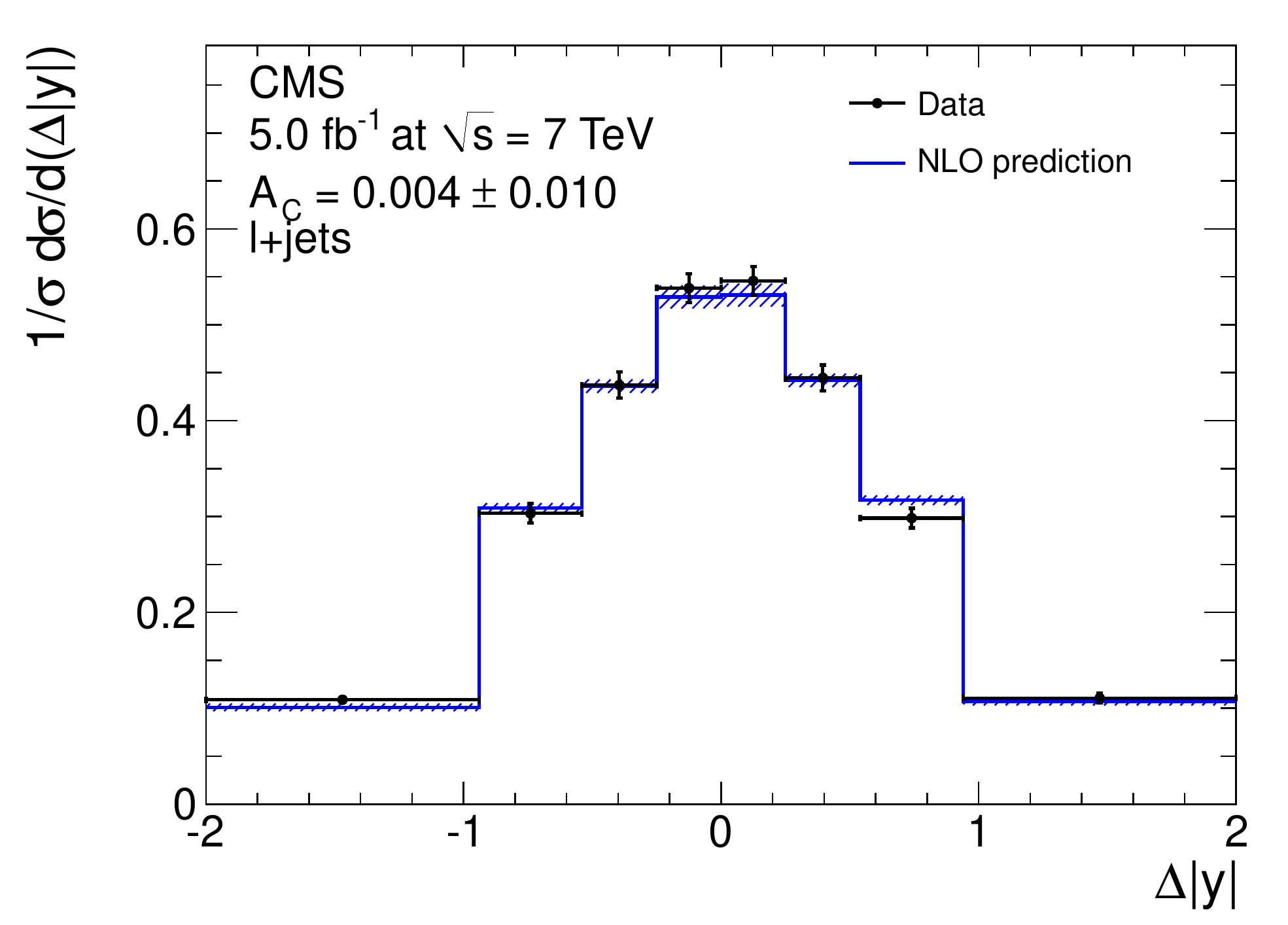}
\end{tabular}
\caption{Top: Reconstructed $\dabsy$ distribution after final selection in the ATLAS measurement. 
Data (dots) is compared to the prediction from the {\sc alpgen} generator (solid line) and its total uncertainty (shaded area).
Bottom: Unfolded $\dabsy$ distribution after final selection in the CMS measurement. 
Unfolded data (dots with error bars representing the total uncertainty) are compared to the SM prediction from~\textcite{Kuhn:2011ri}.
The first and last bins include underflow and overflow events, respectively.
From~\textcite{Aad:2013cea,Chatrchyan:2012cxa}. \label{fig:dabsy}}
\end{center}
\end{figure}
Using simulated $\ttb$ events, the fraction of events
with correctly reconstructed $\dabsy$ sign is $\approx 75\%$, corresponding to a dilution
factor $D = 2 \times 0.75-1 = 0.5$. Such dilution results in a reduction by a factor of two of the
measured asymmetry relative to the parton-level asymmetry, which is effectively corrected for by the 
unfolding procedure. After subtracting the background, the measured $\dabsy$ distribution is unfolded to 
the parton level and the charge asymmetry computed, yielding $\ac = 0.006 \pm 0.010\,{\rm (stat)} \pm 0.005\,{\rm (syst)}$. 
This measurement is in agreement with the SM prediction of $\ac \simeq 0.0115$ (see Sec.~\ref{sec:2c}).

Similarly, the CMS Collaboration has performed a measurement of the inclusive charge asymmetry in
the $\ell$+jets channel using the full dataset collected at $\sqrt{s}=7$~TeV, corresponding to 
an integrated luminosity of $5~\ifb$~\cite{Chatrchyan:2012cxa}. Events were collected using triggers
requiring a single lepton together with at least three jets.
The offline selection requires exactly one isolated electron or muon with $\pt>30$~GeV and
$\pt>20$~GeV, respectively. In contrast to the ATLAS measurement, no minimum requirement on
$\met$ is made. In addition, it is required that the event has at least four jets reconstructed with the anti-$k_t$ algorithm
with a radius parameter $R=0.5$ and satisfying $\pt>30$~GeV and $|\eta|<2.5$. Similar to the ATLAS measurement,
a requirement of at least one $b$-tagged jet is made using an algorithm with a typical $b$-tagging efficiency 
of 60\% and a light-jet mistag rate of 1\%.
The corresponding selected data sample has approximately $58000$ events, with an estimated $\ttb$ purity of also $80\%$.  In the case of the CMS measurement, simulated $\ttb$ events are modeled using 
the NLO event generator {\sc powheg} interfaced with {\sc pythia} for the simulation
of showering and fragmentation. The background composition is similar to the one in the ATLAS measurement,
and similar strategies are used in the estimation of the background normalization and the modeling of its
kinematics. The reconstruction of the 4-momenta of the top and antitop quarks is also based on a
likelihood technique, in this case using as inputs the reconstructed invariant masses for the
leptonic and hadronic top quarks, the hadronically-decaying $W$ boson, and the $b$-tagging information
of the jets assigned to the final state quarks. Also in this case the reconstruction hypothesis
leading to the highest likelihood value is selected. The performance of this reconstruction technique is
comparable to that of the ATLAS analysis. After subtracting the background, the measured asymmetry is unfolded to 
the parton level. Figure~\ref{fig:dabsy}~(bottom) compares the unfolded $\dabsy$ distribution to the theoretical
prediction. The resulting measured charge asymmetry is  $\ac = 0.004 \pm 0.010\,{\rm (stat)} \pm 0.011\,{\rm (syst)}$, also
in agreement with the SM prediction and the ATLAS measurement.

The ATLAS and CMS measurements at $\sqrt{s}=7$~TeV discussed above have been combined~\cite{CMS:2014jua}.
The combination has been performed using the Best Linear Unbiased Estimate (BLUE) method~\cite{Lyons:1988rp,Valassi:2003mu},
taking into account a detailed categorization of systematic uncertainties and their correlation between both
experiments. The combined result is $\ac = 0.005 \pm 0.007\,{\rm (stat)} \pm 0.006\,{\rm (syst)}$,
representing a 40\% (18\%) improvement with respect to the CMS (ATLAS) measurement.

\begin{figure}[htb]
\begin{center}
\includegraphics[width=8cm, clip]{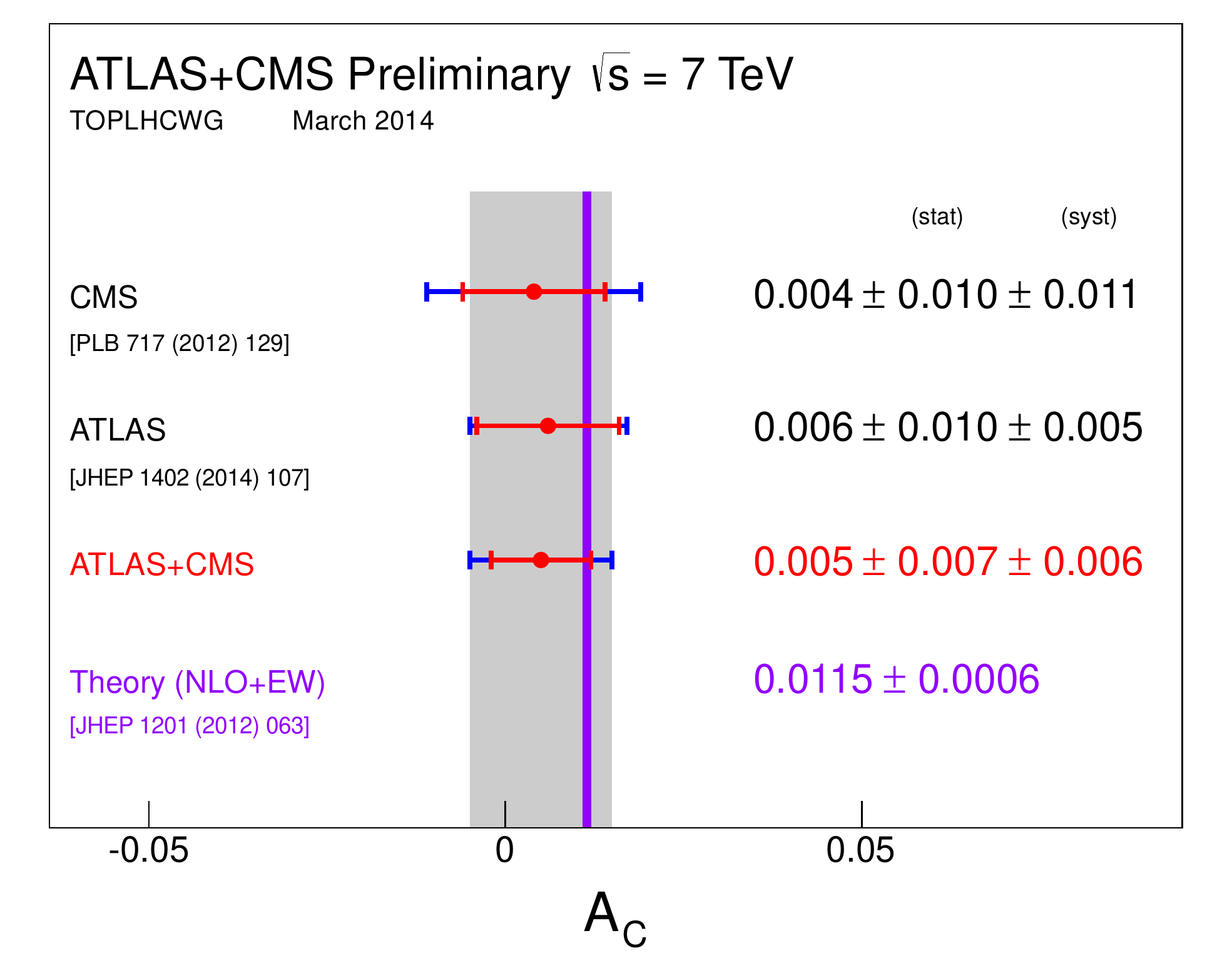}
\caption{Summary of the measurements of the inclusive charge asymmetry by the ATLAS and CMS
Collaborations in the $\ell$+jets channel at 7 TeV, as well as their combination, compared to the theoretical prediction. For each measurement, 
the outer blue (inner red) error bar indicates the total (statistical) uncertainty. The grey band illustrates the
total uncertainty of the combined result. From~\textcite{CMS:2014jua}. \label{fig:lhc_combo}}
\end{center}
\end{figure}

The CMS Collaboration has also performed a measurement of the inclusive charge asymmetry in
the $\ell$+jets channel using the full dataset collected in 2012 at $\sqrt{s}=8$~TeV, corresponding to 
an integrated luminosity of $20~\ifb$~\cite{CMS:2013nfa}. This measurement follows closely the
strategy for event selection, signal and background modeling, kinematic reconstruction, and unfolding
technique used at the previous $\sqrt{s}=7$~TeV measurement  discussed above. The higher $\ttb$ cross section
at $\sqrt{s}=8$~TeV and the 4-fold increase in the integrated luminosity results in a very large sample of approximately $375000$ events, 
with an estimated $\ttb$ purity of $80\%$. After subtracting the background, the measured asymmetry is unfolded to 
the parton level. The resulting charge asymmetry is  $\ac = 0.005 \pm 0.007\,{\rm (stat)} \pm 0.006\,{\rm (syst)}$.
It is worth noting that the statistical uncertainty of this measurement does not scale as expected given the increased number of 
$\ttb$ candidate events compared to the measurement at $\sqrt{s}=7$~TeV: this is due to the increased number of bins used
in the unfolding as well as the smaller improvement resulting from the use of regularization in presence of such large statistics in the data.
This measurement has comparable precision to the combination of ATLAS and CMS measurements at $\sqrt{s}=7$~TeV 
discussed above and is also in agreement with the SM prediction of $\ac \simeq 0.0102$. 

\subsubsection{Inclusive asymmetry in the dilepton channel}
\label{sec:4a2}

The ATLAS Collaboration has performed a measurement of the inclusive charge asymmetry in
the dilepton channel using the full dataset at $\sqrt{s}=7$~TeV, corresponding to $4.6~\ifb$~\cite{Aad:2015jfa}. 
Events were collected using single-lepton triggers, and required to have exactly two opposite-sign leptons
($ee$, $e\mu$, or $\mu\mu$), and at least two jets. The lepton and jet requirements are similar to those
used in the measurement in the $\ell$+jets channel. In order to suppress background from $Z/\gamma^*$+jets 
production, in the same-flavor dilepton channels the dilepton invariant mass is required to be more than
10~GeV away from the $Z$ boson mass, and $\met>60$~GeV is required.
The resulting selected data sample contains approximately $8000$ events, with an estimated $\ttb$ purity of $86\%$.  
Simulated $\ttb$ events are modeled using the NLO Monte Carlo generator {\sc powheg}
interfaced with {\sc pythia} for the simulation of showering and fragmentation.
After selection, the background is dominated by processes with prompt leptons, including 
$Z$+jets, single top and diboson production, which are estimated from the simulation.
In addition, non-negligible contributions arise from processes with one or two jets misidentified as a lepton,
resulting from QCD multijets or $W$+jets production, which are estimated in situ using data-driven techniques.
The reconstruction of the $\ttb$ kinematics is performed using the neutrino weighting technique~\cite{Abbott:1997fv}.
This technique scans different hypotheses for the values of the pseudorapidities of the two neutrinos in the final state.
For each hypothesis, it calculates the full event kinematics assuming the $W$ boson and top quark masses, and
then assigns a weight to the resulting solution based on the level of agreement between the calculated and measured
missing transverse momentum. Jet energy measurements are accounted for by fluctuating the jet energies within the
expected resolutions, and all possible lepton-jet associations are considered. Finally, the solution corresponding to
the maximum weight is chosen to represent the event.
Figure~\ref{fig:dabsy_dil}~(top) displays the reconstructed $\dabsy$ distribution in the $e\mu$  channel.
\begin{figure}[htb]
\begin{center}
\begin{tabular}{r}
\includegraphics[height=2.5in, width=0.42\textwidth, clip]{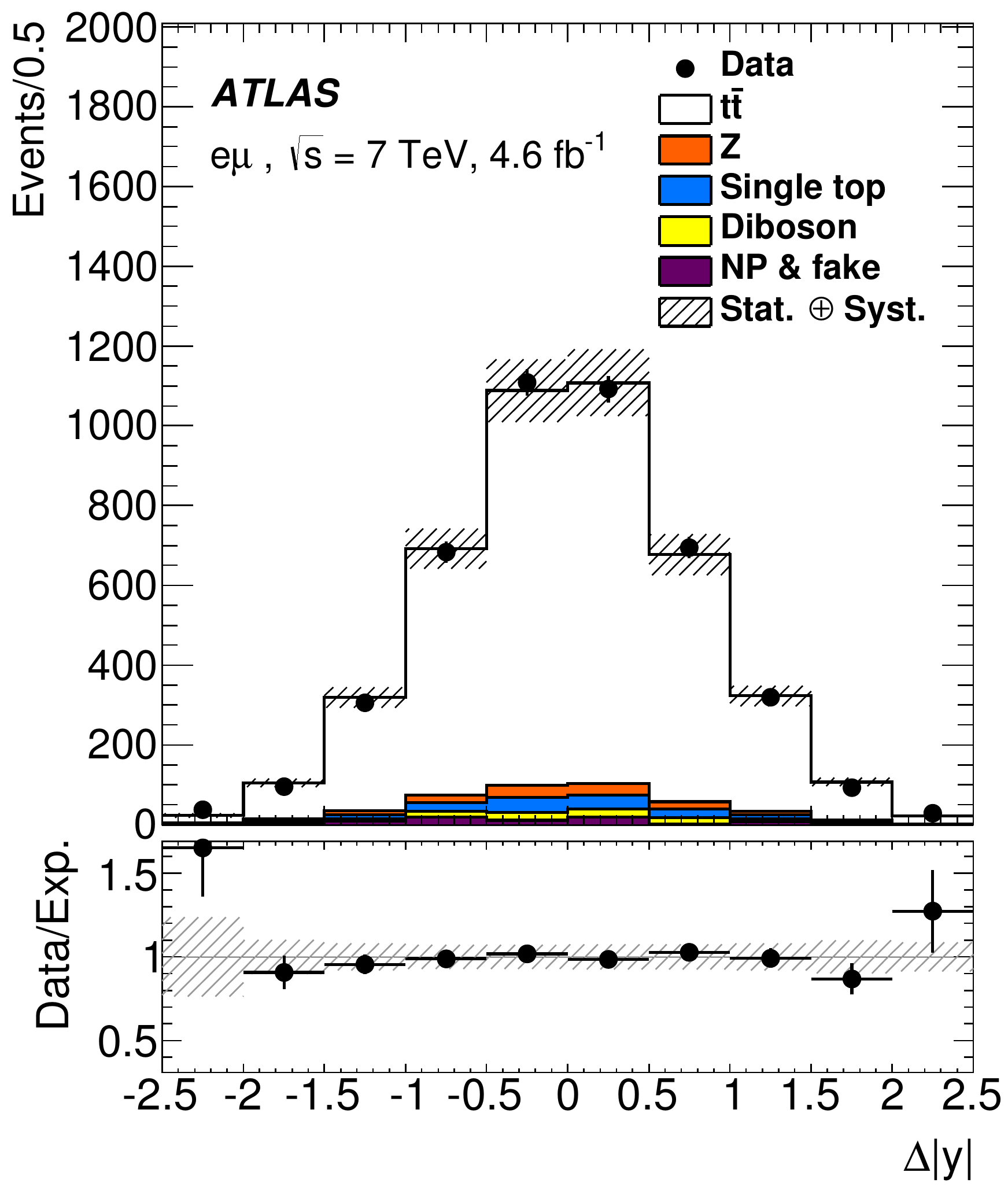} \\
\includegraphics[height=3.0in, width=0.42\textwidth, clip]{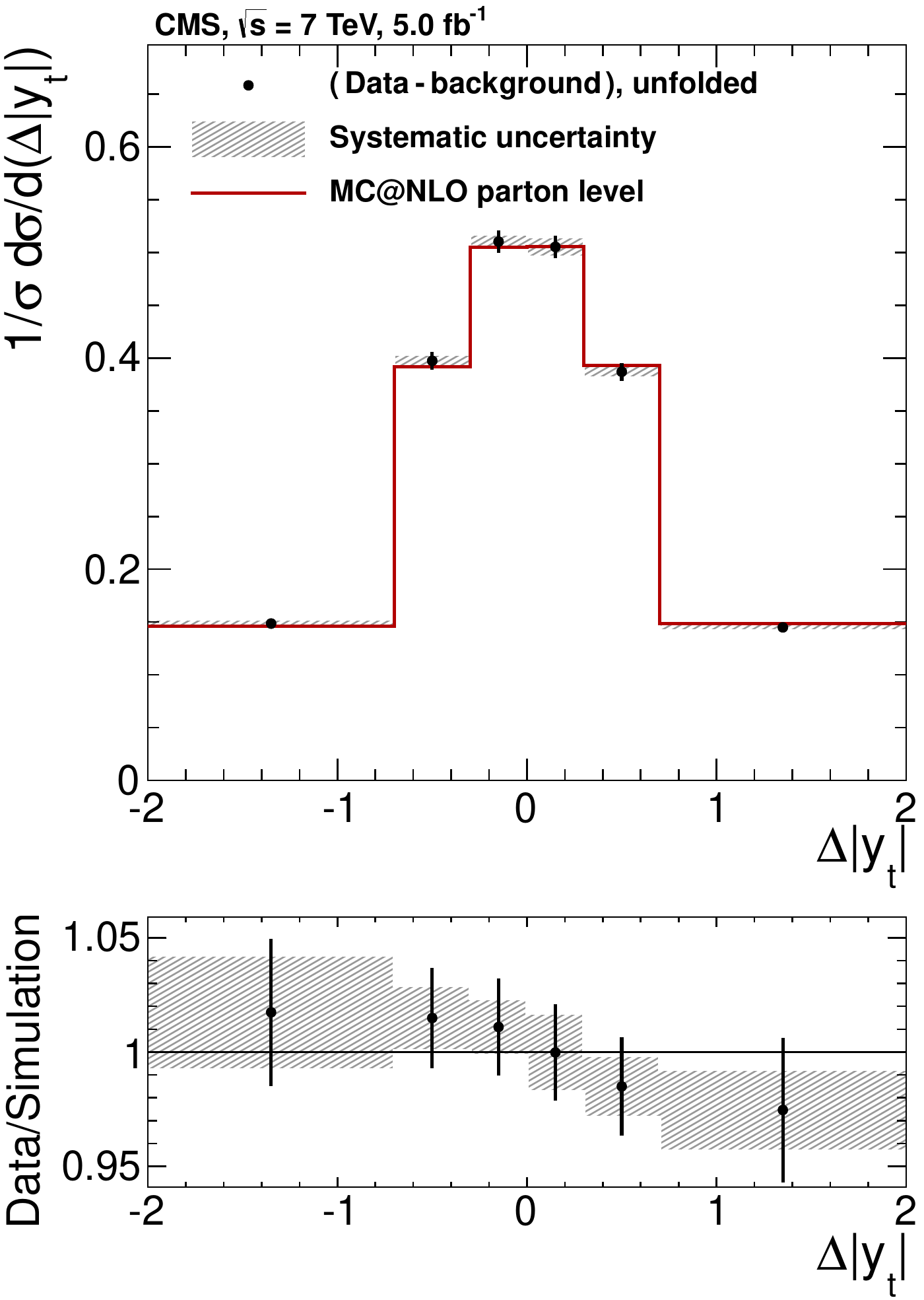}
\end{tabular}
\caption{Top: Reconstructed $\dabsy$ distribution in the $e\mu$ channel in the ATLAS measurement. 
Data (dots) is compared to the prediction from the {\sc powheg}  generator (solid line) and its total uncertainty (shaded area).
Bottom: Unfolded $\dabsy$ distribution after final selection in the CMS measurement. 
Unfolded data (dots with error bars representing the stat uncertainty and hatched bands representing the
systematic uncertainty) are compared to the {\sc mc@nlo} prediction. The bottom panel displays the ratio 
between the data and the {\sc mc@nlo} prediction. The first and last bins include underflow and overflow events, respectively.
From~\textcite{Aad:2015jfa,Chatrchyan:2014yta}. \label{fig:dabsy_dil}}
\end{center}
\end{figure}
After subtracting the background, the measured $\dabsy$ distributions in the $ee$, $e\mu$ and $\mu\mu$ channels 
are unfolded to the parton level and the corresponding charge asymmetries computed. The combination of the measurements
in the three channels using the BLUE method yields $\ac = 0.021 \pm 0.025\,{\rm (stat)} \pm 0.017\,{\rm (syst)}$. 

Similarly, the CMS Collaboration has performed a measurement of the inclusive charge asymmetry in
the dilepton channel using the full dataset at $\sqrt{s}=7$~TeV, corresponding to $5~\ifb$~\cite{Chatrchyan:2014yta}. 
Events were collected using dilepton triggers, and required to have exactly two opposite-sign leptons
($ee$, $e\mu$, or $\mu\mu$) with $\pt>20$~GeV and at least two jets.
The jet requirements are the same as in the measurement in the $\ell$+jets channel. In contrast with the ATLAS
measurement, at least one jet is required to be $b$-tagged. The typical per-jet 
$b$-tagging efficiency and light-jet mistag rate are 70\% and 1.5\%, respectively. 
In the same-flavor dilepton channels the dilepton invariant mass is required to be above 20~GeV and more than 15~GeV away 
from the  $Z$ boson mass  in order to suppress background from $Z/\gamma^*$+jets and heavy-quarkonium 
resonance production. In addition, a requirement of $\met>40$~GeV is made.
The resulting selected data sample contains approximately $10000$ events, with an estimated $\ttb$ purity of $92\%$. The increased purity, compared to the ATLAS measurement, results from the $b$-tagging requirement.
Simulated $\ttb$ events are modeled using {\sc mc@nlo} interfaced with {\sc herwig}.
After selection, the background is dominated by single top production, followed by Drell-Yan and 
$\ttb$ non-dileptonic backgrounds.  
The reconstruction of the $\ttb$ kinematics is performed using a weighting technique, referred to as
Analytical Matrix Weighting Technique~\cite{Chatrchyan:2011nb}, similar in spirit to the one used in the ATLAS measurement. 
Each event can have up to 8 possible solutions for the $\ttb$ system, each of which is assigned a weight based
on the probability of observing the given configuration. The solution with the highest weight is selected to 
reconstruct the $\ttb$ kinematics. About 14\% of events have no solution, which is taken as an additional 
selection requirement. After subtracting the background, the measured asymmetry is unfolded to 
the parton level. Figure~\ref{fig:dabsy_dil}~(bottom) compares the unfolded $\dabsy$ distribution to the theoretical
prediction. The resulting measured charge asymmetry is  $\ac = -0.010 \pm 0.017\,{\rm (stat)} \pm 0.008\,{\rm (syst)}$.

\subsubsection{Kinematic dependence of the asymmetry}
\label{sec:4a3}

Given the small expected inclusive charge asymmetry at the LHC, comparable to the experimental uncertainties, it is
of particular importance to measure the charge asymmetry differentially, as a function of variables that are suitable
to enhance it in particular kinematic regions, especially those where new physics effects may be more apparent.
The ATLAS and CMS Collaborations have measured the charge asymmetry as a function of $\pTttb$, $\yttb$  and 
$\mttb$, each of which is particularly sensitive to a certain aspect.  
The ATLAS Collaboration has measured the charge asymmetry as a function of the above kinematic variables in the $\ell$+jets channel 
using the full dataset at 7~TeV. The CMS Collaboration has performed similar measurement in 
the $\ell$+jets channel, at both 7~TeV and 8~TeV.
These measurements are based on the analyses described in Sec.~\ref{sec:4a1}.

As discussed in Sec.~\ref{sec:2}, the transverse momentum of the $\ttb$ system provides sensitivity to
the different diagrams contributing to the charge asymmetry with different sign. The low $\pTttb$ region is dominated
by the Born and box diagrams, whose interference results in a positive contribution to the charge asymmetry, while
the high $\pTttb$ region should be dominated by events with an extra jet in the final state, often originating from
initial or final-state radiation diagrams, whose interference results in a negative contribution to the charge asymmetry.
Figure~\ref{fig:ac_pTtt} shows the unfolded $\ac$ measurements as a function of $\pTttb$ from the ATLAS and CMS Collaborations.
Good agreement is found with the SM prediction within the experimental uncertainties.

\begin{figure}[htb]
\begin{center}
\begin{tabular}{r}  
\includegraphics[width=8cm, clip]{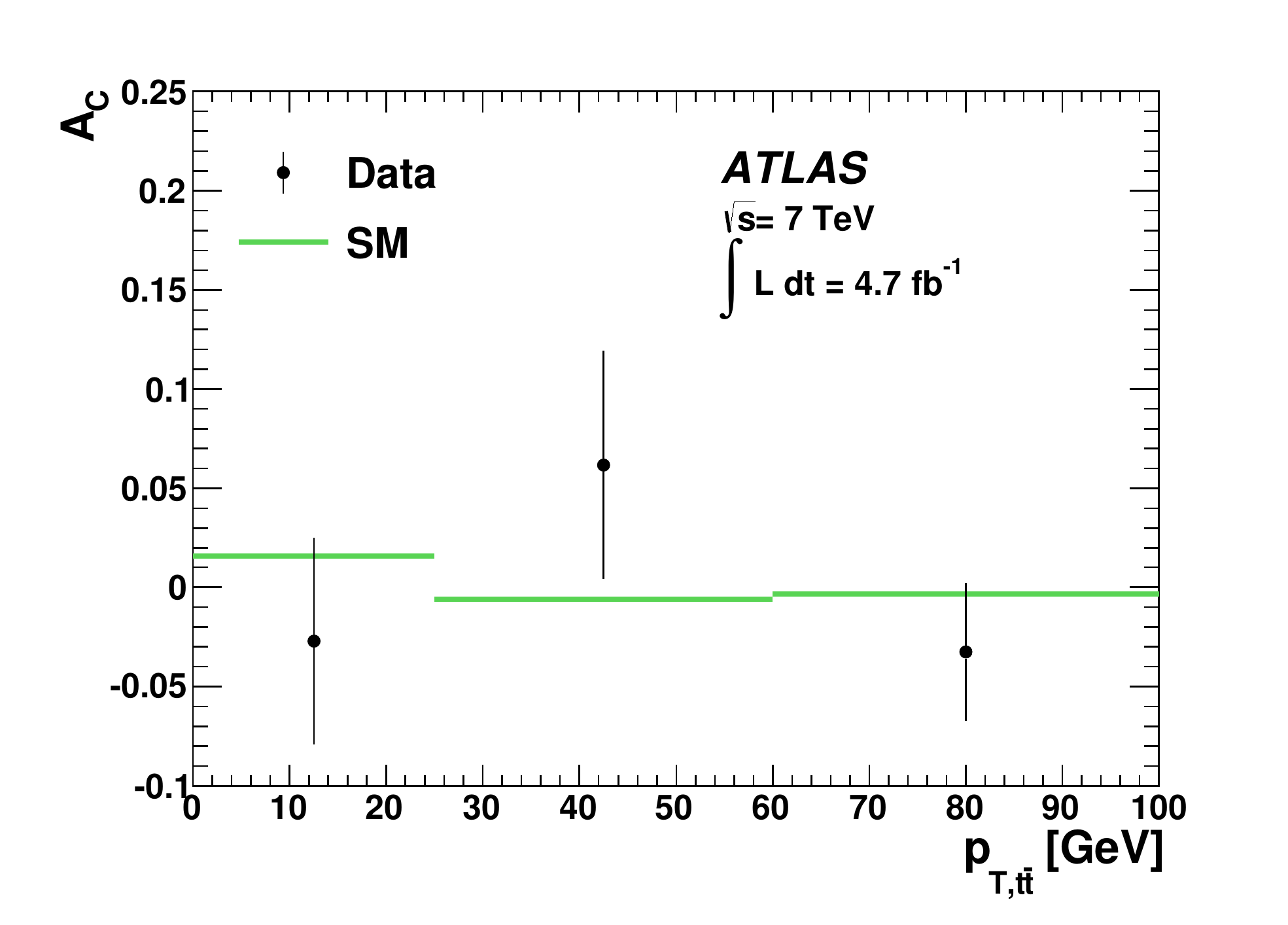} \\
\includegraphics[width=8cm, clip]{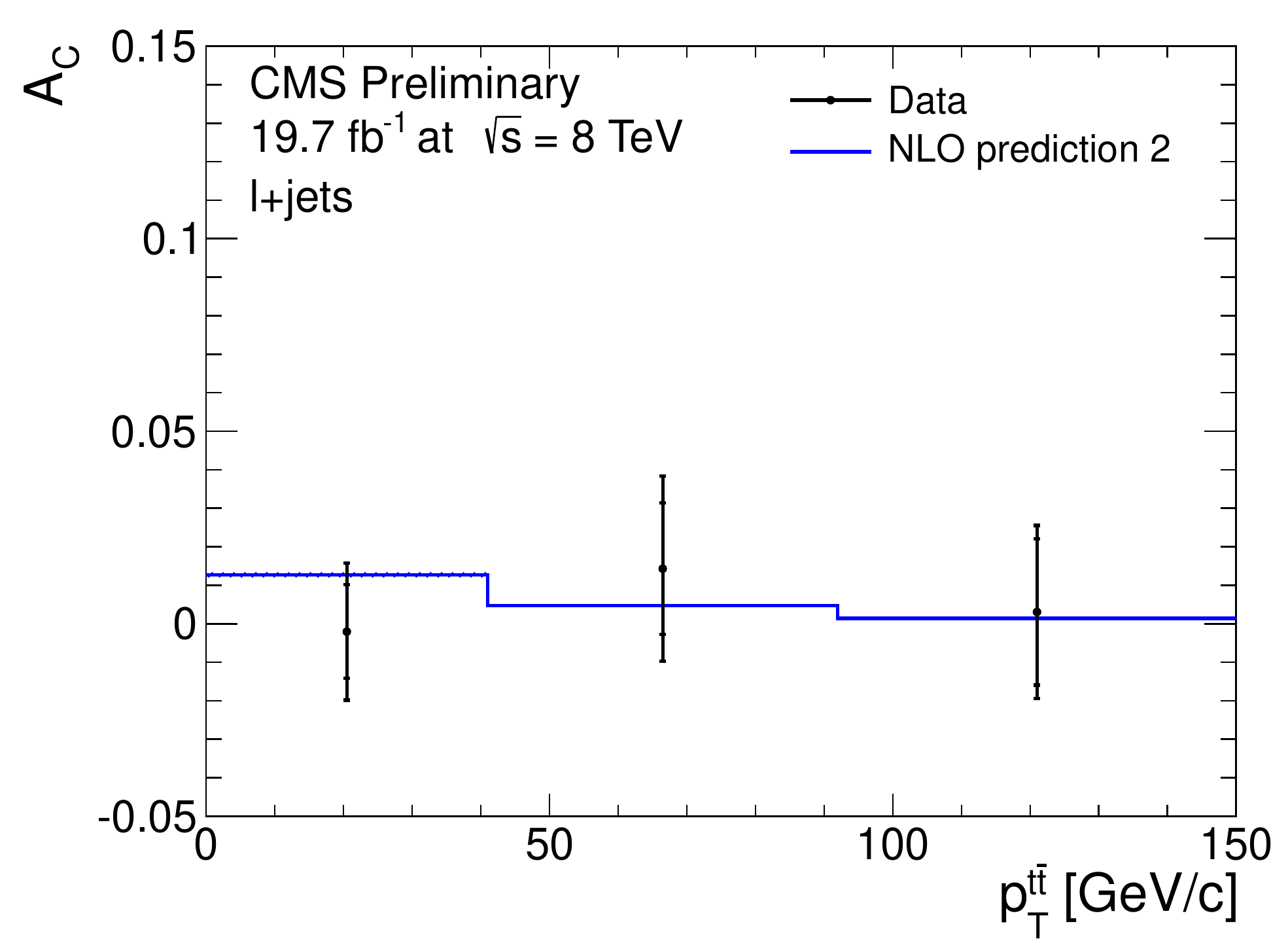}
\end{tabular}
\caption{$\ac$ as a function of $\pTttb$ from the ATLAS measurement (top) and the CMS measurement (bottom).
Unfolded data (dots with error bars representing the total uncertainty) are compared to the SM prediction from~\textcite{Bernreuther:2012sx}.
From~\textcite{Aad:2013cea,CMS:2013nfa}.
 \label{fig:ac_pTtt}}
\end{center}
\end{figure}

\begin{figure}[htb]
\begin{center}
\begin{tabular}{r}  
\includegraphics[width=8cm, clip]{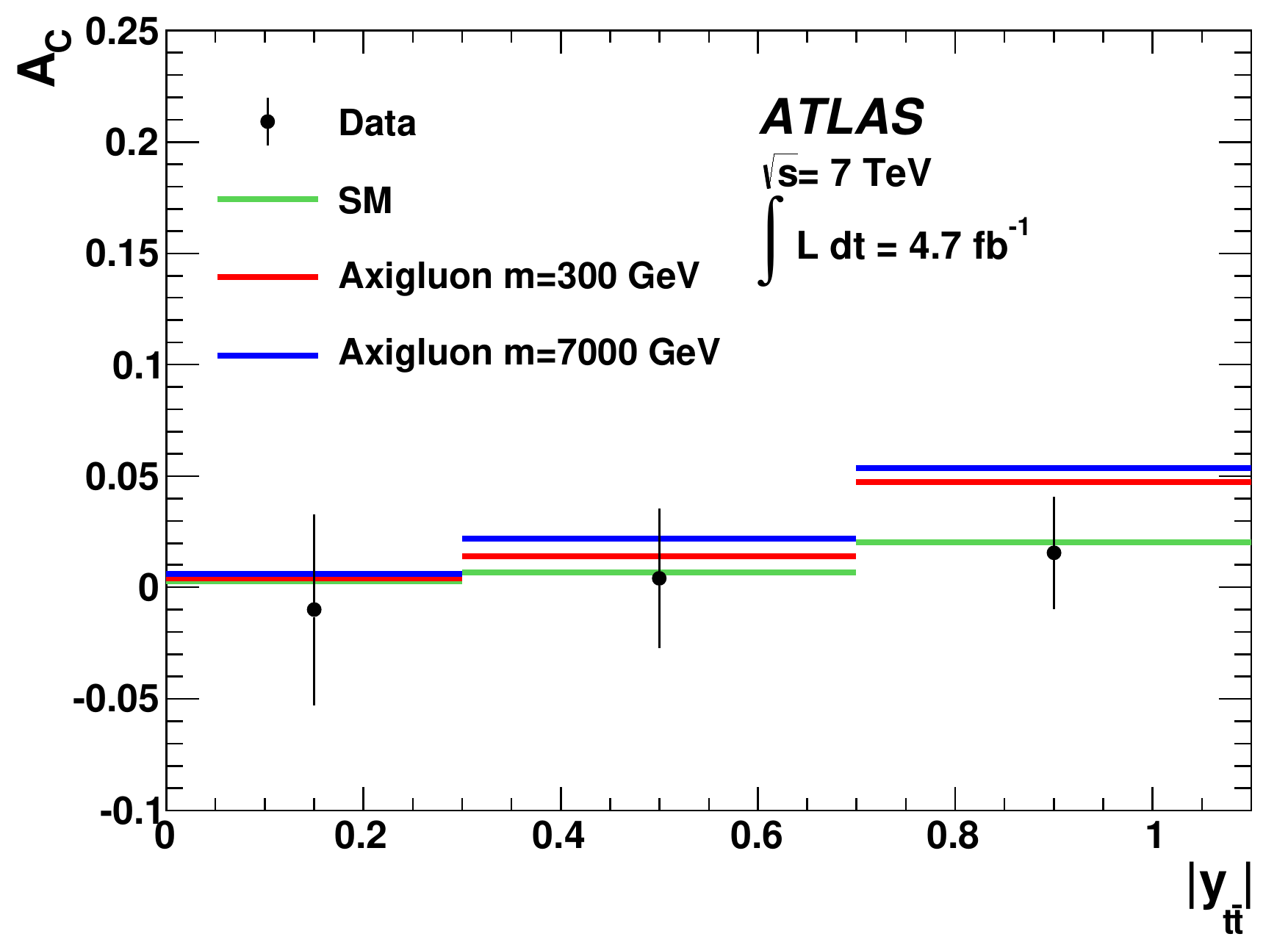} \\
\includegraphics[width=8.3cm, clip]{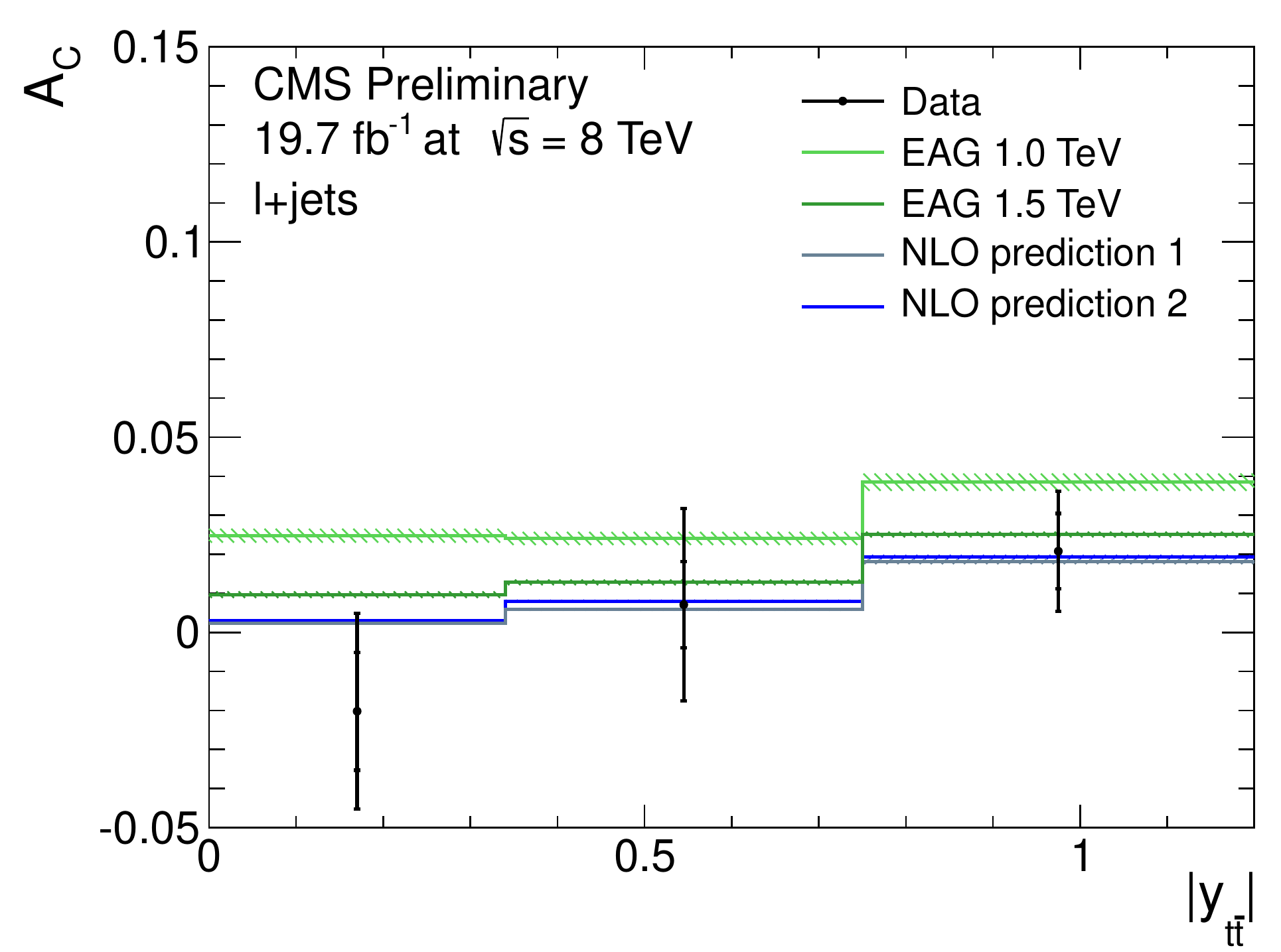}
\end{tabular}
\caption{$\ac$ as a function of $\yttb$ from (top) the ATLAS measurement and (bottom) the CMS measurement. 
Unfolded data (dots with error bars representing the total uncertainty) are compared to the SM predictions from~\textcite{Kuhn:2011ri} (NLO prediction 1) 
and~\textcite{Bernreuther:2012sx} (NLO prediction 2, also referred to as SM in the top figure).
Also shown are the predictions for an axigluon exchanged in the s-channel for two assumed mass 
values~\cite{AguilarSaavedra:2011ci}, as well as for an effective axial-vector coupling of the gluon 
(EAG)~\cite{Gabrielli:2011zw}. From~\textcite{Aad:2013cea,CMS:2013nfa}. \label{fig:ac_ytt}}
\end{center}
\end{figure}

The rapidity of the $\ttb$ system in the laboratory frame, $|\yttb|$, is sensitive to the ratio of contributions from the $q\bar{q}$
and $gg$ initial states to $\ttb$ production, and thus provides a means to enhance the charge asymmetry by increasing
the $q\bar{q}$ fraction~\cite{Kuhn:2011ri}. Indeed, $\ttb$ events produced through $gg$ fusion will tend to populate the central rapidity region,
while $q\bar{q}$-mediated production will typically result in $\ttb$ events boosted along the beam direction and thus
having larger values of $|\yttb|$. A requirement on minimum $|\yttb|$ is equivalent to a requirement on 
the $z$-component of the $\ttb$-system velocity, $\betattb$, since $\yttb=1/2\log [ (1+\betattb)/(1-\betattb) ]$~\cite{AguilarSaavedra:2011cp}.
The ATLAS Collaboration has measured the inclusive charge asymmetry after the requirement of $\betattb>0.6$,
obtaining $\ac = 0.011 \pm 0.017\,{\rm (stat)} \pm 0.007\,{\rm (syst)}$, in good agreement with the SM prediction of 
$\ac = 0.020^{+0.006}_{-0.007}$~\cite{Bernreuther:2012sx}.
Figure~\ref{fig:ac_ytt} shows the unfolded $\ac$ measurements as a function of $\yttb$ from the ATLAS and CMS Collaborations.
Again, good agreement is found with the SM prediction within the experimental uncertainties.

Finally, the dependence of the charge asymmetry on the invariant mass of the $\ttb$ system, $\mttb$, is particularly
interesting because of its sensitivity to new heavy particles mediating $\ttb$ production, whose amplitudes would
interfere with the SM ones, leading to additional contributions (positive or negative) to the charge asymmetry.
The ATLAS Collaboration has measured the inclusive charge asymmetry after the requirement of $\mttb>600$~GeV,
obtaining $\ac = 0.018 \pm 0.021\,{\rm (stat)} \pm 0.005\,{\rm (syst)}$, in good agreement with the SM prediction of 
$\ac = 0.0175^{+0.0005}_{-0.0004}$~\cite{Bernreuther:2012sx}.
More interesting is the differential measurement of the charge asymmetry as a function of $\mttb$, shown in
Figure~\ref{fig:ac_mtt} for both the ATLAS and CMS Collaborations. In addition, the ATLAS Collaboration has
measured this distribution after the requirement of $\betattb>0.6$, shown in Fig.~\ref{fig:ac_mtt_betacut} in an attempt 
to further increase the $q\bar{q}$ fraction, and thus the sensitivity to new physics contributions. 
Such measurement, currently limited by statistical uncertainties, will become more interesting with the full dataset at 8~TeV.
All differential measurements as a function of $\mttb$ are found to be in good  agreement with the SM predictions.

\begin{figure}[htb]
\begin{center}
\begin{tabular}{r} 
\includegraphics[width=8cm, clip]{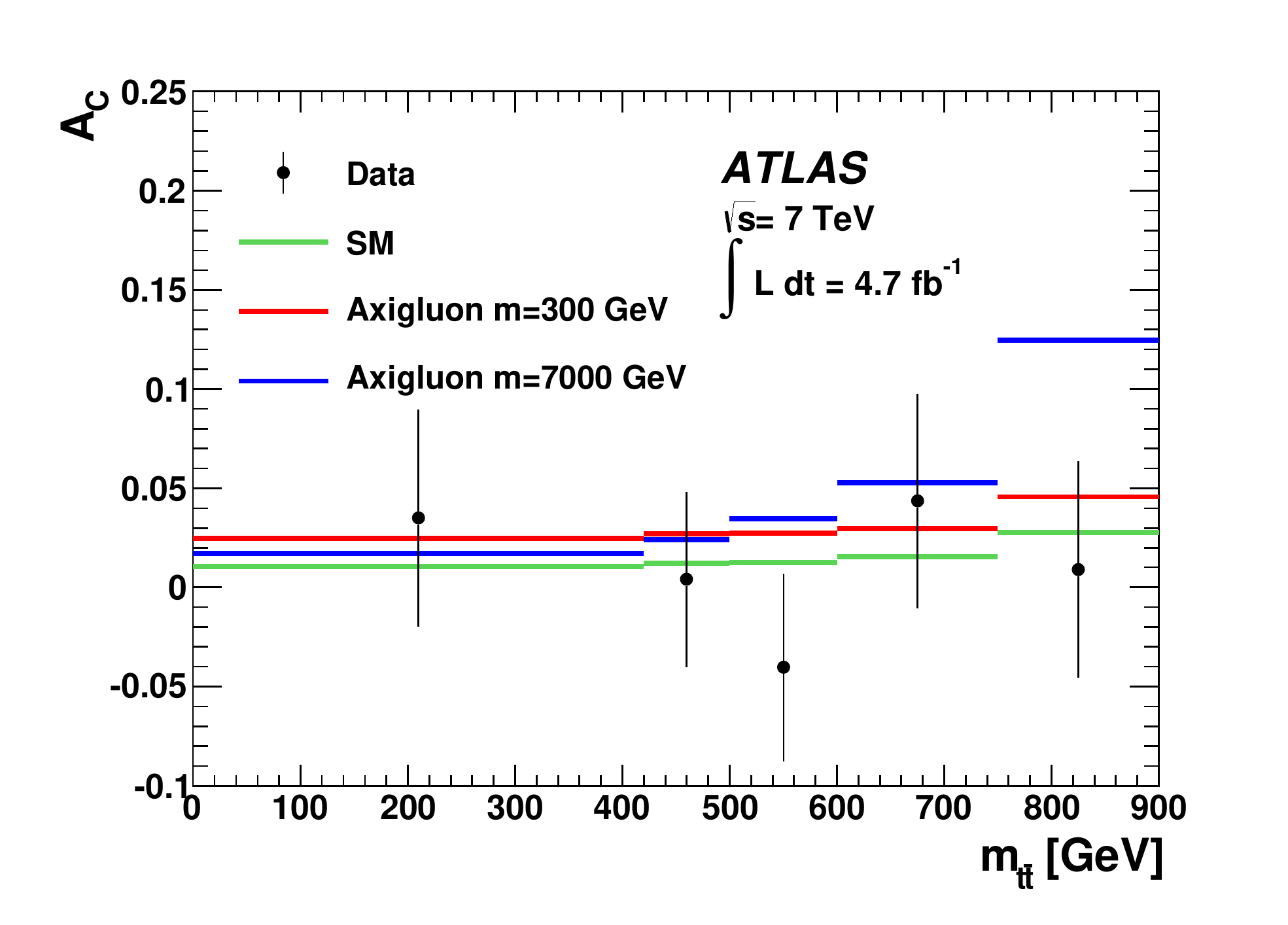} \\
\includegraphics[width=8.3cm, clip]{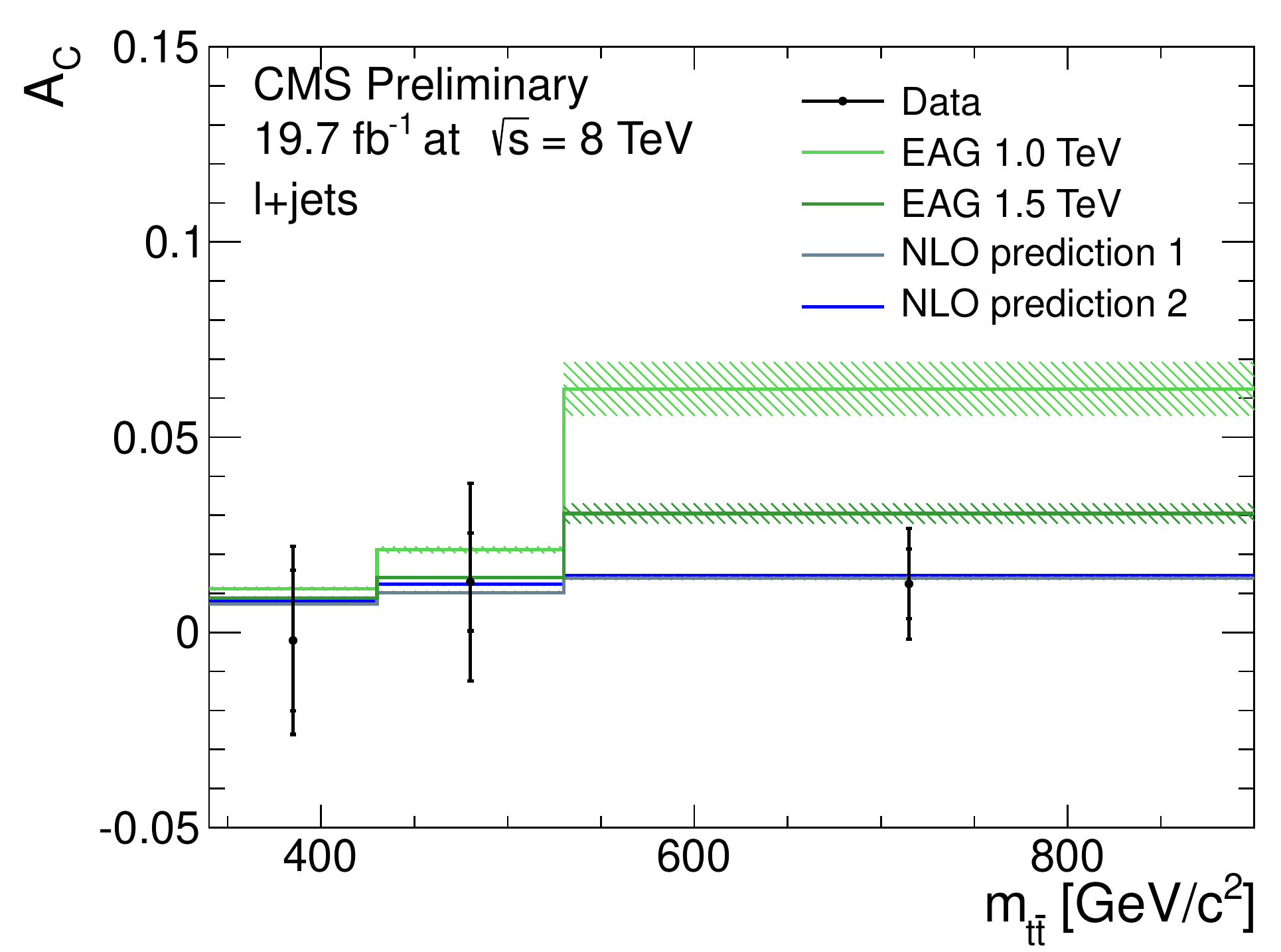} 
\end{tabular}
\caption{$\ac$ as a function of $\mttb$ from (top) the ATLAS measurement and (bottom) the CMS measurement. 
Unfolded data (dots with error bars representing the total uncertainty) are compared to the SM predictions from~\textcite{Kuhn:2011ri} (NLO prediction 1) 
and~\textcite{Bernreuther:2012sx} (NLO prediction 2, also referred to as SM in the top figure).
Also shown are the predictions for an axigluon exchanged in the s-channel for two assumed mass 
values~\cite{AguilarSaavedra:2011ci}, as well as for an effective axial-vector coupling of the gluon 
(EAG)~\cite{Gabrielli:2011zw}. From~\textcite{Aad:2013cea,CMS:2013nfa}. \label{fig:ac_mtt}}
\end{center}
\end{figure}

\begin{figure}[htb]
\begin{center}
\includegraphics[width=8cm, clip]{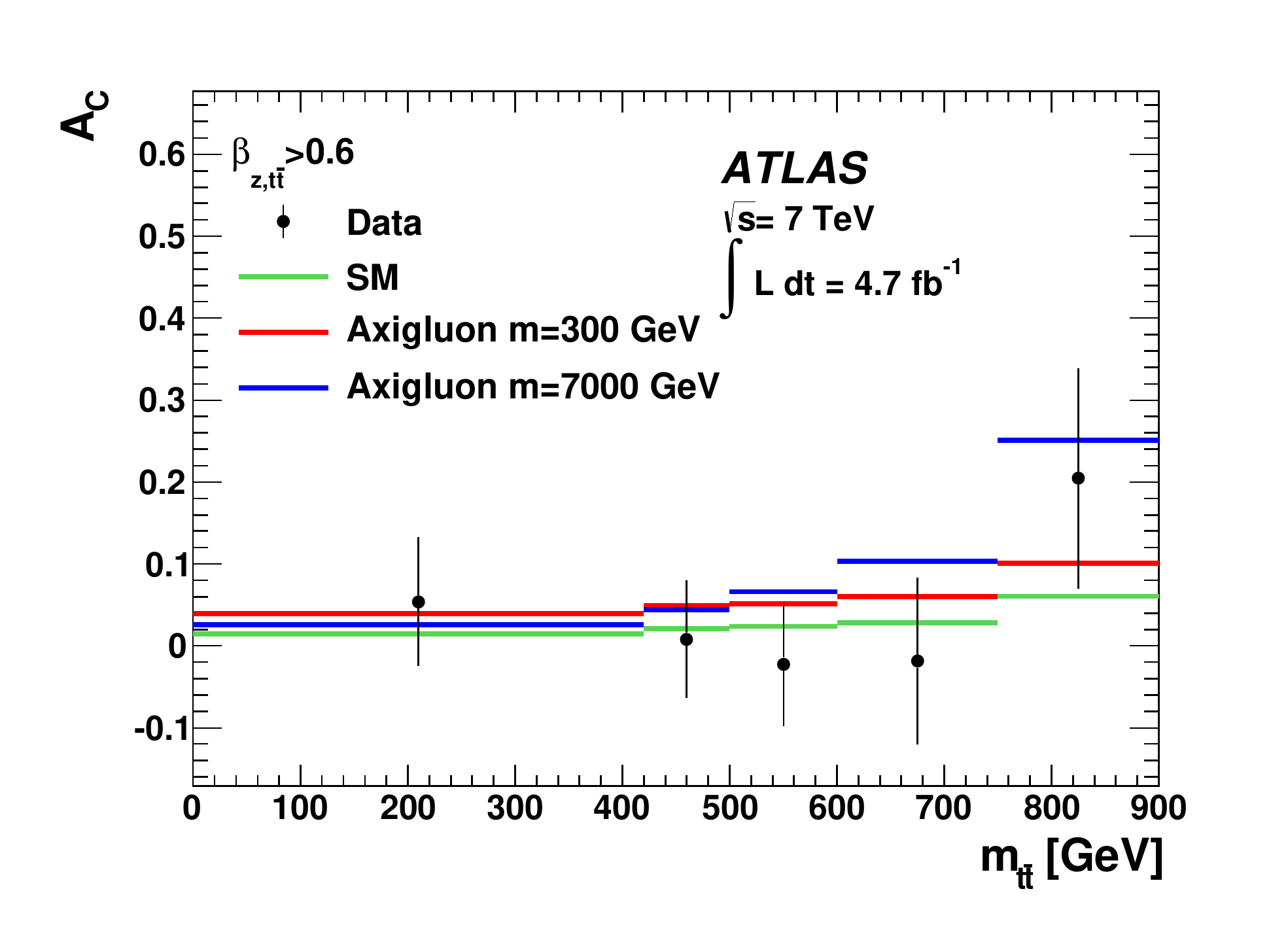}
\caption{$\ac$ as a function of $\mttb$ after the requirement $\betattb>0.6$ from the ATLAS measurement. 
Unfolded data (dots with error bars representing the total uncertainty) are compared to the SM prediction from~\textcite{Bernreuther:2012sx}. 
Also shown are the predictions for an axigluon exchanged in the s-channel for two assumed mass values~\cite{AguilarSaavedra:2011ci}.
From~\textcite{Aad:2013cea}. \label{fig:ac_mtt_betacut}}
\end{center}
\end{figure}

While preliminary combinations of ATLAS and CMS  measurements of the inclusive asymmetry are starting 
to become available, so far only in the $\ell$+jets channel at 7~TeV (see Sec.~\ref{sec:4a1}), combinations of the differential measurements 
have not yet performed, owing to different choices in binning for these kinematic variables adopted by the Collaborations. 
It would be important to harmonize these choices in the near future, in order to be able to maximally exploit the LHC measurements
through their quantitative comparison and eventual combination.

\subsection{Leptonic asymmetry measurements}
\label{sec:4b}

Measurements of the leptonic asymmetry at the LHC have so far only been performed in the dilepton channel.
In this case, the observable used is $\acll$, based on the difference of the absolute values if the pseudorapidities of the
positive and negative leptons (see Eq.~\ref{ec:acll}). Although this asymmetry is diluted by the top quark decay,
it has the advantage that it can be measured without reconstructing the $\ttb$ kinematics, and especially the fact that
it has a very small experimental dilution owing to the precise lepton reconstruction at the LHC experiments.
Existing measurements by the ATLAS and CMS Collaborations are based on the analyses described in Sec.~\ref{sec:4a2}.

Both ATLAS and CMS have measured the inclusive $\acll$ at 7~TeV obtaining 
$\acll = 0.024 \pm 0.015\,{\rm (stat)} \pm 0.009\,{\rm (syst)}$ and  
$\acll = 0.009 \pm 0.010\,{\rm (stat)} \pm 0.006\,{\rm (syst)}$, respectively,
in good agreement with the SM prediction of  $\acll = 0.0070 \pm 0.0003$.
In addition, the CMS experiment has measured $\acll$ differentially as a function of $\pTttb$, $\yttb$  and 
$\mttb$, taking advantage of the kinematic reconstruction performed and discussed in Sec.~\ref{sec:4a2}.
Figure.~\ref{fig:acll_cms} shows the unfolded $\acll$ as a function of $\yttb$ and $\mttb$, compared to the
parton-level predictions from the {\sc mc@nlo} generator. 

\begin{figure}[htb]
\begin{center}
\begin{tabular}{r} 
\includegraphics[width=8cm, clip]{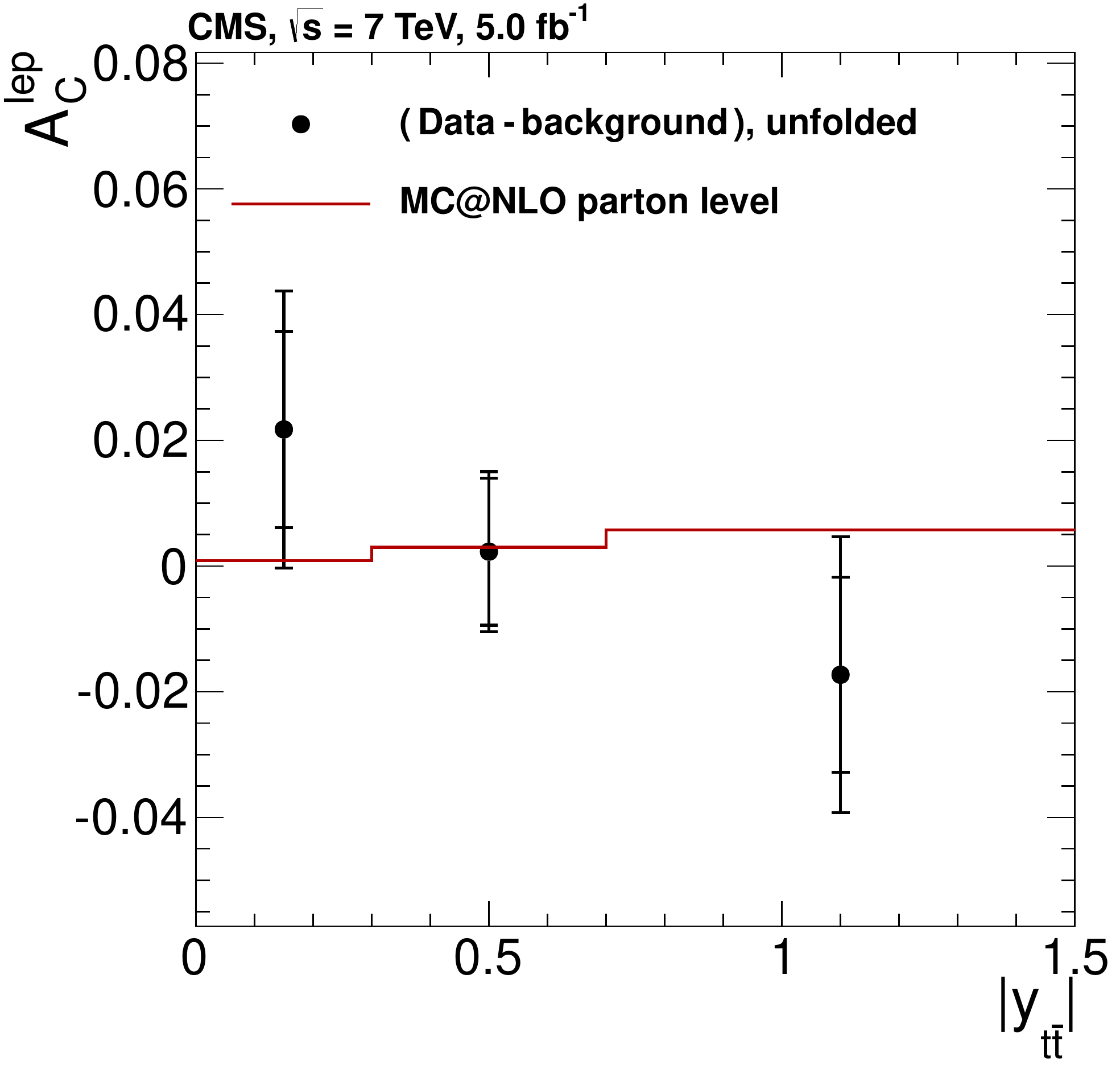} \\
\includegraphics[width=8cm, clip]{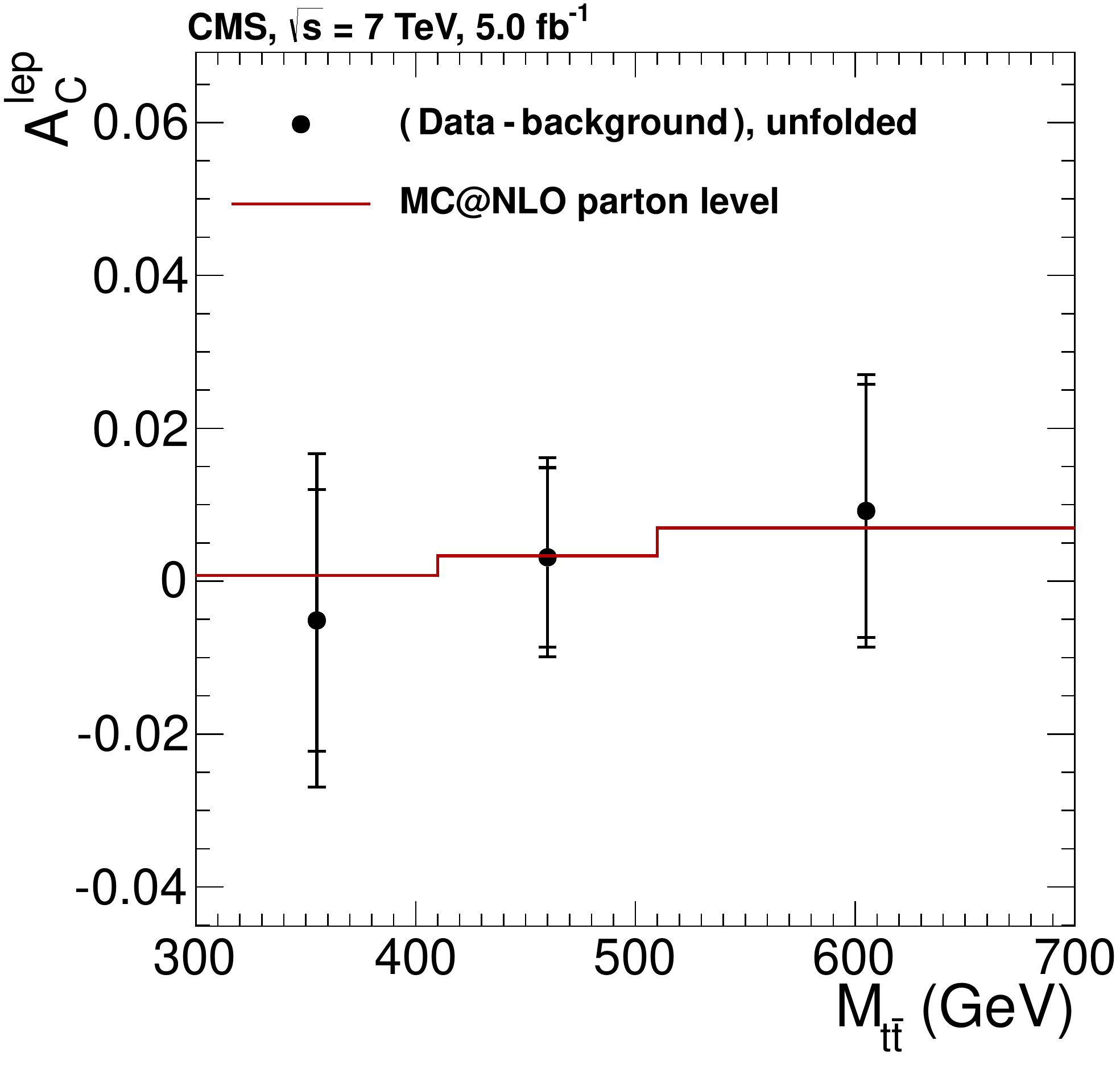} 
\end{tabular}
\caption{Top: $\acll$ as a function of $\yttb$. Bottom: $\acll$ as a function of $\mttb$. 
In these figures $\acll$ are denoted by $A_C^{\rm lep}$. Both measurements are  from the CMS Collaboration. 
Unfolded data (dots with error bars representing the total uncertainty) are compared to the parton-level predictions from {\sc mc@nlo}.
From~\textcite{Chatrchyan:2014yta}. \label{fig:acll_cms}}
\end{center}
\end{figure}

\subsection{LHC summary}
\label{sec:4c}

A summary of the inclusive $\ac$ and $\acll$ measurements by the ATLAS and CMS Collaborations
is shown in Fig.~\ref{fig:ac-summary}. Also shown are simple error-weighted averages for the 7 TeV measurements, 
neglecting correlations in systematic uncertainties between both experiments. All measurements so far are found 
to be consistent with the SM predictions. The complete set of inclusive measurements using the full 8 TeV dataset, 
as well as differential measurements combining multiple channels within and across experiments would be quite 
important for more precise tests.

\begin{figure}[htb]
\begin{center}
\includegraphics[width=7.5cm, clip]{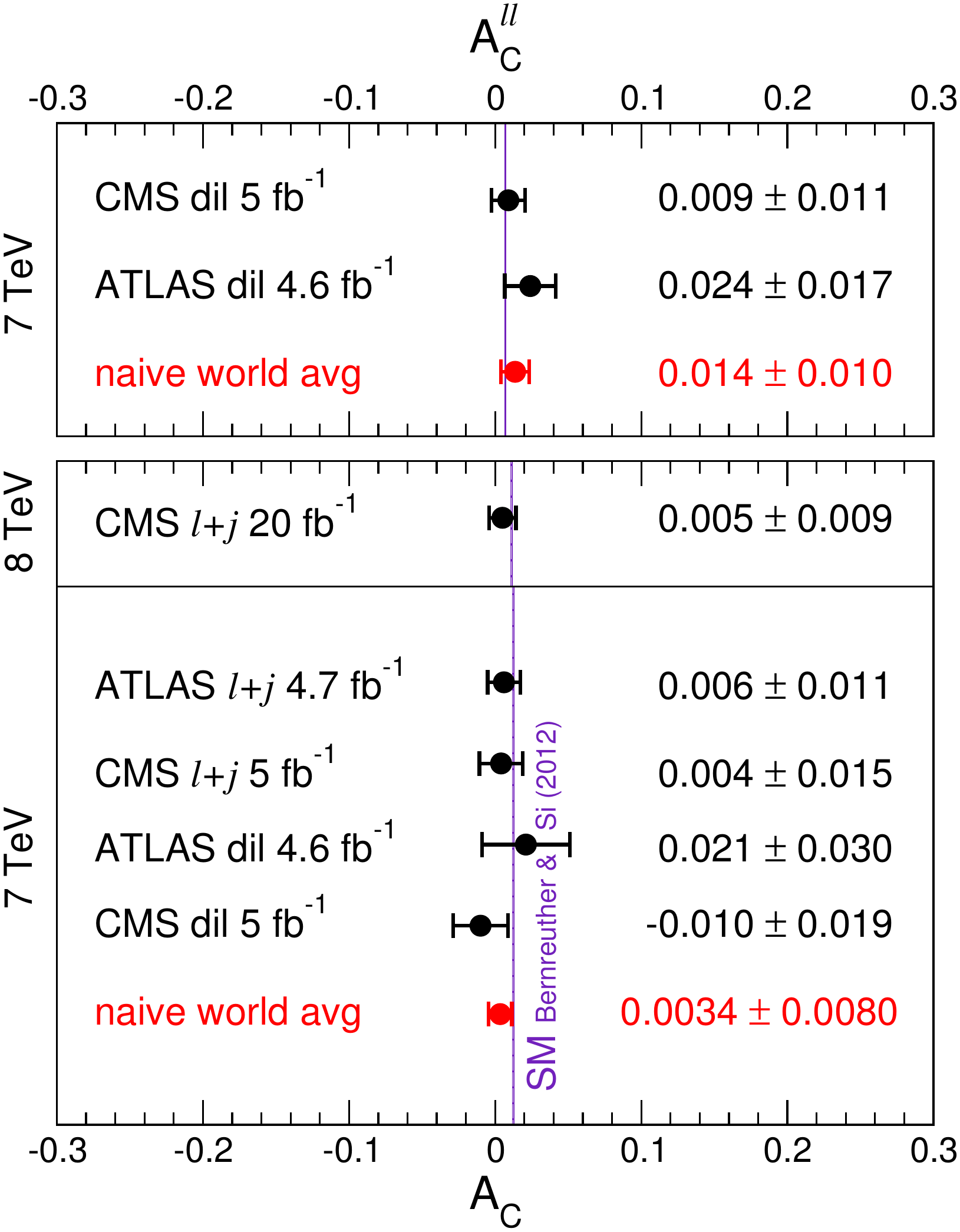}
\caption{Summary of inclusive $\ac$ and $\acll$ measurements by the ATLAS and CMS Collaborations, compared to the
respective SM predictions.  The uncertainties on the experimental measurements include both statistical and systematic contributions. 
Also shown are simple error-weighted averages of ATLAS and CMS measurements at 7 TeV. 
\label{fig:ac-summary}}
\end{center}
\end{figure}

\subsection{Future LHC prospects}
\label{sec:4d}

Because of the increased importance of $gg \to \ttb$ with rising CM energy, the predicted SM asymmetry for the second LHC run is roughly one half of the asymmetry for $7-8$ TeV, {\it i.e.} $\ac = 0.0067 \pm 0.0004$ at 14 TeV~\cite{Bernreuther:2012sx}. The measurement will be difficult and demanding, and, likely, rather unconclusive. If one assumes that the systematic uncertainties will be of the same magnitude as in current measurements (this may be too optimistic due to the increased pile-up), the uncertainty will still be of the same order as the asymmetry itself, with the disadvantage with respect to $7,8$ TeV that potential deviations from the SM are further smeared by $gg$ fusion. On the other hand, one can exploit the high $\ttb$ statistics to make measurements at high $\betattb$, $\mttb$ or $|y_t|$, where the SM prediction is larger. In this respect, an interesting proposal is to exploit the large coverage of the LHCb detector to make measurements in the very forward region $2 \leq |y_t| \leq 5$~\cite{Kagan:2011yx}.

The large luminosity and cross sections at the second LHC run will also allow for  measurements in the production of $\ttb$ pairs in association with a photon~\cite{Aguilar-Saavedra:2014vta} or a $W$ boson~\cite{Maltoni:2014zpa}. A charge asymmetry can be also defined in these processes as in~(\ref{ec:ac}),
\begin{equation}
\ac^{\ttA,\ttb W} = \frac{N(\Delta |y| >0) - N(\Delta |y| <0)}{N(\Delta |y| >0) + N(\Delta |y| <0)} \,.
\label{ec:acA}
\end{equation}
In $\ttA$, the presence of the photon enhances the $\qqb$ fraction with respect to $\ttb$ production, since a photon cannot be emitted from initial gluons (and radiative top decay $t \to Wb\gamma$ can be suppressed with suitable kinematical cuts). Additionally, the extra photon changes the relative importance of $\uub$ and $\ddb$ contributions, since it couples differently to up and down quarks. We show in Fig.~\ref{fig:ttASM} the $\qqb$ fraction $F_u+F_d$ and the ratio $F_d/F_u$ for $\ttb (\gamma)$ production at the LHC with 8 and 14 TeV, as well as for $\ttb$ at the Tevatron. The presence of the extra photon is much more effective to ``approach'' the Tevatron point than kinematical cuts on $\betattb$ or $\mttb$.
\begin{figure}[htb]
\begin{center}
\includegraphics[height=7cm,clip=]{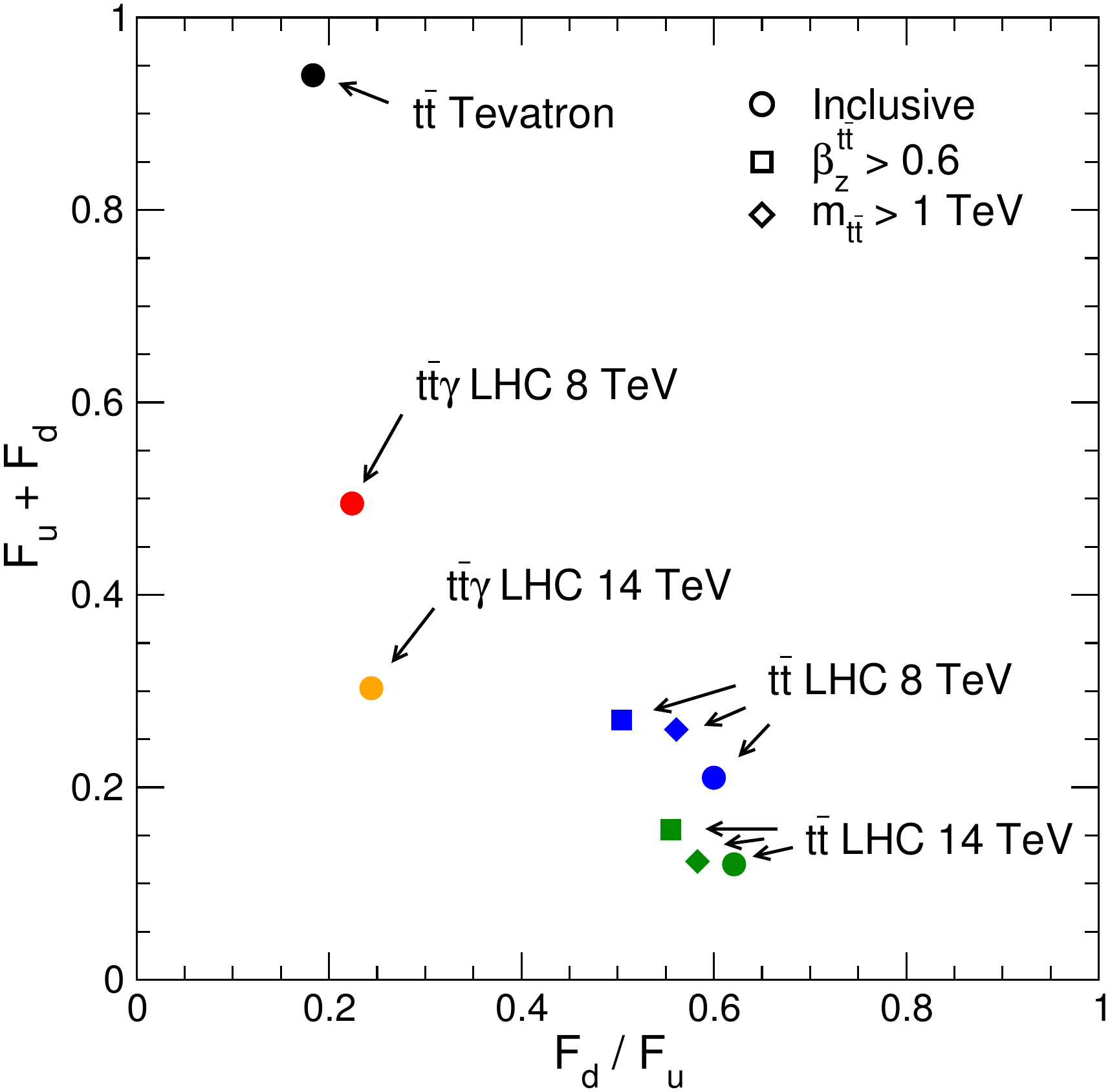}
\end{center}
\caption{$F_d/F_u = \sigma(\ddb) / \sigma(\uub)$ and $q\bar q$ fraction $F_u + F_d$ for $\ttbar$ and $\ttA$ production in the SM. For comparison, we also plot these quantities in $\ttb$ production after imposing high-$\mttb$ or high-$\betattb$ requirements. From~\textcite{Aguilar-Saavedra:2014vta}.}
\label{fig:ttASM}
\end{figure}

Within the SM the asymmetry in $\ttA$, $\acA = -0.038$ at 14 TeV, appears already at the tree level, due to the interference of photon emission in the initial state and from a top quark. New physics can also contribute to this asymmetry. Intriguingly, if there exists some conspiracy between new physics contributions in $\uub$ and $\ddb$ initial states to render a SM-like $\ac$ in $\ttb$ production at the LHC (see Sec.~\ref{sec:2c}), a measurement of the asymmetry in $\ttA$ could uncover it, since the balance between these contributions is different for this process. This is shown in Fig.~\ref{fig:ttANP}, using as new physics benchmark a new color octet with mass $M=250$ GeV and arbitrary couplings to the up, down and top quark that give a good fit to all $t \bar t$ data. It is observed that, even when the asymmetry in $\ttb$ is close to the SM prediction, there can be sizeable deviations in $\ttA$. 

\begin{figure}[htb]
\begin{center}
\includegraphics[height=5.5cm,clip=]{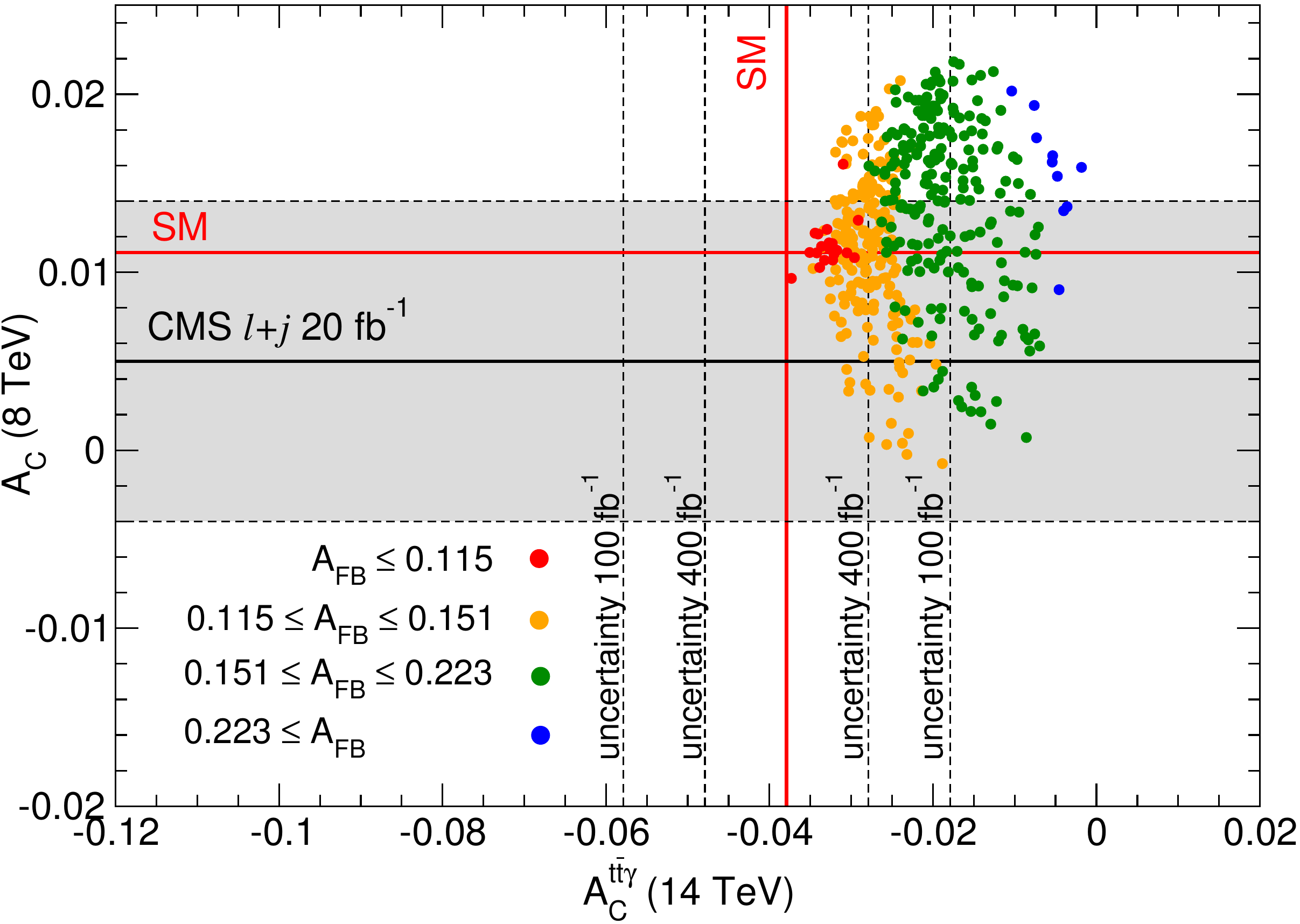}
\end{center}
\caption{Charge asymmetry $\acA$ in $\ttA$ at 14 TeV versus $\ac$ at 8 TeV for points in the parameter space of a new color octet. The horizontal band is the current 8 TeV measurement in~\textcite{CMS:2013nfa} and its uncertainty. The vertical dashed lines represent the expected statistical uncertainty for 100 fb$^{-1}$ and 400 fb$^{-1}$. The SM predictions are also included. From~\textcite{Aguilar-Saavedra:2014vta}.}
\label{fig:ttANP}
\end{figure}

An independent handle is provided by $\ttb W^\pm$ production. At LO, $\ttb W^+$ ($\ttb W^-$) can only be produced from $u\bar d,c\bar s$ ($\bar u d,\bar c s$) states, and symmetric $gg$ fusion only contributes at NNLO. Hence the asymmetry generated is also larger than in the SM, $\acW = 0.022^{+0.0043}_{-0.0033}$ at NLO for 14 TeV~\cite{Maltoni:2014zpa}.\footnote{In this reference, the asymmetry is built using the $t$, $\bar t$ pseudo-rapidities rather than the rapidities; the difference with respect to the asymmetry defined from rapidities is small. Also, NLO denominators are used.} New physics can also contribute to this asymmetry, and the deviations with respect to the SM prediction may be more significant than in $\ttb$ production. This is demonstrated in Fig.~\ref{fig:ttW}, using as new physics benchmark a color octet with mass $M=200$ GeV (labelled as I, II) and $M=2$ TeV (III, IV). The coupling is chosen as left-handed (I, III) or axial (II, IV). (For a purely right-handed coupling to the light quarks the contribution of the octet to the amplitude vanishes and its presence would be unnoticed in $\ttb W^\pm$.) The upper panel corresponds to the asymmetry in $\ttb$, and the lower panel to the asymmetry in $\ttb W^\pm$. 

\begin{figure}[htb]
\begin{center}
\begin{tabular}{c} 
\includegraphics[width=8cm, clip]{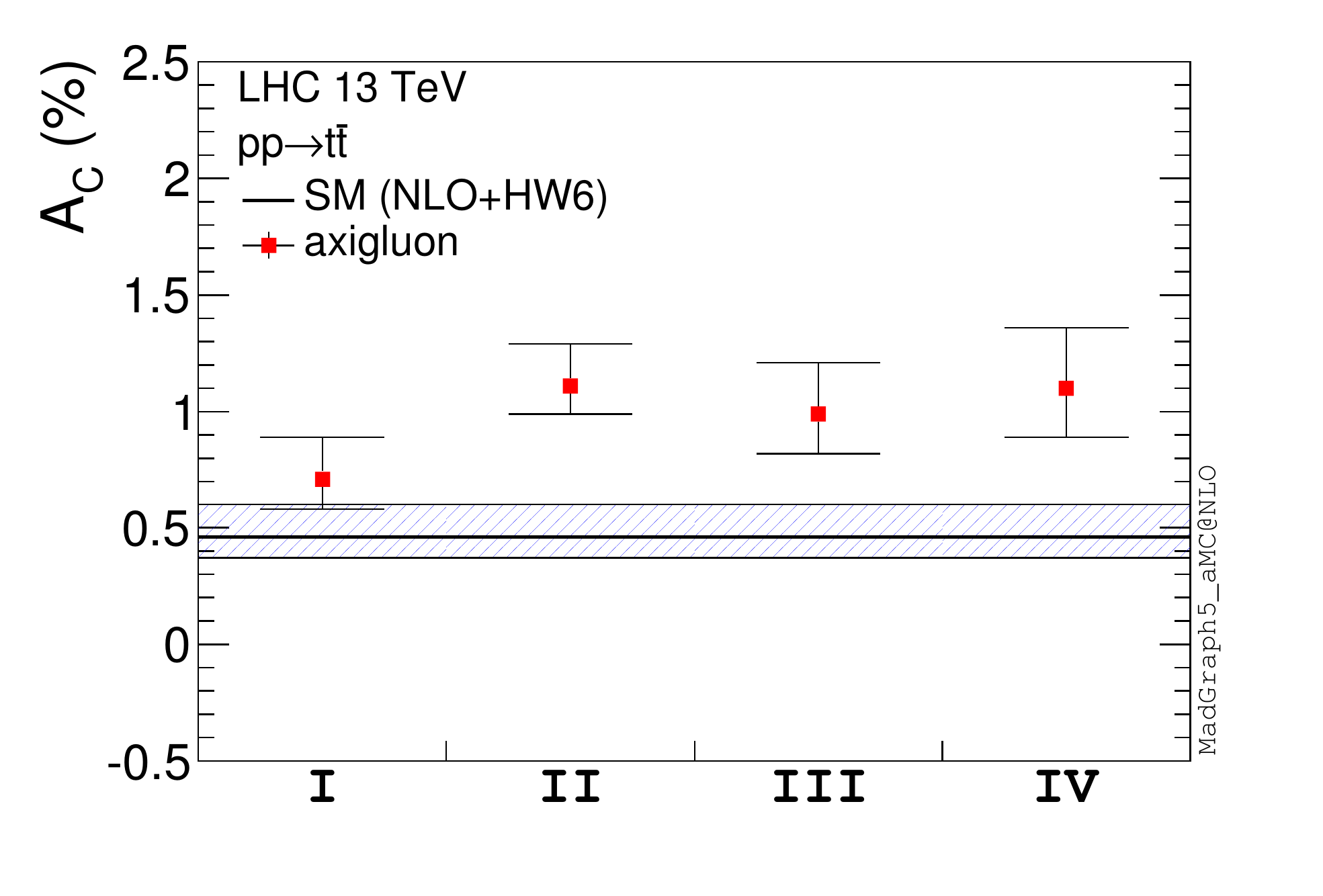} \\[2mm]
\includegraphics[width=8cm, clip]{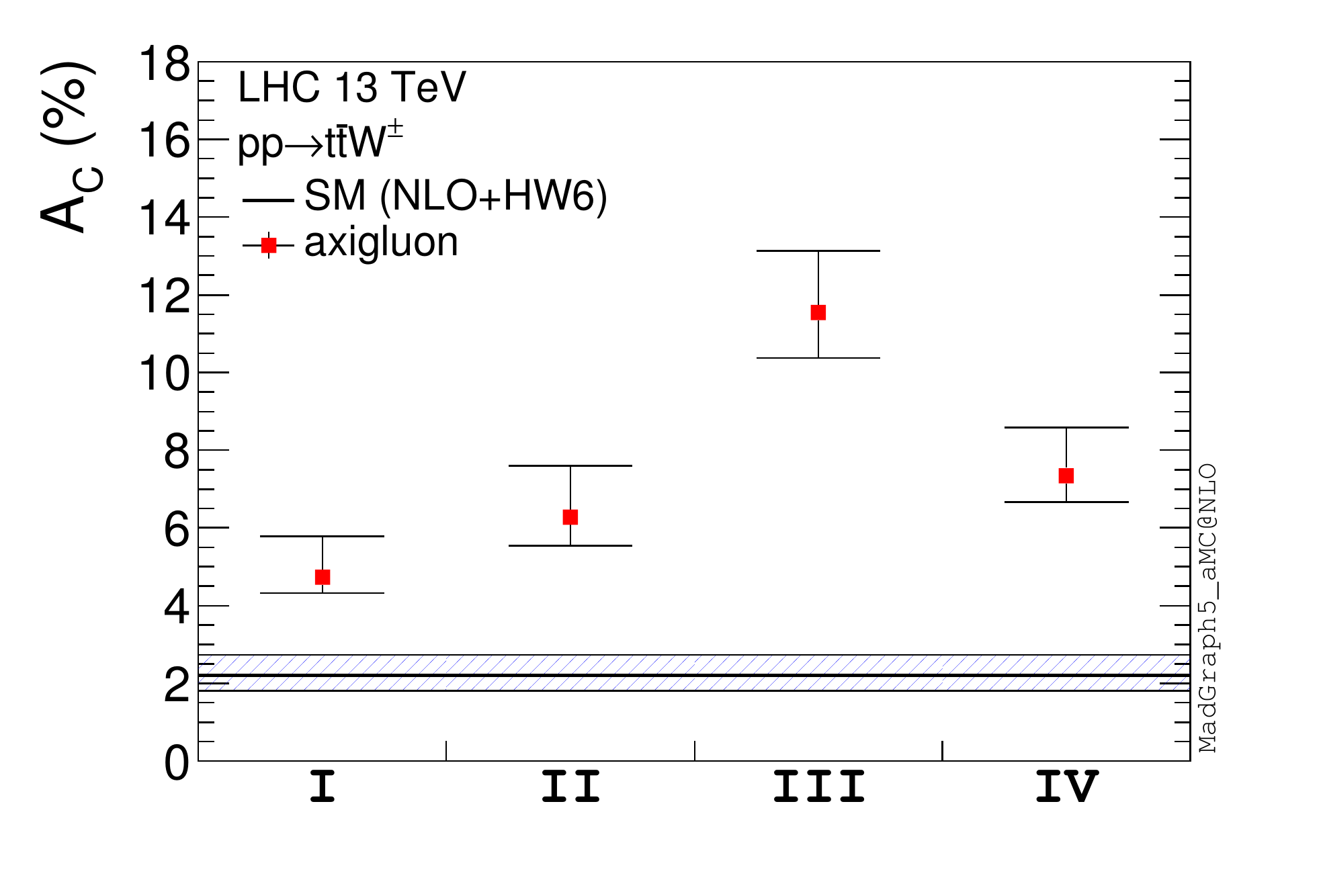} 
\end{tabular}
\caption{Top: Charge asymmetry in $\ttb$ production at 13 TeV for the SM (horizontal band) and four benchmark points of a color octet, labelled as ``axigluon''. Bottom: The same, but in $\ttb W^\pm$ production. 
From~\textcite{Maltoni:2014zpa}. \label{fig:ttW}}
\end{center}
\end{figure}


\section{New physics interpretations}
\label{sec:5}
An appealing possibility is that the deviations between (some of) the experimental results and the SM predictions for $\afb$ are a signal of new physics in $\ttb$ production. This would agree well with the general expectation that the top quark may be particularly sensitive to new physics in the electroweak breaking sector, due to its large Yukawa coupling.\footnote{This is actually the case in many popular explicit models. For instance, in composite-Higgs models, the large Yukawa coupling of the top quark arises from the fact that it is mostly composite. Therefore, it couples strongly to the resonances of the composite sector. These resonances could thus mediate top-quark pair production and contribute to the charge asymmetries. More generally, all natural scenarios of electroweak symmetry breaking introduce new particles associated to the top-quark to partially cancel its large radiative contributions to the Higgs mass. These particles often give rise to new effects in the top-quark sector.} In the last few years, different extensions of the SM have been proposed to explain the excess of the measured $\afb$. To match the experimental central values, the new-physics contribution must be comparable to the SM one. Then, some effects in other observables, either in $\ttb$ production or elsewhere, can be generically expected. We discuss them in sections \ref{sec:6} and \ref{sec:7}, respectively. These effects often translate into strong constraints on viable explanations.

The most obvious constraint is already used in this section as a first filter to select the possible new physics that could explain the discrepancies. It comes from the agreement of the SM prediction of the total cross section at Tevatron, $\sigma_{\mathrm{SM}}=7.5\pm 0.5~\mathrm{pb}$~\cite{Aliev:2010zk}, and the measured value, $\sigma_{\mathrm{exp}}=7.60\pm0.41~\mathrm{pb}$~\cite{Aaltonen:2013wca}. Writing the cross section in the presence of new physics as $\sigma = \sigma_\mathrm{SM} + \delta \sigma_\mathrm{int} + \delta \sigma_\mathrm{quad}$, this agreement implies that the interference between SM and new-physics amplitudes, $\delta \sigma_\mathrm{int}$, and the modulus square of the new-physics amplitude, $\delta \sigma_\mathrm{quad}$, must satisfy the condition 
\begin{equation}
\delta \sigma_\mathrm{int} + \delta \sigma_\mathrm{quad} \simeq 0. \label{ec:sigmacancel}
\end{equation}
This equation requires $\sigma_\mathrm{int}\lesssim 0$, since $\delta \sigma_\mathrm{quad}$ is positive semidefinite. To obtain a tighter bound, let us decompose $\delta \sigma = \delta \sigma^F+\delta \sigma^B$, where
$\delta \sigma^{F}$ and $\delta \sigma^{B}$ represent the contributions of the forward and backward hemispheres, respectively, to the corresponding terms in the cross section. Eq.~(\ref{ec:sigmacancel}) allows us to approximate the new-physics contribution to the FB asymmetry, 
$\Delta \afb = \afb- \afb^\mathrm{SM}$, as
\begin{equation}
\Delta \afb \simeq \frac{\delta \sigma_\mathrm{int}^F - \delta \sigma_\mathrm{int}^B + \delta \sigma_\mathrm{quad}^F - \delta \sigma_\mathrm{quad}^B}{\sigma_\mathrm{SM}}.
\end{equation}
Because the term $\delta \sigma_\mathrm{quad}^B$ is positive semidefinite, the model-independent condition~(\ref{ec:sigmacancel}) implies, in particular, $\delta\sigma_\mathrm{int}^B \lesssim - \frac{1}{2} \Delta \afb \sigma_\mathrm{SM}$. A sizable positive $A_\mathrm{FB}^\mathrm{new}$ thus requires a sizable and negative $\delta\sigma_\mathrm{int}^B$~\cite{Grinstein:2011yv}. So, we learn that new physics interfering with the SM amplitudes---with the tree-level ones for a significant effect---is preferred.  Conversely, incoherent new physics (see~\onlinecite{Isidori:2011dp}  for an example) cannot generate a large asymmetry. 

The restrictions on new physics stemming from  Eq.~(\ref{ec:sigmacancel}) are much stronger than the plain requirement of interference, which in practice is quite mild. We can distinguish two scenarios:
\begin{itemize}
\item {\em Linear new physics.} If $\delta \sigma_\mathrm{int} \simeq 0$, the quadratic terms $\delta \sigma_\mathrm{quad}^{F,B}$ must be suppressed, so sizable corrections to the asymmetry come from the interference terms only. The suppression of quadratic terms is natural when the scale of new physics is large or its couplings to the SM fields small. 
\item {\em Quadratic new physics.} If $\delta \sigma_\mathrm{int}$ is sizable (and necessarily negative), $\Delta \afb$ can be produced by interference and/or quadratic contributions, with the same or opposite sign. In this scenario, the cancellation (\ref{ec:sigmacancel}) is non-trivial. This has important consequences: First, the parameters of the theory have to be tuned, generically, as they appear with different powers in both terms. Second, to give rise to sizable quadratic contributions, the new physics must either be present at low scales (below 1~TeV) or couple strongly to the SM. Third, this cancellation, imposed at the Tevatron relevant energies, needs not hold at higher energies. As discussed in Sec.~\ref{sec:6}, this typically leads to an excess in the LHC $\ttb$ cross section at high values of $\mttb$.
\end{itemize}
As customary, we neglect in the following the interference of new physics with NLO QCD amplitudes. The error in the asymmetry in this approximation can be estimated to be smaller than or comparable to $\afb^\mathrm{SM}$, in the linear and quadratic scenarios, respectively.

\subsection{Heavy new physics}
\label{sec:5a}
If the new degrees of freedom are heavy in comparison with Tevatron and LHC energies, their effect can be parametrized model-independently by an effective Lagrangian that involves only the SM fields. In view of the recent experimental results in Higgs physics~\cite{ATLAS-CONF-2014-009,Chatrchyan:2014vua}, we include the Higgs doublet explicitly as a SM field, and assume that electroweak symmetry is broken by a vacuum expectation value of this field, just as in the SM.  In the electroweak symmetric phase, the effective Lagrangian must be invariant under $SU(3)_C\times SU(2)_L \times U(1)_Y$ gauge transformations. It can expanded as
\begin{equation}
\Lcal_\mathrm{eff} = \Lcal_\mathrm{SM} + \sum_{n=1}^\infty \frac{1}{\Lambda^n} \sum_{i=1}^{r_n} \left[C^{(n)}_{i} \Ocal^{(n)}_{i} + \mathrm{h.c.} \right], \label{ec:effLag}
\end{equation}
where $\Lcal_\mathrm{SM}$ is the SM Lagrangian, $\Lambda$ is the characteristic scale of new physics, $C^{(n)}_i$ are dimensionless coefficients and $\Ocal^{(n)}_i$ are local gauge-invariant operators of dimension $4+n$. The expansion~(\ref{ec:effLag}) is appropriate for decoupling new physics, which can be either weakly or strongly interacting. This is related to the renormalizability of $\Lcal_\mathrm{SM}$. The first terms are expected to be give a good approximation for processes with typical energy $E \ll \Lambda$. In the case of $\ttb$ production, we have $E \lesssim \mttb$.

The $\ttb$ cross section and asymmetry calculated with $\Lcal_\mathrm{eff}$ inherit this perturbative structure. The first corrections to the SM predictions appear at order $1/\Lambda^2$, from the interference of the SM amplitude with an amplitude that has one insertion of a dimension-6 operator. These are the only effects that need to be taken into account in the linear scenario. For the quadratic scenario,  on the other hand, we also need to consider $1/\Lambda^4$ corrections. They can arise in three different manners: (i) from the interference of the SM amplitude with an amplitude with two insertions of dimension-6 operators; (ii) from the modulus square of an amplitude with one insertion of a dimension-6 operator; (iii) from the interference of the SM amplitude with an amplitude with one insertion of a dimension-8 operator. 

The contributions of the third kind depend on many free parameters and have always been neglected in the literature. Several arguments have been given to justify this approximation \cite{Delaunay:2011gv,AguilarSaavedra:2011vw}. Let us consider a fixed scale $\Lambda$, much larger than the $\ttb$ invariant masses that the Tevatron and the LHC can produce. If the series is to converge rapidly, it is plausible that the coefficients $C_i^{(4)}$ are not larger than the coefficients $C_i^{(2)}$, to ensure that $C_i^{(4)} \mttb^2/\Lambda^2 \ll C_i^{(2)}$. Because it turns out that some coefficients $C_j^{(2)}$ relatively larger than 1 are needed in the quadratic scenario, this implies that the coefficients $C_i^{(4)}$ are smaller than $[{C_j^{(2)}}]^2$. Then, the contributions from dimension-8 operators will be  suppressed with respect to the other $1/\Lambda^4$ contributions. This behavior has been confirmed in explicit weakly-coupled models and it is also expected, from naive dimensional analysis, in strongly-coupled theories. On the other hand, the interfering dimension-8 operators will have a structure similar to the one of interfering dimension-6 operators (typically, with additional covariant derivatives), so to a large extent their effects on inclusive observables can be absorbed into corrections to the coefficients of the dimension-6 operators.
All this suggests that considering only the corrections from dimension-6 operators does not entail a significant loss of generality, even in a quadratic scenario. A direct consequence of this approximation is that all the observables will depend only on ratios $C_i^{(2)}/\Lambda^2$. Let us add, nevertheless, a word of caution: the hierarchy of coefficients we have assumed above may be spoiled by the elimination of redundant dimension-6 operators, as the necessary field redefinitions can induce dimension-8 operators with $C_i^{(4)} \sim [{C_i^{(2)}}]^2$. For this reason, we consider in the following a complete basis of dimension-6 operators, including the ones that vanish by the dimension-4 equations of motion. Of course, when working to order $1/\Lambda^2$ the redundant operators can be safely eliminated. 

We focus on the operators contributing to  $q \bar{q} \to \ttb$ partonic processes, $q=u,d$, which are the most relevant at the Tevatron. We can distinguish operators with one quark current, which modify the trilinear vertices of the SM amplitudes, and operators with two fermionic currents, which produce contact four-fermion interactions. Oblique corrections do not generate an asymmetry, and are furthermore very restricted by gauge symmetry and electroweak precision tests.

Let us first discuss operators with one quark current, which can modify the vertices in the SM diagrams with exchange of a gauge boson. Amplitudes with $s$-channel exchange of a $\gamma$ or $Z$ boson with anomalous couplings will give contributions suppressed by $\alpha^2/ \Lambda^2$, since they do not interfere with the SM gluon amplitude, due to the different color structure. On the other hand, flavor-changing $dtW$ and $utZ$ anomalous couplings, which would give rise to $t$-channel amplitudes and $\alpha \alpha_s/\Lambda^2$ interfering contributions, must be very small, due to flavor-physics constraints. We can thus focus on quark-quark-gluon vertices in diagrams with an exchanged gluon. There are eight non-flavor-changing  operators of this type:
\begin{align}
& \Ocal_{uG\phi} = \bar{q}_L \lambda^a \sigma^{\mu\nu} u_R \tilde{\phi} G_{\mu\nu}^a, ~ &  \Ocal_{tG\phi} = \bar{Q}_L \lambda^a \sigma^{\mu\nu} t_R \tilde{\phi} G_{\mu\nu}^a, \nonumber \\
& \Ocal_{dG\phi} = \bar{q}_L \lambda^a \sigma^{\mu\nu} d_R \phi G_{\mu\nu}^a  \label{ec:chromomoments}
\end{align}
and
\begin{align}
& \Ocal_{qG} = \bar{q}_L \lambda^a D^\nu q_L G_{\mu\nu}^a, ~ 
           & \Ocal_{QG} = \bar{Q}_L \lambda^a D^\nu Q_L G_{\mu\nu}^a, \nonumber \\
& \Ocal_{uG} = \bar{u}_R\lambda^a \gamma^\mu D^\nu u_R G_{\mu\nu}^a, 
           & \Ocal_{tG} = \bar{t}_R\lambda^a \gamma^\mu D^\nu t_R G_{\mu\nu}^a, \nonumber \\
& \Ocal_{dG} = \bar{d}_R\lambda^a \gamma^\mu D^\nu d_R G_{\mu\nu}^a . \label{ec:vertexcorrections}
\end{align}
Here, $\lambda^a$ are Gell-Mann matrices, $D_\mu$ is the covariant derivative, $q_L$ and $Q_L$ represent, respectively, first and third-generation left-handed quark doublets, and $u_R$, $d_R$, $t_R$ are the right-handed up, down and top quarks, respectively. These operators arise only at the loop level, so their coefficients will be suppressed by $1/16\pi$ if the fundamental theory is weakly coupled. The Hermitian (anti-Hermitian) parts of the chirality-flipping operators in (\ref{ec:chromomoments}) give chromomagnetic (chromoelectric)  dipole moments to the involved quarks. None of them generates a FB asymmetry, but the Hermitian parts contribute to the cross section at the interfering level \cite{Atwood:1994vm,Haberl:1995ek}. 
More relevant for us are the operators in (\ref{ec:vertexcorrections}). Their axial combinations generate derivative axial-vector couplings of the gluon to the quarks. As shown in~\textcite{Gabrielli:2011jf}, the gluon-exchange diagram with two of these axial couplings interferes with the SM amplitude, producing a FB asymmetry. This is a $1/\Lambda^4$ contribution of the first kind. Quite large coefficients or a low scale are needed to obtain a sizable effect. On the other hand, these couplings give only $1/\Lambda^4$  quadratic contributions to the cross section. Therefore, an explanation of the Tevatron anomaly with these axial operators is as disfavored as in incoherent scenarios.  This problem might always  be mitigated by negative contributions to $\delta \sigma_\mathrm{int}$ from other operators, in particular, the vector combinations of the operators in (\ref{ec:vertexcorrections}).

Let us next consider the impact of four-quark operators on top pair production. A complete basis has been given in~\textcite{AguilarSaavedra:2010zi}. In contrast to the two-quark operators considered above, they can be generated at tree level.  Seven of these operators can produce non-negligible interfering $1/\Lambda^2$ contributions to $\ttb$ observables~\cite{Degrande:2010kt} :
\begin{align}
 \Ocal_{Qq}^{(1)} &= \ohs \left(\bar{Q}_L\gamma^\mu \lambda^a Q_L\right) \left(\bar{q}_L\gamma_\mu \lambda^a q_L \right), \nonumber \\
 \Ocal_{Qq}^{(3)} &= \ohs \left(\bar{Q}_L\gamma^\mu \lambda^a \tau^I Q_L\right) \left(\bar{q}_L\gamma_\mu \lambda^a \tau^I q_L \right), \nonumber \\
 \Ocal_{tu} & = \ohs \left(\bar{t}_R \gamma^\mu \lambda^a t_R\right)\left(\bar{u}_R\gamma_\mu \lambda^a u_R \right), \nonumber \\
 \Ocal_{td} & = \ohs \left(\bar{t}_R \gamma^\mu \lambda^a t_R\right)\left(\bar{d}_R\gamma_\mu \lambda^a d_R \right), \nonumber \\
 \Ocal_{Qu} & = \ohs \left(\bar{Q}_L\gamma^\mu \lambda^a Q_L\right) \left(\bar{u}_R\gamma_\mu \lambda^a u_R \right), \nonumber \\
 \Ocal_{Qd} & = \ohs \left(\bar{Q}_L \gamma^\mu \lambda^a Q_L\right) \left(\bar{d}_R\gamma_\mu \lambda^a d_R \right), \nonumber \\
 \Ocal_{qt} & = \ohs \left(\bar{q}_L \gamma^\mu \lambda^a q_L\right) \left(\bar{t}_R\gamma_\mu \lambda^a t_R \right), 
\label{ec:4F}
\end{align}
with $\tau^I$ the Pauli matrices. We have not written operators that are very constrained by flavor physics, nor an operator that only interferes with the QCD amplitude after a down-quark mass insertion. The $1/\Lambda^2$ corrections to the $u\bar{u}\to \ttb$ and $d \bar{d} \to \ttb$ cross sections only depend, respectively, on the ``vector-vector'' combinations of coefficients
\begin{align}
& C_{Vv}^u=C_{qt}+C_{tu}+C_{Qu}+C_{Qq}^{(1)}+C_{Qq}^{(3)}, \nonumber \\ 
& C_{Vv}^d= C_{qt}+C_{td}+C_{Qd}+C_{Qq}^{(1)}- C_{Qq}^{(3)},
\end{align}
whereas the $1/\Lambda^2$ corrections to the charge asymmetries in these processes depend on the ``axial-axial'' combinations 
\begin{align}
C_{Aa}^u=-C_{qt}+C_{tu}-C_{Qu}+C_{Qq}^{(1)}+C_{Qq}^{(3)}, \nonumber \\ 
C_{Aa}^d = -C_{qt}+C_{td}-C_{Qd}+ C_{Qq}^{(1)}-C_{Qq}^{(3)},
\label{ec:CAa}
\end{align}
which contribute, respectively, to the collider-independent asymmetries $A_u$ and $A_d$.
In the absence of other corrections, the linear scenario is realized, for $u\bar{u}$ and $d\bar{d}$ initial states separately, when $C^{u}_{Vv}=C^{d}_{Vv}=0$. If this condition is met, $O(1/\Lambda^4)$ corrections must be subleading and the correction to the inclusive FB asymmetry is
\begin{equation}
\Delta \afb^{\mathrm{linear}} = [0.093 \, C^u_{Aa} + 0.014 \, C_{Aa}^d] \times \left( \frac{1~\mathrm{TeV}}{\Lambda} \right)^2 ,
\end{equation}
We see that at the Tevatron the asymmetry (and also the cross section) is significantly more sensitive to the operators involving the $u$ quark, due to the larger $u\bar{u}$ fraction $F_u$. The coefficients have to be relatively large to reproduce a large asymmetry, e.g. to match the central value of  the CDF measurement of $\afb$. For instance if $C^d_{Aa}=0$, we need $C_{Aa}^u > 0.8$ when $\Lambda>1\,\mathrm{TeV}$ and $C_{Aa}^u > 3$ when $\Lambda>2\,\mathrm{TeV}$. 

If the cross section is modified at order $1/\Lambda^2$, the quadratic $1/\Lambda^4$ terms are important and other four-quark operators must be taken into account, in addition to the ones in (\ref{ec:4F}). In this quadratic scenario, $C^{u}_{Vv}$ (or $C^d_{Vv}$) must be sizable and negative, to compensate for the quadratic terms. General analyses with all the dimension-6 operators to $O(1/\Lambda^4)$ have been performed in \textcite{Delaunay:2011gv,AguilarSaavedra:2011vw}. They show that the Tevatron cross section and FB asymmetry can be well fitted in large regions of the space of operator coefficients, not necessarily obeying the condition $C^{u}_{Vv}=C^{d}_{Vv}=0$. 

A general feature of dimension-6 operators, and therefore of heavy new physics, is that they affect 
the $\ttb$ observables more significantly at high $\mttb$. This agrees, at least qualitatively, with the mass dependence of the FB asymmetry observed by the CDF collaboration. On the other hand, the cross section is also distorted at large invariant masses, and the deviations could be observable at the LHC. This leads to strong constraints, discussed in Sec.~\ref{sec:6}. The energy sensitivity is much more dramatic for the $1/\Lambda^4$ corrections.

The same operators and combinations of coefficients are relevant for the charge asymmetry at the LHC. Already at the $1/\Lambda^2$ level, it is clear that heavy new physics allows in principle for different sizes and signs of $A_C$, consistent with $\afb>0$. Indeed, the coefficients $C^{u}_{Aa}$  and $C^d_{Aa}$, which are independent in this formalism, can be adjusted to reproduce the required values of $A_u$ and $A_d$ in Eqs.~(\ref{ec:AuAd}). As we will see below,  these two coefficients are actually generated by independent couplings in explicit models that realize the linear scenario.

\subsection{Light extra particles}
\label{sec:5b}
Large contributions to the FB asymmetry can be most naturally produced by tree-level exchanges of new particles. Lorentz invariance and the renormalizability of the corresponding SM extensions---which avoids extra higher-scale suppressions---limit their spin to be either 0 or 1.\footnote{Spin 2 particles have been considered in \textcite{Grinstein:2012pn}. Their derivative couplings increase their effects with energy and lead to strong LHC constraints.} These new particles can be exchanged in the $s$, $t$ or $u$ channels, depending on their precise interactions with quarks. The corresponding forms of the propagator have a significant impact on the rapidity and invariant-mass distributions of the asymmetries and cross section:
\begin{itemize}

\item $s$ channel: The propagator itself does not modify the angular distributions, so any charge asymmetry must be produced by chiral couplings. To avoid a visible peak in the differential cross section, these particles must be either heavier than the available energies at the LHC, lighter than the $\ttb$ threshold or very broad. In the first case, the cross sections and asymmetries increase faster than in the infinite-mass limit, especially when $\mttb$ gets close to the mass of the new particle. In the second case, the dependence with $\mttb$ is rather mild. In the third case, their behavior will depend on the precise mass and width of these particles. In all cases, the conservation of angular momentum, as imposed in~Eq.~(\ref{ec:ampsq}), implies that the expansion in Legendre polynomials~(\ref{ec:Legendre_expansion}) of the non-standard contributions to the differential cross section has only a few terms: for scalars only the Legendre momenta $a_0$ and $a_1$ can be modified---the latter only if there is interference with the gluon-exchange amplitude,  whereas vectors can contribute to $a_0$, $a_1$ and $a_2$ at most. 

\item $t$ channel: The propagator favors forward top quarks, so it alone can generate a positive FB asymmetry. The asymmetries are increased at high rapidities and invariant mass. In this case, since $\cos{\theta}$ appears in the denominator, higher-order Legendre momenta will accompany the lower-order ones. This angular dependence is disfavored by the corresponding CDF results shown in Fig.~\ref{fig:costh}.

\item $u$ channel: The propagator prefers to send the top quarks backwards. Hence, it favors a negative asymmetry. To obtain a positive FB asymmetry, the numerator of the amplitude has to counteract this effect. However, as the invariant mass increases, the influence of the propagator becomes more significant, and eventually the asymmetries become negative. Another problem of $u$-channel exchanges is that they also contribute to higher Legendre momenta.

\end{itemize}
These different behaviors become milder as the mass of the exchanged particle increases, relative to the Mandelstam variable in the denominator of the propagator. In the heavy-particle limit, the propagator approaches a constant and the effect is described, in all three cases, by a four-fermion operator. The coefficients of these operators are given by ratios $g_1 g_2/M^2$, with $g_{1,2}$ trilinear couplings and $M$ the mass of the new boson. They have been calculated explicitly, for arbitrary scalars and vector bosons, in~\textcite{AguilarSaavedra:2011vw}.

The possible quantum numbers and interactions of the new particles are strongly restricted by the requirement of $SU(3)_C\times SU(2)_L \times U(1)_Y$ invariance of the SM extension, in the electroweak symmetric phase. In particular, the extra fields must furnish complete representations of this symmetry. 
There are ten possible irreducible representations of new vector bosons and eight irreducible representations of scalars contributing to $q\bar{q} \to \ttb$. They are collected in Table~\ref{t:multiplets}, together with the relevant interaction Lagrangian. We also indicate the symmetry properties, if any, of the coupling matrices $g_{ij}$. 
\begin{table*}
\begin{center}
\caption{Vector bosons and scalar representations mediating $q \bar q \to t \bar t$. The notation is standard, with left-handed doublets $q_{Li}$, right-handed singlets $u_{Ri}$, $d_{Ri}$,  $\tilde \phi = \epsilon \phi$ and $\psi^c = C \bar \psi^T$, where $\epsilon=i\tau^2$ and $C$ is the charge conjugation matrix. The indices $a,b,c$ represent color,  with $\varepsilon_{abc}$ the totally antisymmetric tensor, and the indices $i,j$ denote the family number in the interaction basis.}
\label{t:multiplets}
\begin{tabular}{cclc}
\hline
\hline
Label & Rep. & \multicolumn{1}{c}{Interaction Lagrangian} & Sym. \\
$Z^\prime_\mu$ & $(1,1)_0$ 
  & $-\left( g_{ij}^q \bar q_{Li} \gM q_{Lj} 
  + g_{ij}^u \bar u_{Ri} \gM u_{Rj} 
  + g_{ij}^d \bar d_{Ri} \gM d_{Rj} \right) Z^\prime_\mu $ & $g=g^\dagger$ \\[1mm]
$\mathcal{W}_\mu$ & $(1,3)_0$
  & $- g_{ij} \bar q_{Li}  \gM \tau^I q_{Lj} \, \mathcal{W}_\mu^I$
  & $g=g^\dagger$ \\[1mm]
$W^\prime_\mu$ & $(1,1)_1$ 
  & $- g_{ij} \bar d_{Ri} \gM u_{Rj} \, {W^\prime_\mu}^{\dagger} + \text{h.c.}$ & -- \\[1mm]
$G_\mu$ & $(8,1)_0$
  & $- \left( g_{ij}^q \bar q_{Li} \gM \frac{\lambda^a}{2} q_{Lj} 
  + g_{ij}^u \bar u_{Ri} \gM \frac{\lambda^a}{2} u_{Rj} 
  + g_{ij}^d \bar d_{Ri} \gM \frac{\lambda^a}{2} d_{Rj} \right) G_\mu^a$ & $g=g^\dagger$ \\[1mm]
$H_\mu$ & $(8,3)_0$
  & $- g_{ij} \bar q_{Li}  \gM \tau^I \frac{\lambda^a}{2} q_{Lj} \, H_\mu^{aI}$ & $g=g^\dagger$ \\[1mm]
$G^\prime_\mu$ & $(8,1)_1$ 
  & $- g_{ij} \bar d_{Ri} \gM \frac{\lambda^a}{2} u_{Rj} \, {G^\prime_\mu}^{a\dagger} + \text{h.c.}$ & -- \\[1mm]
$Q_\mu$ & $(3,2)_{\frac{1}{6}}$
  & $-g_{ij} \varepsilon_{abc} \bar d_{Rib} \gM \epsilon q_{Ljc}^c \, Q_\mu^{a\dagger} + \text{h.c.}$ & -- \\[1mm]
$Q^\prime_\mu$ & $(3,2)_{-\frac{5}{6}}$
  & $-g_{ij} \varepsilon_{abc} \bar u_{Rib} \gM \epsilon q_{Ljc}^c \, {Q^\prime_\mu}^{a\dagger} + \text{h.c.}$ & -- \\[1mm]
$Y_\mu$ & $(\bar 6,2)_{\frac{1}{6}}$
  & $-g_{ij} \oh \left[ \bar d_{Ria} \gM \epsilon q_{Ljb}^c + 
  \bar d_{Rib} \gM \epsilon q_{Lja}^c \right] Y_\mu^{ab\dagger}  + \text{h.c.}$ & -- \\[1mm]
$Y^\prime_\mu$ & $(\bar 6,2)_{-\frac{5}{6}}$
  & $-g_{ij} \oh \left[ \bar u_{Ria} \gM \epsilon q_{Ljb}^c + 
  \bar u_{Rib} \gM \epsilon q_{Lja}^c \right] {Y^\prime_\mu}^{ab\dagger}  + \text{h.c.}$ & -- \\[1mm]
$\phi$ & $(1,2)_{\frac{1}{2}}$
  & $- g_{ij}^u \bar q_{Li} u_{Rj} \, \tilde{\phi} - g_{ij}^d \bar q_{Li} d_{Rj} \, \phi  + \text{h.c.}$ & -- \\[1mm]
$\Phi$ & $(8,2)_{\frac{1}{2}}$
  & $- g_{ij}^u \bar q_{Li} \frac{\lambda^a}{2} u_{Rj} \, \tilde{\Phi}^a - g_{ij}^d \bar q_{Li} \frac{\lambda^a}{2} d_{Rj} \, \Phi^a  + \text{h.c.}$ & -- \\[1mm]
$\omega^\prime$ & $(3,1)_{-\frac{1}{3}}$
  & $- g_{ij} \varepsilon_{abc} \bar d_{Rib} u_{Rjc}^c \, {\omega^\prime}^{a\dagger} + \text{h.c.}$ & -- \\[1mm]
$\Omega^\prime$ & $(\bar 6,1)_{-\frac{1}{3}}$
  & $-g_{ij} \oh \left[ \bar d_{Ria} u_{Rjb}^c + 
  \bar d_{Rib} u_{Rja}^c \right] {\Omega^\prime}^{ab\dagger} + \text{h.c.}$ & -- \\[1mm]
$\omega$ & $(3,1)_{-\frac{4}{3}}$
  & $- g_{ij} \varepsilon_{abc} \bar u_{Rib} u_{Rjc}^c \, \omega^{a\dagger} + \text{h.c.}$ & $g=-g^T$ \\[1mm]
$\Omega$ & $(\bar 6,1)_{-\frac{4}{3}}$
  & $-g_{ij} \oh \left[ \bar u_{Ria} u_{Rjb}^c + 
  \bar u_{Rib} u_{Rja}^c \right] \Omega^{ab\dagger} + \text{h.c.}$ & $g=g^T$ \\[1mm]
$\sigma$ & $(3,3)_{-\frac{1}{3}}$
  & $- g_{ij} \varepsilon_{abc} \bar q_{Lib} \tau^I \epsilon q_{Ljc}^c \, \sigma^{a\dagger} + \text{h.c.}$ & $g=-g^T$ \\[1mm]
$\Sigma$ & $(\bar 6,3)_{-\frac{1}{3}}$
  & $-g_{ij} \oh \left[ \bar q_{Lia} \tau^I \epsilon q_{Ljb}^c + 
  \bar q_{Lib} \tau^I \epsilon q_{Lja}^c \right] \Sigma^{Iab\dagger} + \text{h.c.}$ & $g=g^T$ \\
\hline
\hline
\end{tabular}
\end{center}
\end{table*}

Allowing for general couplings, the relevant components of the $Z^\prime$, $\mathcal{W}$, $G$, $H$, $\phi$ and $\Phi$ multiplets can be exchanged in either the $s$ or the $t$ channels in $q\bar{q} \to t\bar{t}$ processes, whereas those of $W^\prime$ and $G^\prime$ can be exchanged in the $t$ channel only, and those of the other ten multiplets, in the $u$ channel only. Obviously, $t$- and $u$-channel exchanges require flavor-changing couplings. All these fields can produce interfering contributions $\delta \sigma_\mathrm{int}^{F,B}$. 

Any new physics contributing at the tree level to $\ttb$ production can be characterized by these multiplets and their interactions. In practice, to perform explicit analyses it is necessary to choose particular directions in this multi-dimensional space. Although scenarios with several multiplets can be interesting, most of the models that have been proposed to explain the anomaly in the FB asymmetry are extensions of the SM with just one of these multiplets. Among these simple models, the following ones have been studied in greater detail: 

 {\em Color-octet vector $G$\/}  \cite{Ferrario:2008wm}.\footnote{The specific couplings and mass range studied in this paper led to the prediction of a negative FB asymmetry, but arbitrary signs can be obtained in general, as explained below.} Exchanged in the $s$ channel via flavor-diagonal couplings, it gives an amplitude that interferes with the SM gluon-exchange diagram. The corresponding contribution to the charge asymmetries in $q\bar{q} \to \ttb$ is proportional to the product of axial couplings $g_A^{u,d} g_A^t$, where $g_A^{u,d}= g_{11}^{u,d}-g_{11}^q$ and $g_A^t=g_{33}^u-g_{33}^q$.  When $\mttb<M_G$ ($\mttb>M_G$), $g_A^q$ and $g_A^t$ must have opposite (same) sign for a positive contribution. In the heavy-mass limit, it is described by a set of operators with $C_{Aa}^q/\Lambda^2= - g_A^q g_A^t\, /\, 4 M_G^2$. This multiplet is particularly promising for several reasons. The main one is that $\delta \sigma_\mathrm{int}=0$ (linear new physics) when the vector couplings to either the light quarks or the top vanish, i.e. either $g_V^u=g_{11}^q+g_{11}^u=0$ and $g_V^d=g_{11}^q+g_{11}^d=0$, or $g_V^t=g_{33}^q+g_{33}^u=0$ (when all vector couplings vanish, the octet is an axigluon). The octet $G$ is actually the only multiplet that can produce, on its own, positive charge asymmetries without interfering contributions to the cross section. Another welcomed feature of an octet vector boson in the $s$ channel is that it reproduces well the observed values of the Legendre momenta, since it only contributes with $J=1$ to the amplitude. The main issue is, as in any $s$-channel model, to hide the resonant peak in the cross section, produced by the quadratic terms. The solutions have already been mentioned, and are further discussed in the next section. On the other hand, dijet and four-top-quark data constrain the possible values of the couplings of the octet to the light quarks and the top, respectively. These bounds are discussed in Sec.~\ref{sec:7}. This multiplet appears naturally in extensions of the SM with a $SU(3)\times SU(3) \to SU(3)_C$ symmetry-breaking pattern~\cite{Frampton:1987dn}. In particular, it can emerge as the lightest Kaluza-Klein excitation of the gluons in extra-dimensional theories with gauge fields in the bulk~\cite{Djouadi:2009nb}. 

{\em Neutral $Z^\prime$ boson\/} \cite{Jung:2009jz}. This SM singlet is particularly interesting when exchanged in the $t$ channel via flavor-changing $t u$ couplings. A negative asymmetry is produced at the interfering level, so significant quadratic contributions are required to obtain a positive correction to the FB asymmetry, and also to cancel $\delta \sigma_\mathrm{int}$. As we will see below in this section and in the following ones, this popular model is strongly disfavored by different Tevatron and LHC observables. From the model-building point of view, these vectors could be the gauge bosons of an extra local flavor symmetry~\cite{Jung:2011zv}.

{\em Charged $W^\prime$ boson\/} \cite{Cheung:2009ch}. This  isosinglet couples to right-handed quarks and contributes in the $t$ channel to the partonic process $d\bar{d} \to \ttb$. Larger couplings are needed to compensate the lower $d\bar{d}$ luminosity. This field also produces negative $\Delta \afb$ at the interfering level and, similarly to the $Z^\prime$, is disfavored by available data. Right-handed $W^\prime$ bosons appear in left-right extensions of the SM gauge group.

 {\em Scalar isodoublet $\phi$\/} \cite{Nelson:2011us}. This Higgs-like doublet works best when exchanged in the $t$ channel. It gives positive asymmetry at the interfering level. For small masses, the required couplings to achieve a sizable $\Delta \afb$ and a cancellation of $\delta \sigma_\mathrm{int} + \delta \sigma_\mathrm{quad}$ are relatively small. The particular flavor-changing couplings in such a two-doublet model can be justified with flavor symmetries.

{\em Color-triplet scalar $\omega$\/} \cite{Shu:2009xf}. This isosinglet of charge 4/3 can only be exchanged in the $u$ channel, with flavor-violating right-handed $tu$ interactions. Once again, its interference contribution to the asymmetry is negative, so large couplings and a significant cancellation are required. Moreover, masses $M_\omega>220~\mathrm{GeV}$ are necessary to soften the effect of the $u$ channel propagator. These fields are included in the scalar sector of many Grand Unified models~\cite{Dorsner:2009mq}.

{\em Color-sextet scalar $\Omega$\/} \cite{Shu:2009xf}. Another isosinglet of charge 4/3, it also contributes in the $u$ channel via flavor-violating right-handed $tu$ couplings and, again, intermediate masses are preferred. However, in this case the interference contribution to $\Delta \afb$ is positive.\footnote{A wrong sign in~\textcite{Shu:2009xf} was corrected in~\textcite{Arhrib:2009hu}.} These fields also appear in models of Grand Unification.

We show in Fig.~\ref{fig:AvsAH} the predictions of these models for the inclusive and high-mass values of the FB asymmetry~\cite{AguilarSaavedra:2011ug}. The colored regions showing these predictions are obtained by a parameter-space scan, subject to some loose constraints from the total $\ttb$ cross section at the Tevatron and the high-mass tail at the LHC. We only consider positive contributions of the new particles to the asymmetry. For the color octet, we use a very heavy axigluon, represented by the corresponding four-fermion operators.
\begin{figure*}[htb]
\begin{center}
\begin{tabular}{ccc}
\includegraphics[height=5.5cm]{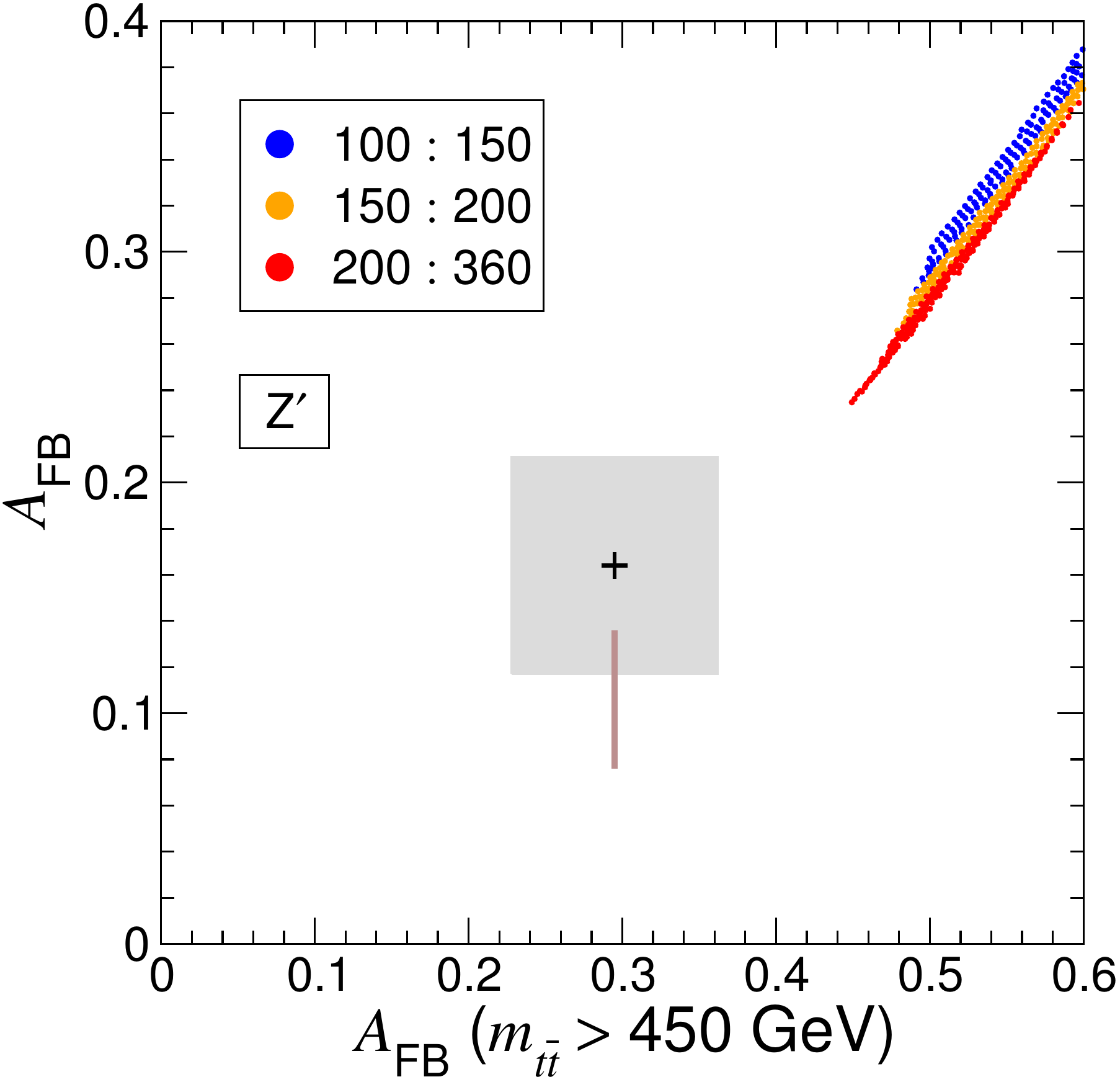} &
\includegraphics[height=5.5cm]{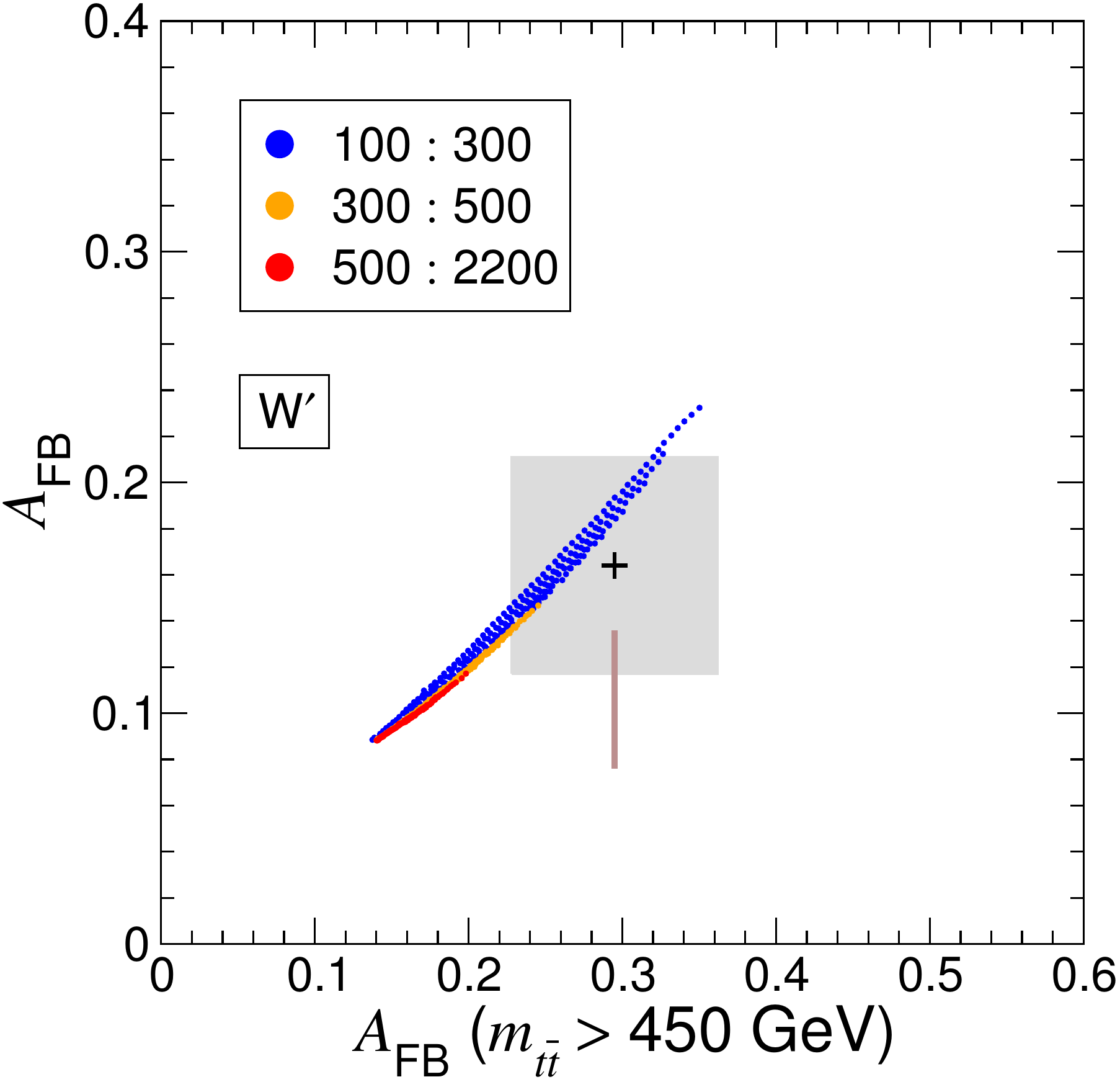} &
\includegraphics[height=5.5cm]{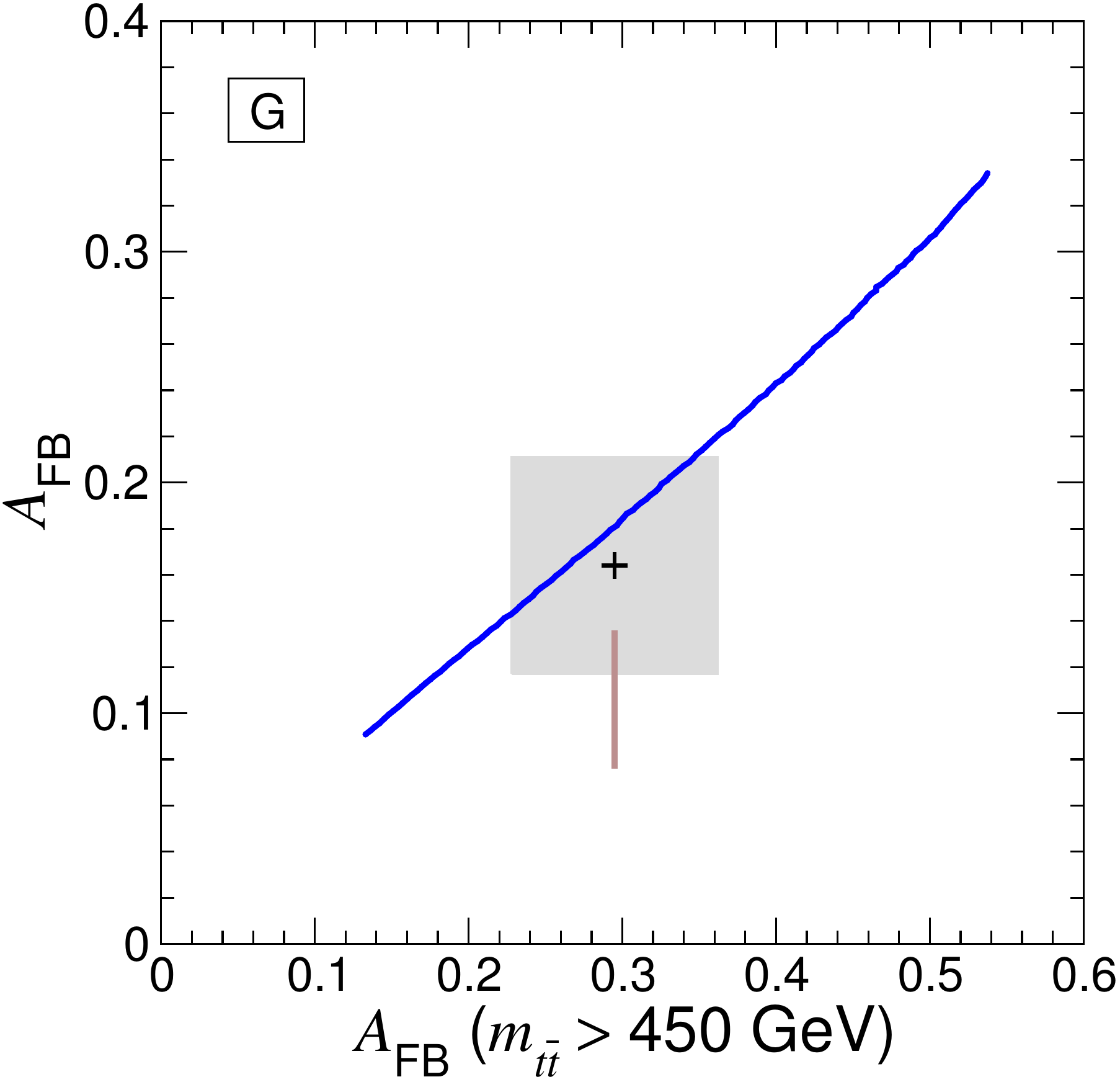} \\
\includegraphics[height=5.5cm]{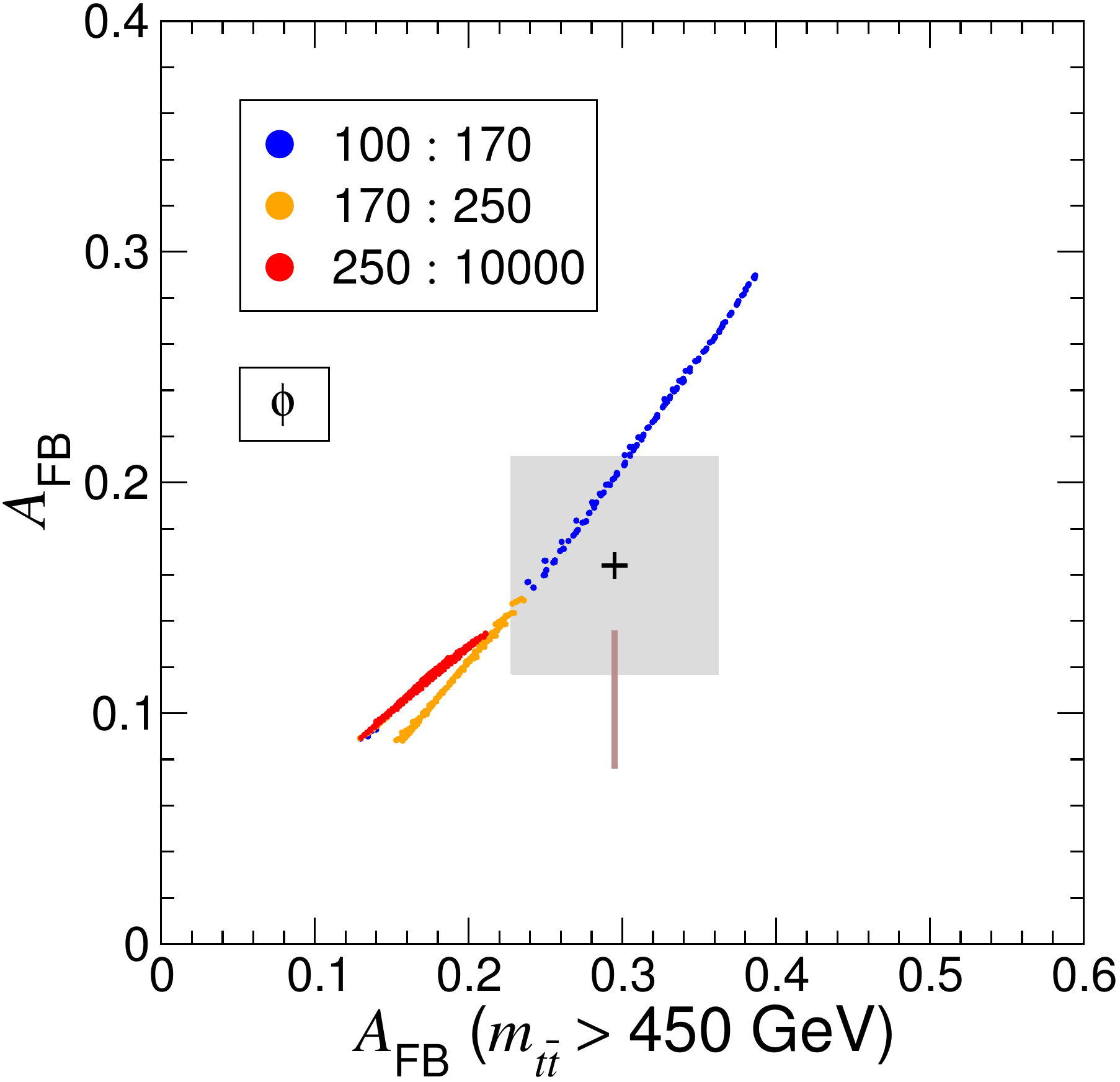} &
\includegraphics[height=5.5cm]{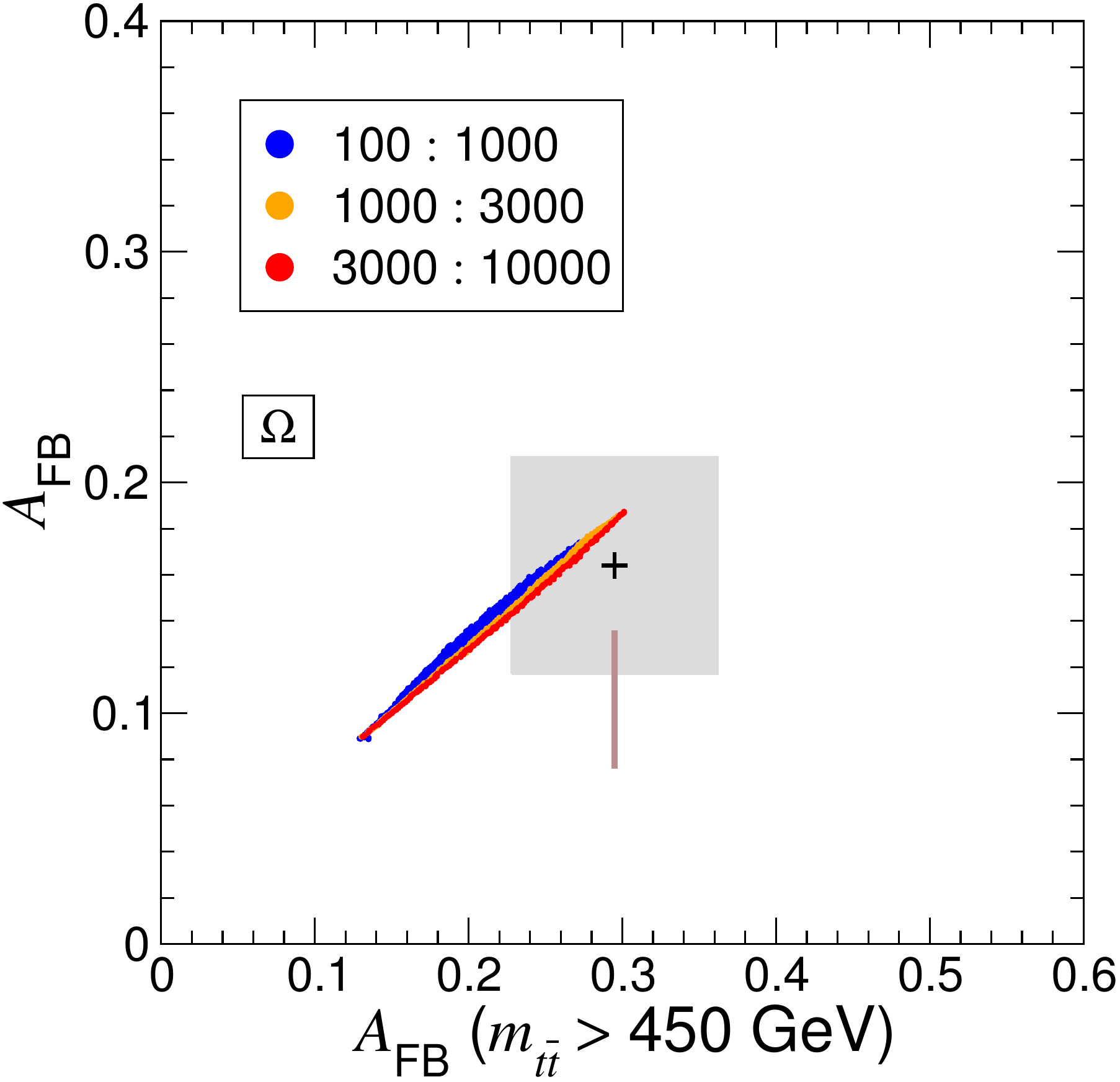} &
\includegraphics[height=5.5cm]{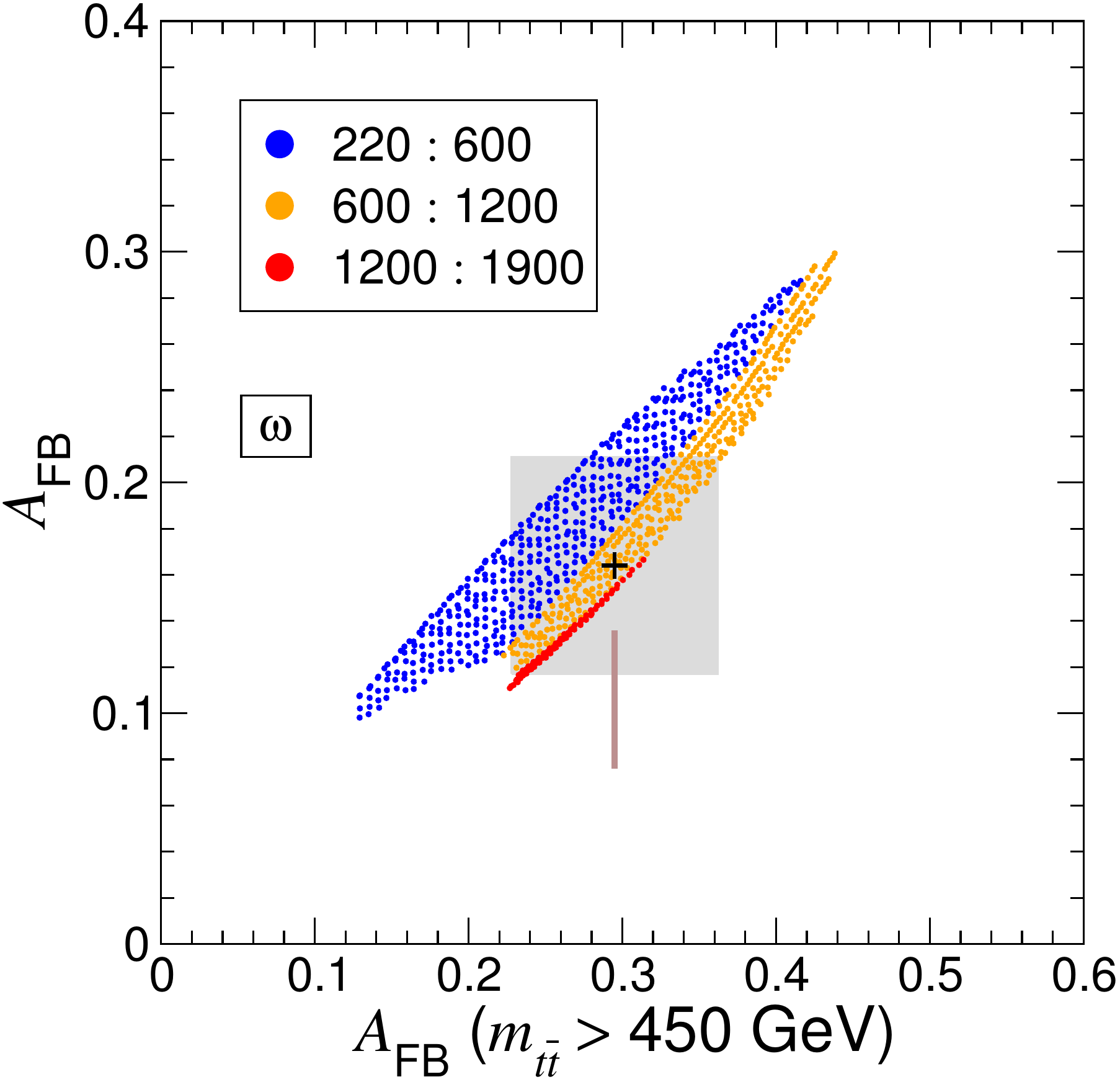} 
\end{tabular}
\end{center}
\caption{\label{fig:AvsAH}Inclusive versus high-mass asymmetries at the Tevatron, for several new physics models. The numbers in the legends indicate the mass range for the new particle, in GeV. The crosses correspond to the CDF measurements, with the shaded boxes indicating the $1\sigma$ uncertainty. The vertical lines corresponds to the D0 measurement of the inclusive FB asymmetry, with the corresponding $1\sigma$ uncertainty (the position in the horizontal axis is arbitrary). From~\textcite{AguilarSaavedra:2011ug}, updated.}
\end{figure*}
We see that most of these simple models can reproduce simultaneously the CDF excess in the inclusive and high-mass FB asymmetries. The exception is the $Z^\prime$ boson, which overpredicts them, especially at high $\mttb$. The reason is that large couplings are necessary in this case to ensure the cancellation of interference and quadratic terms, in the region with a positive asymmetry. A conclusion one can draw from these plots is that the mass dependence observed by the CDF collaboration can be explained naturally by new physics, without the need of contrived models. Let us, nevertheless, point out that a different mass dependence results from light octets $G$. In particular, for octets with mass under the $\ttb$ threshold the invariant-mass distributions of the FB and charge asymmetries are flatter, so they agree better with the findings of the D0 collaboration~\cite{AguilarSaavedra:2011ci}. On the other hand, the precise dependence on the polar angle measured by the CDF collaboration is best reproduced among these simple models by the color octet $G$, since the $t$ and $u$ exchanges generate higher-order Legendre momenta.

In Sec.~\ref{sec:2c} we have argued that $\afb$ and $\ac$ are independent in general. However, actual models give correlated predictions for both. Let us then consider the predictions of the new particles for the charge asymmetry $\ac$ at the LHC. As implied by Eq.~(\ref{ec:AuAd}), the relative contributions of a given model to $\afb$ and $\ac$ depend on their relative contributions to $u$ and $d$ initiated processes.  In Fig.~\ref{fig:AFBvsAC} we plot the predicted values of $\afb$ and $\ac$ for a parameter scan in the same simple models considered above, except the $Z^\prime$, which as explained above cannot reproduce the Tevatron data. 
\begin{figure}[htb]
\begin{center}
\includegraphics[width=8cm]{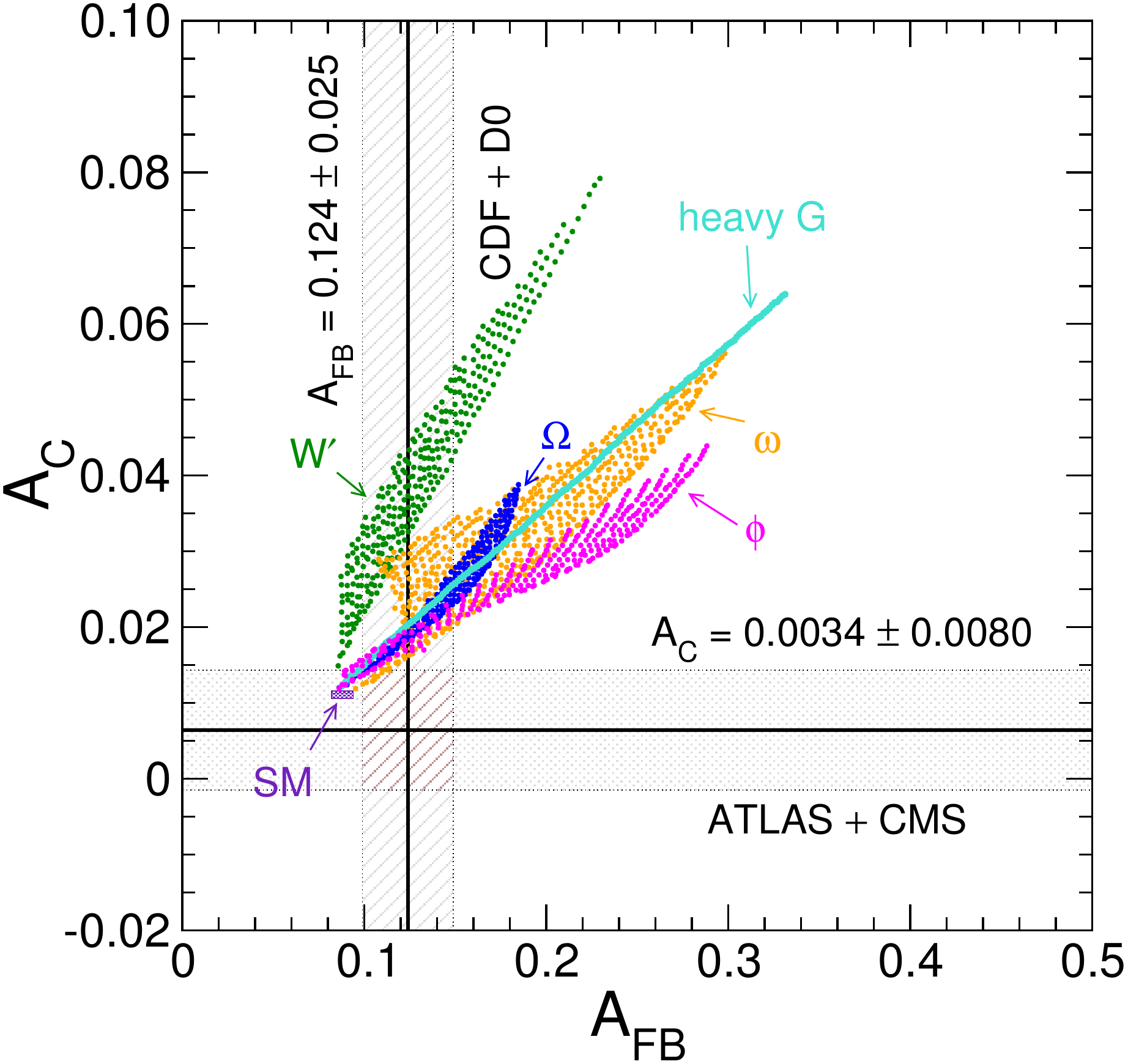}
\caption{\label{fig:AFBvsAC}Comparison of predictions for the inclusive asymmetries $\afb$ and $\ac$ for several simple models, together with the experimental measurements. From~\textcite{AguilarSaavedra:2011hz}, updated.}
\end{center} 
\end{figure}
These models follow a similar slope, except in the case of the $W^\prime$ boson. This particle leads to twice the slope of the others because the $d\bar{d}\to \ttb$  process has higher relative importance at the LHC than at the Tevatron. As a result, the $W^\prime$ boson is disfavored, since the agreement of its prediction, within one sigma, with the average of the measured $\afb$ leads to a two sigma disagreement in $\ac$. The other models in this set cannot reproduce the central values of the $\afb$ and $\ac$ measurements either, but they are consistent with them at the one sigma level. More extreme behaviors, including different signs for $\afb$ and $\ac$, are also possible. For instance, in the explicit octet model proposed in \textcite{Drobnak:2012cz} this is achieved with light-quark axial couplings $g_A^u$ and $g_A^d$ of opposite sign and with $|g_A^d|>|g_A^u|$. In particular, it is possible to accomodate the central values of the Tevatron and LHC asymmetries (see also \onlinecite{Ko:2012ud,Drobnak:2012rb,Alvarez:2012ca}).
This requires a cancellation of new physics effects in $\ac$, which could be uncovered by the measurement of a charge asymmetry in $t\bar{t}\gamma$ production at the LHC (see Sec.~\ref{sec:4d}).

Finally, let us note that, in order to discriminate between different models, the analysis of the $\mttb$ dependence of $A_C$ (see Sec.~\ref{sec:4a3}) could be useful~\cite{AguilarSaavedra:2011hz}. We display the predictions for three models in Fig.~\ref{fig:AcMtt}. For illustration we include the point corresponding to the ATLAS measurement $\ac = 0.018 \pm 0.022$ for $\mttb > 600$ GeV~\cite{Aad:2013cea}.
\begin{figure}[htb]
\begin{center}
\includegraphics[width=8cm]{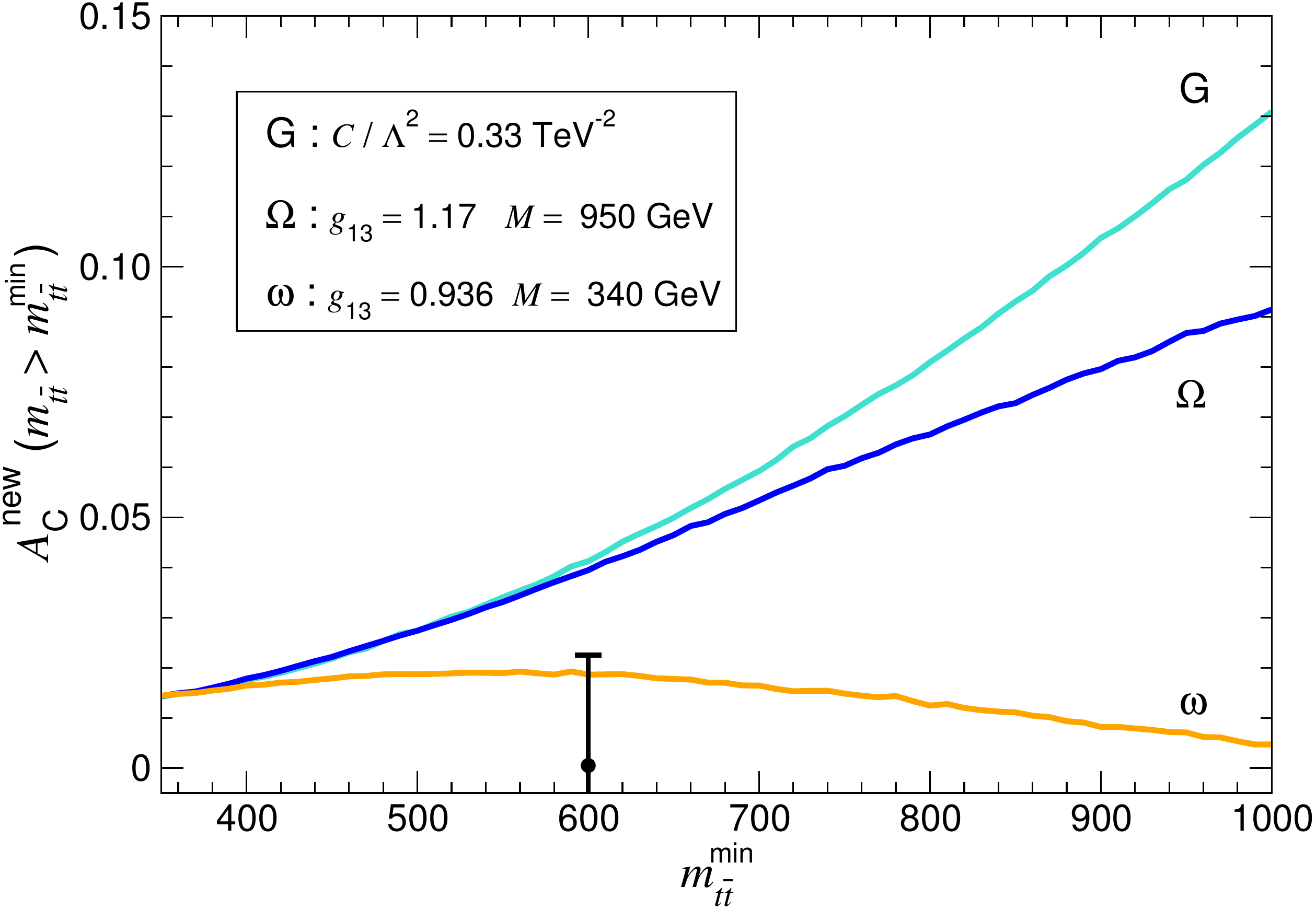}
\caption{\label{fig:AcMtt} Dependence of the charge asymmetry on an $\mttb$ lower cut in three simple models, for a point with $\Delta \afb\simeq 0.13$, $\ac^\mathrm{new} \simeq 0.016$. The ATLAS measurement for $\mttb > 600$ GeV is also included. From~\textcite{AguilarSaavedra:2011hz}, adapted.}
\end{center} 
\end{figure}
The discrimination power will be much higher with 8 TeV data, when the available statistics allows to measure $\ac$ at higher $\ttb$ invariant mass.


\section{Correlated effects in $\ttb$ production}
\label{sec:6}

\subsection{Enhancement of the high-$\mttb$ tail}
\label{sec:6a}

The distortion of the $\mttb$ differential distribution with respect to the SM prediction is a rather general consequence of hypothetical new physics contributions to $\ttb$ production~\cite{Delaunay:2011gv,AguilarSaavedra:2011vw}, especially in the quadratic new physics scenario. Then, since the bulk of the $\ttb$ cross section results from moderate $\mttb$ not far from the threshold, the agreement of the predicted Tevatron cross section with the experimental measurements---which is a basic requirement for realistic models---has the almost unavoidable consequence that deviations appear in the high-$\mttb$ tail. 
\begin{figure}[htb]
\begin{center}
\begin{tabular}{c}
\includegraphics[height=5.5cm,clip=]{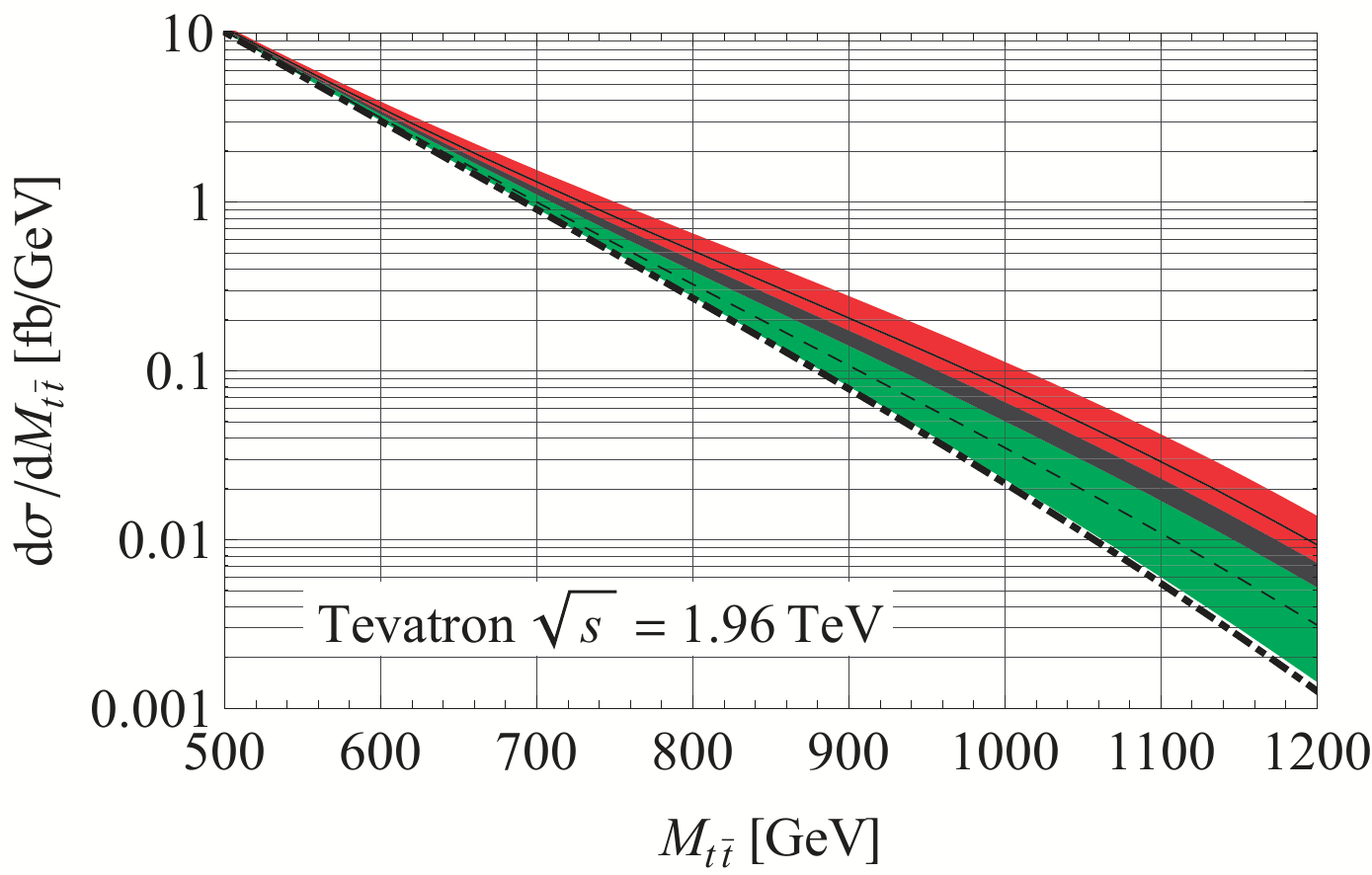} \\[1mm]
\includegraphics[height=5.5cm,clip=]{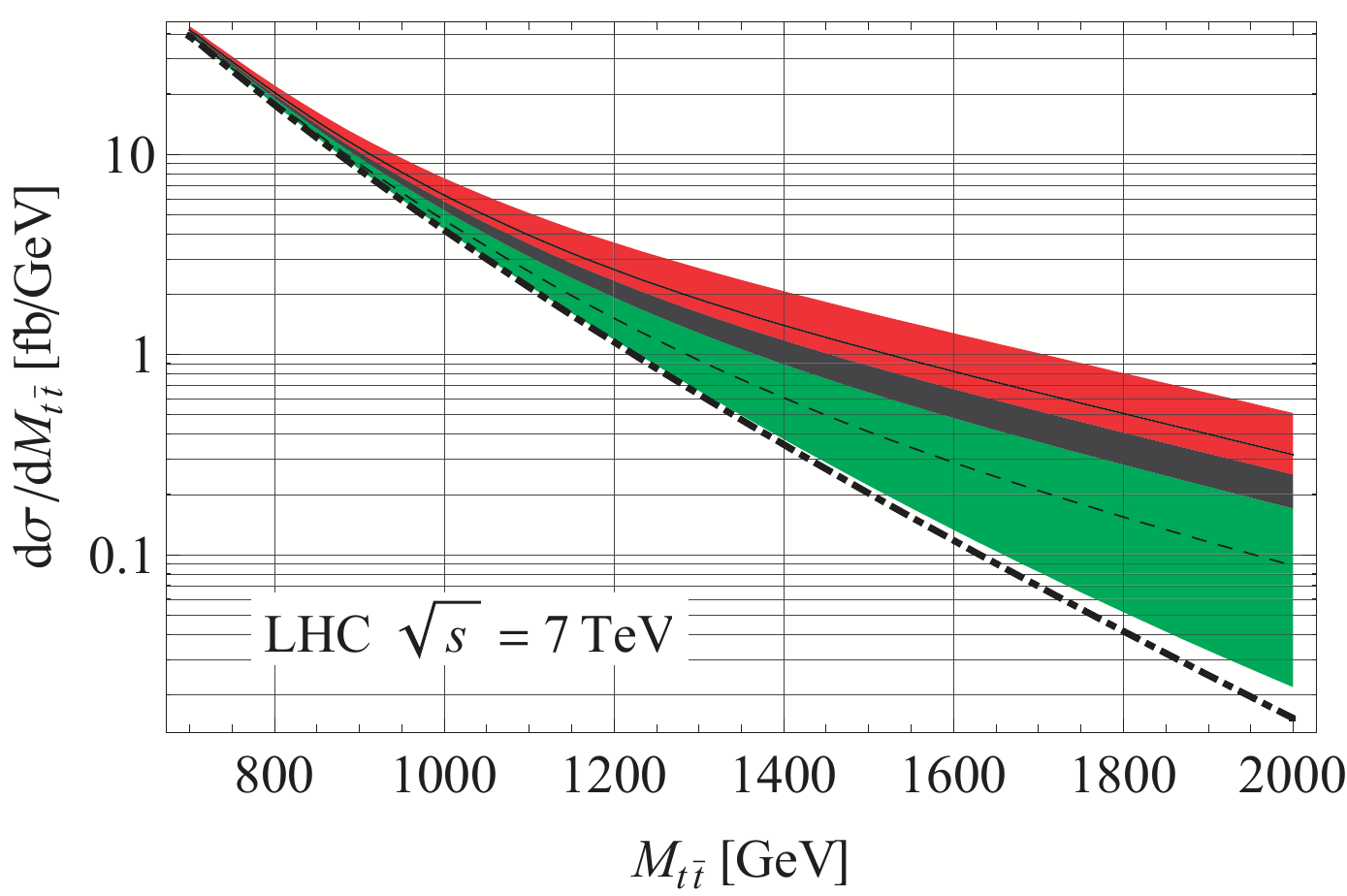}
\end{tabular}
\end{center}
\caption{Enhancement of the high-$\mttb$ tail at the Tevatron (top) and the LHC (bottom) for heavy new physics. The lower dot-dashed line is the SM LO prediction. The dashed and solid lines correspond to two non-zero $C_{Aa}^u$ values chosen to have inclusive $\afb = 0.158$ and $\afb = 0.475$ for $\mttb > 450$ GeV, respectively. The bands represent the variation around these values (see text for details). From~\textcite{Blum:2011up}.}
\label{fig:6A0}
\end{figure}
These deviations are illustrated in Fig.~\ref{fig:6A0}, for the Tevatron (top panel) and the LHC with 7 TeV (bottom panel), for linear heavy new physics parametrized by a non-zero $C_{Aa}^u$ in Eq.~(\ref{ec:CAa}) that fits the former CDF measurements from~\textcite{Aaltonen:2011kc}, $\afb = 0.158 \pm 0.075$ (dashed line and green area) and $\afb = 0.475 \pm 0.114$ for $\mttb > 450$ GeV (solid line and red area). For quadratic new physics scenarios the tail enhancements are larger than the corresponding ones in Fig.~\ref{fig:6A0}, which are ``minimal'' for heavy new physics.

For light mediators the tail enhancements are much less pronounced~\cite{AguilarSaavedra:2011ug,AguilarSaavedra:2011ci}. Moreover, at the Tevatron the potential deviations in the high-$\mttb$ tail may remain hidden if the new physics contributions concentrate in the forward region, in which the detection efficiency is small due to the detector coverage~\cite{Gresham:2011pa}. This is the case for example when light $Z'$ or $W'$ particles are exchanged in the $t$ channel. Also, the statistical uncertainties in the high-$\mttb$ tail are large at the Tevatron. But at the LHC the detectors have a larger rapidity coverage and the analyzed datasets have much higher statistics, allowing for precise measurements of the $\mttb$ spectrum over a wide range. In general they exhibit a good agreement with the SM prediction, as illustrated in Fig.~\ref{fig:6A1}. (Electroweak Sudakov corrections slightly reduce the high-$\mttb$ tail with respect to the fixed-order Monte Carlo predictions, see~\onlinecite{Manohar:2012rs}.)
\begin{figure}[htb]
\begin{center}
\includegraphics[height=7cm,clip=]{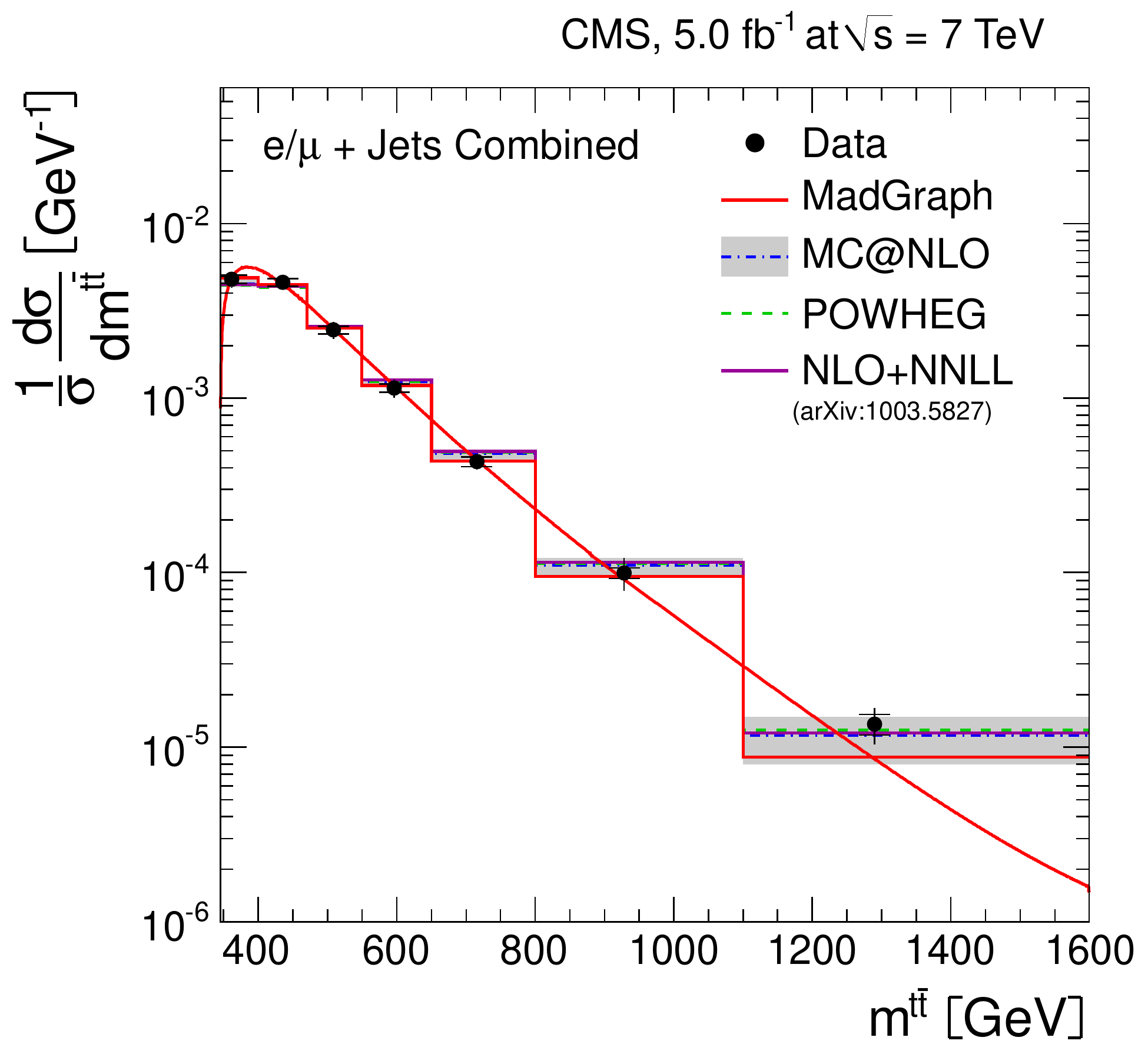}
\end{center}
\caption{Normalized $\ttb$ invariant mass distribution at the LHC with 7 TeV, measured in the semileptonic decay channel, and SM predictions from several Monte Carlo generators. From~\textcite{Chatrchyan:2012saa}.}
\label{fig:6A1}
\end{figure}
This imposes severe constraints on quadratic new physics models that accommodate an excess $\afb$.
Although a precise statement requires a dedicated analysis taking into account the possibly different $\ttb$ acceptance in the presence of new contributions, the tail enhancements in the $Z'$ and $W'$ models are so pronounced that they are eventually excluded as candidates to yield an $\afb$ excess. On the other hand, for $u$-channel color sextets and triplets, as well as for a scalar isodoublet, an asymmetry excess is compatible with the observed differential $\mttb$ spectrum.

In the linear new physics models where $\delta \sint$ vanishes (also when considered differentially as a function of $\mttb$) and $\delta \squa$ is small---for example, a color octet $G$ exchanged in the $s$ channel---the $\mttb$ distribution is preserved except at the resonance, where a potentially large enhancement results from $\delta \squa$. As it has been previously mentioned, this enhancement can be hidden if the octet is wide or if it is lighter than the $\ttb$ threshold. In the former case, new particles may be required to yield extra $G$ decay modes that account for its large width, for example new quarks~\cite{Barcelo:2011vk,Barcelo:2011wu} or colored scalars~\cite{Tavares:2011zg}. Figure~\ref{fig:6A2} shows the $\mttb$ distribution for the SM and when an extra octet $G$ with mass $M=850$ GeV, is included, without and with extra decay modes that yield a large width $\Gamma/M = 0.7$.
In case that $G$ is lighter than the $\ttb$ threshold~\cite{AguilarSaavedra:2011ci} the $\mttb$ spectrum is preserved independently of $\Gamma$. Nevertheless, a large width may be required in order to comply with other collider constraints (see Sec.~\ref{sec:7}). If $G$ is very heavy, the tail enhancement corresponds to the one shown in Fig.~\ref{fig:6A0}.

\begin{figure}[htb]
\begin{center}
\includegraphics[height=7cm,clip=]{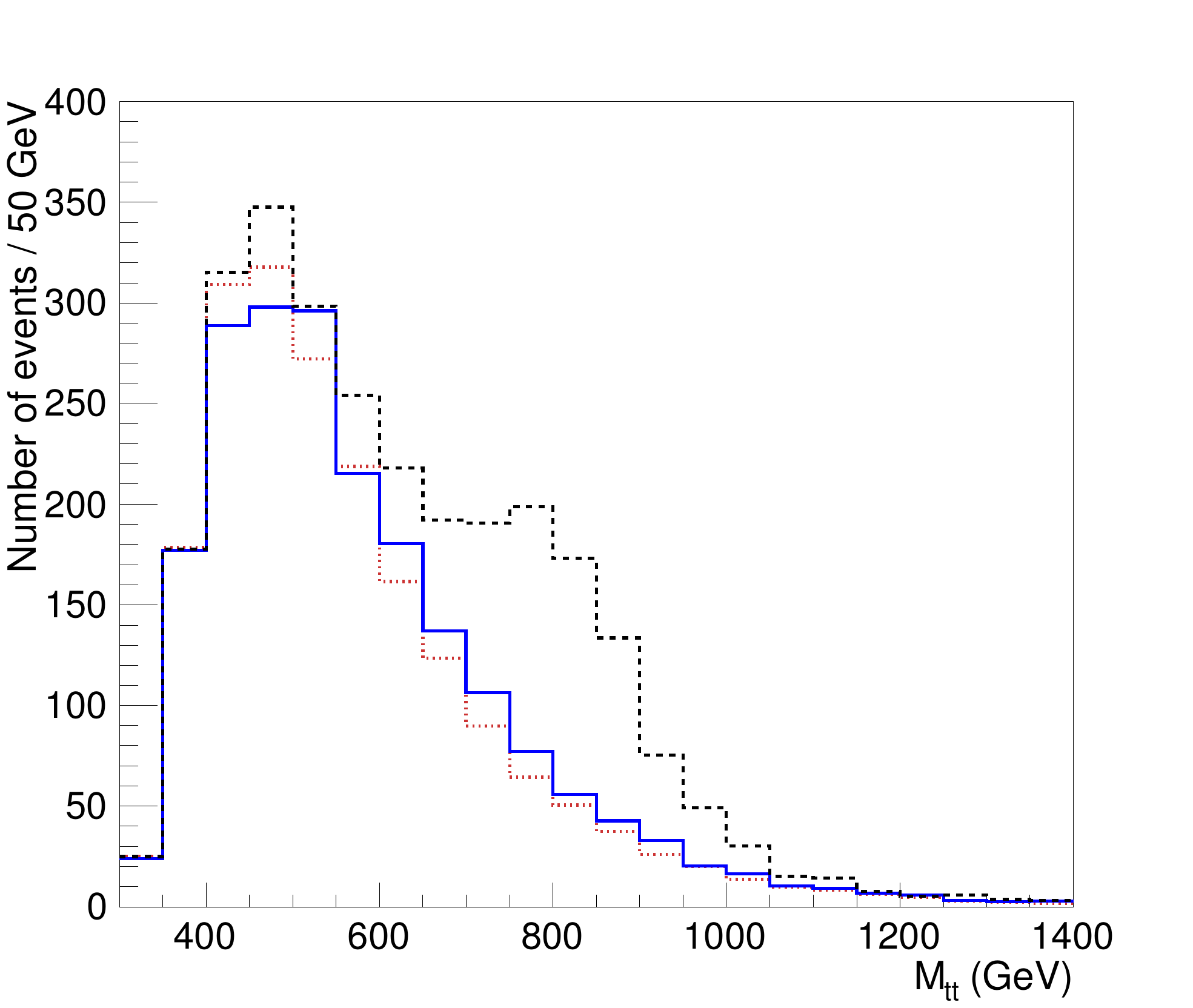}
\end{center}
\caption{Normalized $\ttb$ invariant mass distribution at the LHC with 7 TeV, for the SM (dotted line), the SM plus a narrow $G$ (dashed line) and the SM plus a wide $G$ (solid line). The predictions are normalized to an integrated luminosity of 1~\fbin. From~\textcite{Barcelo:2011vk}.}
\label{fig:6A2}
\end{figure}

\subsection{Top quark polarization and $\ttb$ spin correlations}
\label{sec:6b}

New $\ttb$ production mechanisms in general modify the top quark and antiquark polarizations, as well as their spin correlation, especially if the coupling of the top quark to the new states exchanged is chiral~\cite{Krohn:2011tw}. The color triplet and sextet scalars $\omega$, $\Omega$ have a right-handed coupling to the top quark as a consequence of gauge symmetry, as described in Sec.~\ref{sec:5}.  On the other hand, the coupling of the top quark to a color octet $G$ and scalar doublet $\phi$ can have any chirality.

In order to discuss angular distributions in the decay of the $\ttb$ pair, let us fix a reference system $(x,y,z)$ in the top quark rest frame and another one $(x',y',z')$ in the top antiquark rest frame, and consider the decay products $X=\ell^+,\nu,u,\bar d,\dots$, $X' = \ell^-,\bar \nu,\bar u,d,\dots$ from the top quark and antiquark, respectively. Then, the double-differential polar angle distribution (see for example~\onlinecite{Bernreuther:2001rq}) is
\begin{eqnarray}
\frac{1}{\sigma}\frac{d^2\sigma}{d\cos \theta_X \, d\cos \theta_{X'}} & = & \frac{1}{4} \left[ 1 + \Pt\alpha_X \cos \theta_X  \right. \notag \\
& & + \Ptb \alpha_{X'} \cos \theta_{X'} \notag \\
& & \left. + C \alpha_X \alpha_{X'} \cos \theta_X \cos \theta_{X'} \right] \,,\quad\quad
\end{eqnarray}
with $\theta_X$, $\theta_{X'}$ being the polar angles of the $X$, $X'$ 3-momenta in their respective reference systems. The coefficients $\Pt$, $\Ptb$ are the polarizations of the top quark and antiquark in the $\hat z$ and $\hat z'$ axes, respectively. The coefficient $C$ measures the spin correlation between the top quark and antiquark, namely
\begin{eqnarray}
C = \frac{N(\uparrow \uparrow)+N(\downarrow \downarrow)-N(\uparrow \downarrow)-N(\downarrow \uparrow)}{N(\uparrow \uparrow)+N(\downarrow \downarrow)+N(\uparrow \downarrow)+N(\downarrow \uparrow)} \,,
\end{eqnarray}
where the up and down arrows indicate spins in the $\pm \hat z$, $\pm \hat z'$ directions for the top quark and antiquark, respectively. The quantities $\alpha_{X,X'}$ are the so-called {\it spin analyzing powers}  of the decay products~\cite{Jezabek:1994qs}, and have opposite sign for particles and antiparticles. For the charged leptons $\alpha_{\ell^+} = -\alpha_{\ell^-} = 1$ in the SM at the tree level, with small QCD corrections~\cite{Bernreuther:2004jv}. Because $|\alpha_{X}| \leq 1$ in general, the charged lepton distributions have the maximum possible dependence on the top polarization and spin correlation.

For $\ttb$ pairs produced via QCD interactions the polarizations $\Pt$, $\Ptb$ vanish for $\hat z$, $\hat z'$ in the production plane, and a small polarization orthogonal to that plane arises at the loop level. The spin correlations are non-zero in general. At the Tevatron, there are no measurements of the top polarization\footnote{In~\textcite{Abazov:2012oxa} the $\cos \theta_{\ell}$ distributions are investigated but results are not presented at the production level.} 
but the CDF and D0 Collaborations have measured the spin correlations using the beamline basis, that is, with $\hat z$ and $\hat z'$ in the proton direction~\cite{CDF10211,CDF10719,Abazov:2011gi}. At the LHC, the top polarization has been measured in the helicity basis~\cite{Aad:2013ksa,Chatrchyan:2013wua}, that is, selecting $\hat z$ in the direction of the top momentum in the $\ttb$ CM frame $\vec p_t$ and $\hat z' = -\hat z$. (The measurements assume  assuming $CP$ conservation so that $P_z = -P_{z'} \equiv P$.) The spin correlation in this basis has also been measured by the ATLAS and CMS Collaborations~\cite{Chatrchyan:2013wua,TheATLAScollaboration:2013gja}. The naive averages of these measurements can be found in Table~\ref{tab:spin} together with the SM predictions~\cite{Bernreuther:2010ny,Bernreuther:2013aga}.

The measurement of $C$ in the beamline basis at the Tevatron does impose some constraints on the parameter space of the models explaining the Tevatron anomalies~\cite{Fajfer:2012si}. But more restrictive are the precise measurements obtained at the LHC. The measurement of $P_z$ excludes at the $2\sigma$ level the color triplet as a viable candidate to explain the anomalies, and also disfavors the color sextet, as it can be seen in the upper panel of Fig.~\ref{fig:6B1}, where the curves represent the allowed values of the new physics contribution $\Delta \afb$ and $P_z$ for each model, resulting from a fit.
Note that for an axigluon one has $P_z=0$, as depicted in Fig.~\ref{fig:6B1}, but this is no longer the case if either the coupling to the light quarks or to the top quark is not purely axial.
The measurement of $C$ in the bottom panel is in some tension with the predictions of all the four models as well as with the SM prediction, at the $1.5\sigma$ level. (A lower spin correlation can be accommodated with general color octets, see~\textcite{Aguilar-Saavedra:2014nja}.) With 8 TeV data the measurements of $P$ and $C$ are not expected to be much more precise than the current 7 TeV ones in Table~\ref{tab:spin}, whose uncertainties are nearly dominated by systematics. Instead, to take advantage of the higher statistics at 8 TeV the most interesting possibility would be to measure the polarization and spin correlation at high $\mttb$ and/or high $\betattb$ where the effect of new physics may be larger.

\begin{table}[t]
\begin{center}
\caption{Summary of the most precise polarization and spin correlation measurements at the Tevatron and at the LHC with 7 TeV, compared to the corresponding SM predictions\label{tab:spin}.}
\begin{tabular}{cccc}
\hline
\hline
Collider & basis & measurement & SM prediction \\
Tevatron & beamline & $C=0.58 \pm 0.20$ & $0.791^{+0.013}_{-0.014}$ \\
LHC7 & helicity & $P = -0.014 \pm 0.029$ & 0 \\
LHC7 & helicity & $C = 0.17 \pm 0.09$ & $0.310 \pm 0.006$ \\
\hline
\hline
\end{tabular}
\end{center}
\end{table}

\begin{figure}[t]
\begin{center}
\begin{tabular}{c}
\includegraphics[height=5cm,clip=]{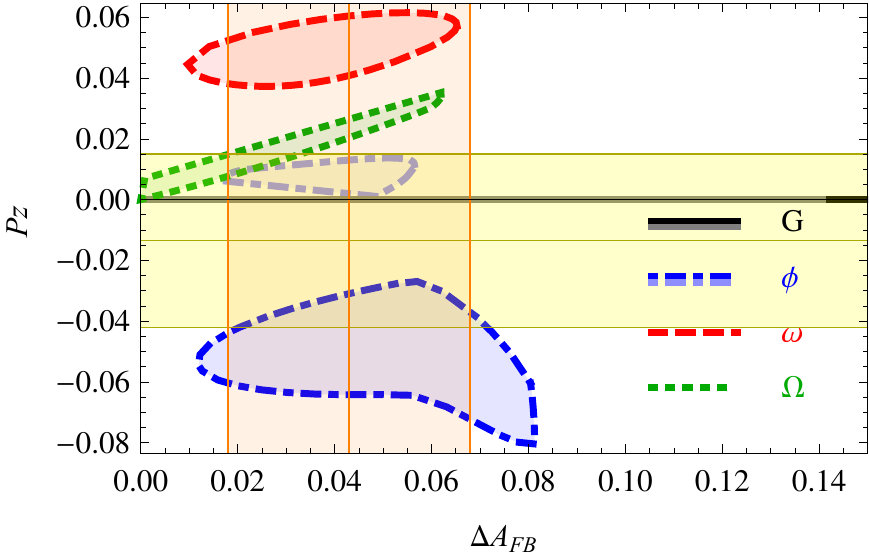} \\[3mm]
\includegraphics[height=5cm,clip=]{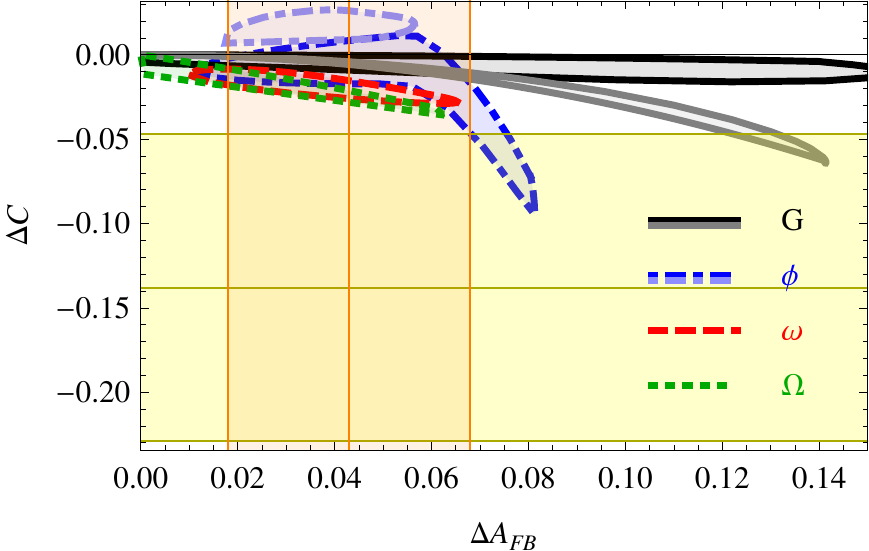}
\end{tabular}
\end{center}
\caption{Top quark `longitudinal' polarization $P_z$ (upper panel) and new physics contributions to $C$ at the LHC (lower panel) versus new physics contributions to $\afb$, for four models: (i) light (black) and heavy (gray) axigluon $G$; (ii) light (dark blue) and heavy (light blue) scalar doublet $\phi$; (iii) color triplet $\omega$; and (iv) color sextet $\Omega$. The shaded bands correspond to the central value and $1\sigma$ uncertainty for the corresponding measurement. From~\textcite{Fajfer:2012si}.}
\label{fig:6B1}
\end{figure}

In addition to the ``longitudinal'' helicity axis $\hat z$, there are two other independent directions in which the top polarization can be investigated. We can specify them by choosing $\hat y$ perpendicular to the production plane, and $\hat x$ orthogonal to the $\hat z$ and $\hat y$,
\begin{equation}
\hat z = \frac{\vec p_t}{|\vec p_t|} \,,\quad
\hat y = \frac{\vec p_t \times \vec p_p}{|\vec p_t \times \vec p_p|} \,,\quad
\hat x = \hat y \times \hat z \,,
\label{ec:axes}
\end{equation}
with $\vec p_p$ the proton momentum in the top quark rest frame. (At the LHC, one can use the motion of the $\ttb$ pair in the laboratory frame to select a preferred direction among the two protons,
 see~\onlinecite{Baumgart:2013yra}.) The $\hat x$ and $\hat y$ directions are usually denoted as ``transverse''  and ``normal'', respectively. The transverse polarization can be non-zero, for example in $s$-channel color octet models~\cite{Baumgart:2011wk,Aguilar-Saavedra:2014yea}.
A polarization in the normal direction requires a complex phase in the amplitude, which may be provided by the propagator of the octet if it is produced on-shell~\cite{Baumgart:2013yra}. Neither the transverse nor the normal polarizations have been measured at the Tevatron nor the LHC.

\subsection{Interplay of asymmetries and top polarization}
\label{sec:6c}

The several new physics models proposed to explain the $\afb$ anomaly give different predictions for the leptonic asymmetries, as it was soon noticed~\cite{Krohn:2011tw}. New particles that couple to $t_R$ produce larger leptonic asymmetries than those coupling to $t_L$. This is illustrated in Fig.~\ref{fig:6C1}, which depicts the relation between new physics contributions to the asymmetries $\Delta \afb$ and $\Delta \afbl$, for a  color octet with mass $M=250$ GeV exchanged in the $s$-channel. The relation between the asymmetries is given for three chiralities of the $\bar q q G$ coupling (axial, right-handed and left-handed) chosen such that $g_A^u = g_A^d > 0$, and a continuous variation of the chirality of the $\bar t t G$ coupling along the curves, including vector, axial, left-handed and right-handed couplings. The sign of $\Delta \afbl$ in each case is explained by the threshold behavior (see Sec.~\ref{sec:2d}). The two asymmetries are {\it de facto} uncorrelated, and their combined measurement can give information of the chirality of the couplings of the new particle to the top quark.

\begin{figure}[htb]
\begin{center}
\includegraphics[width=7.5cm,clip=]{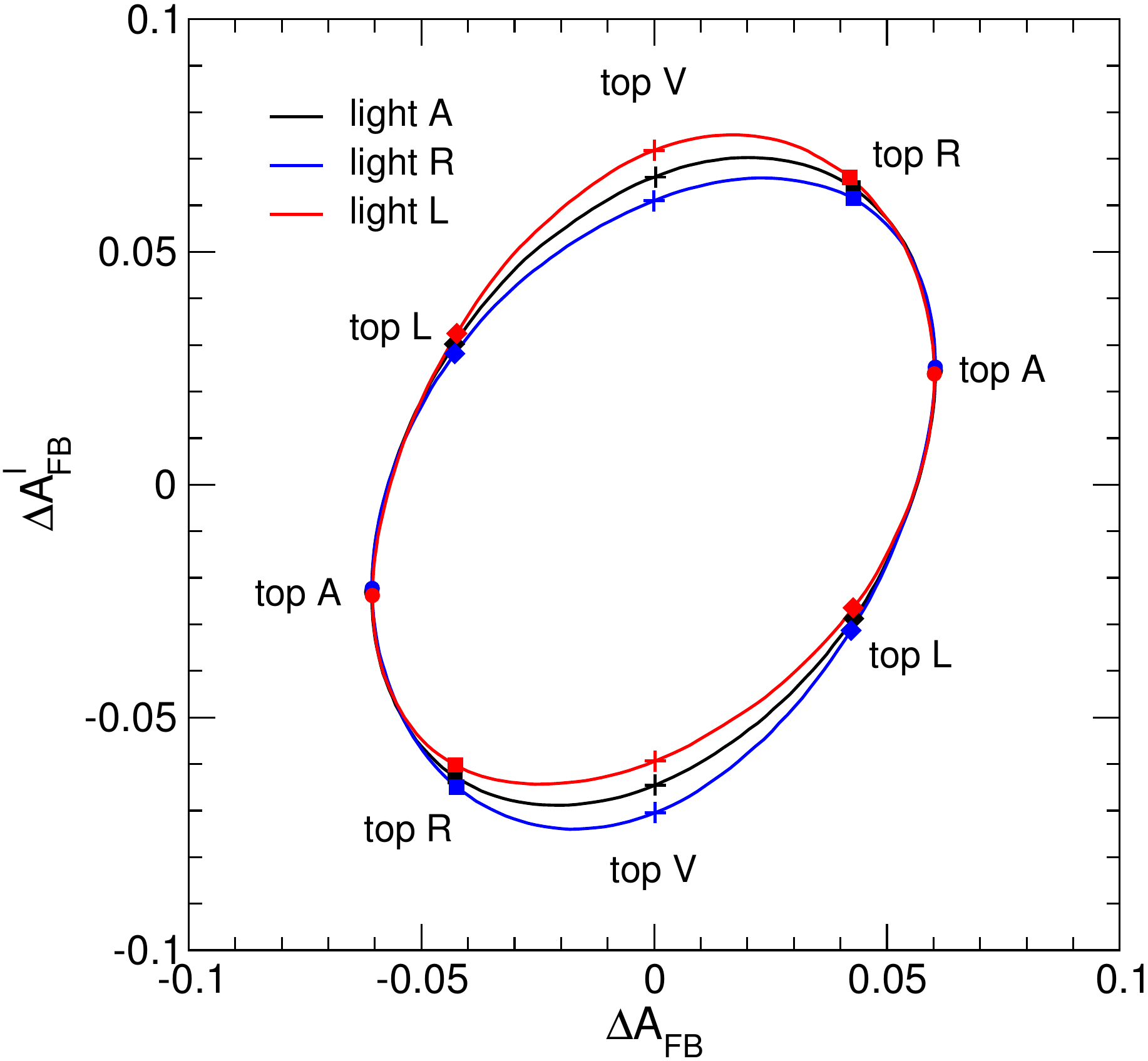}
\end{center}
\caption{Relation between $\Delta \afb$ and $\Delta \afbl$, for three choices of light quark couplings and continuous variation of the chirality of top quark couplings to a color octet $G$. The abbreviations refer to axial (A), vector (V), left-handed (L) and right-handed (R). From~\textcite{Aguilar-Saavedra:2014yea}.}
\label{fig:6C1}
\end{figure}

The longitudinal polarization $P_z$ (in the helicity basis) is also uncorrelated from the asymmetries, and in general it can be positive, negative, or nearly zero. This can also be illustrated with the color octet model, where $P_z$ depends not only on the coupling to the top quark but also on the light quark couplings. Figure~\ref{fig:6C2} shows the polarization at the Tevatron as a function of the continuous parameter
\begin{equation}
\phi_h = \arg (g_A^t + i g_V^t) \in [0,2\pi] \,,
\label{ec:phi}
\end{equation}
 for three choices of light quark couplings, all with $g_A^{u,d}>0$. (We do not consider vector couplings to $u,d$ since in this case the interference with the SM amplitudes does not generate any asymmetry $\afb$.) One can see that $P_z = 0$ if the top quark coupling is either vectorial or axial. On the other hand, $P_x = 0$ only when the top quark coupling is axial. In particular, one can see that for an axigluon $P_x=P_z = 0$; $P_y$ is also small unless it is produced on its mass shell.

\begin{figure}[htb]
\begin{center}
\includegraphics[width=7.5cm,clip=]{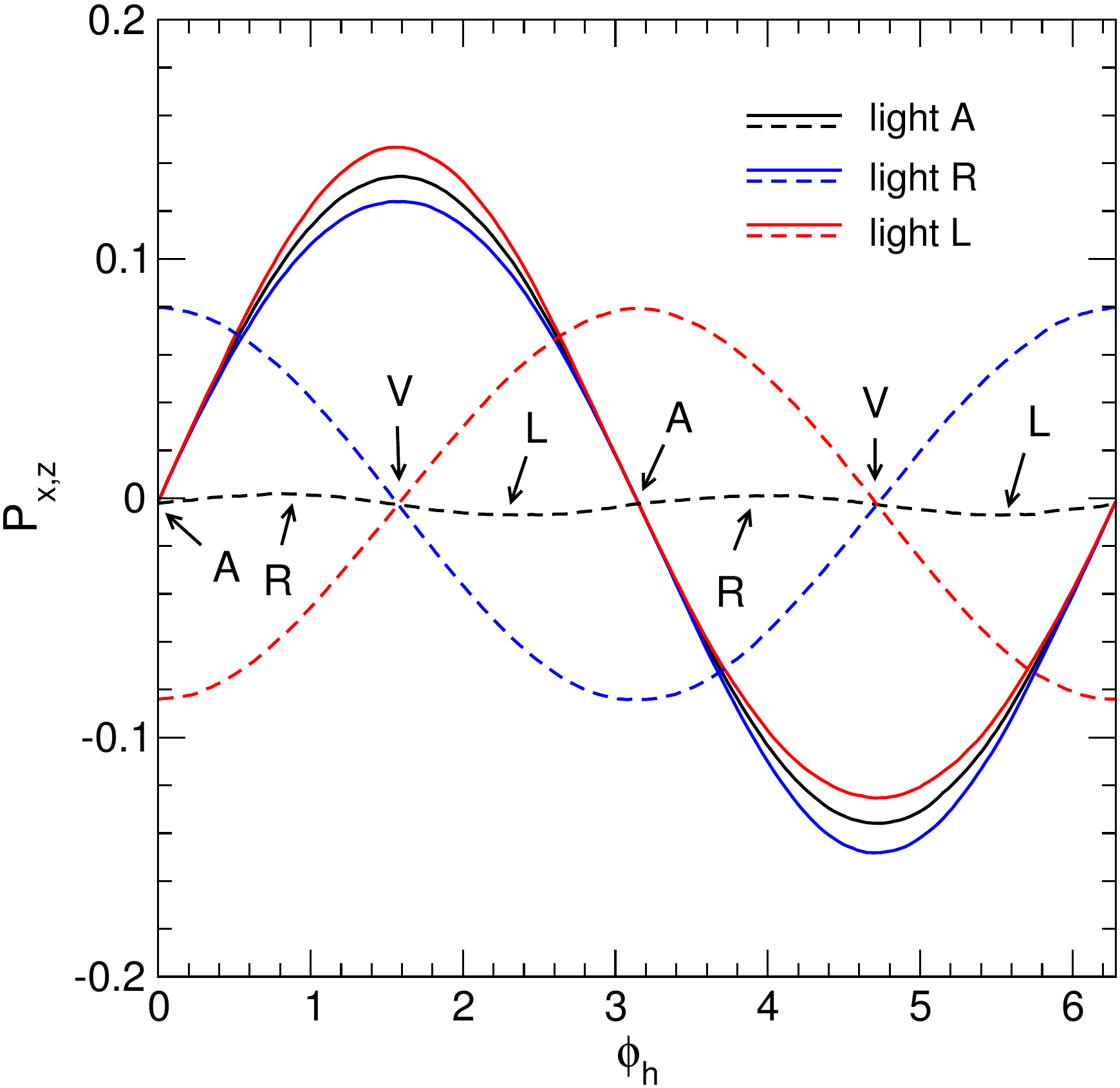}
\end{center}
\caption{Longitudinal (dashed lines) and transverse polarization (solid lines), for three choices of light quark couplings and continuous variation of the chirality of top couplings (see Eq.~(\ref{ec:phi})). The abbreviations refer to axial (A), vector (V), left-handed (L) and right-handed (R). From~\textcite{Aguilar-Saavedra:2014yea}.}
\label{fig:6C2}
\end{figure}

Once it is established that $\afb$ and $\afbl$ (and also $\ac$) are in general independent, one can attempt to fit these asymmetries---as well as other $\ttb$ observables---in the context of any new physics model that explain the Tevatron anomalies. A light color octet is the best suited candidate for this since, as we have mentioned throughout Sec.~\ref{sec:5} and Sec.~\ref{sec:6}, it can reproduce the Tevatron and LHC asymmetries while keeping good agreement with the remaining $\ttb$ data. A fit including the Tevatron and LHC cross sections, asymmetries and polarization observables has been performed in~\textcite{Aguilar-Saavedra:2014nja}. The global agreement of the SM is $\chi^2/\text{d.o.f.} = 15.8/10$ for 10 measurements ($1.3\sigma$). While the overall consistency with data is good within the SM, the agreement can be improved to $\chi^2/\text{d.o.f.} = 8.1/8$ ($\chi^2/\text{d.o.f.} = 6.4/6$) for an octet with a reference mass $M=250$ GeV, $\Gamma/M = 0.2$, and right-handed (general) couplings to quarks. Color octets with pure axial or left-handed couplings to quarks do not improve the fit with respect to the SM, and are thus disfavored on a purely statistical basis.

Finally, one can go a step further diagnosing potential new physics and study the relation between $\afb$ and $\afbl$ differentially, for example as a function of the charged-lepton transverse momentum $\pTl$~\cite{Falkowski:2012cu}. Different SM extensions predict not only different ratios $\afbl / \afb$ but also quite a different dependence on $\pTl$. This is shown in Fig.~\ref{fig:6C3}, for $\pTl$ (in GeV) in the intervals $[0,20[$, $[20,40[$, $[40,60[$, $[60,100[$, $[100,150[$, $[150,\infty[$. The direction of the curves is such that for the $L$ and $R$ benchmarks $\afb$ grows with $\pTl$, and for the SM and $A$ benchmark $\afbl$ increases. This information could be used to distinguish these models from SM-based explanations of the asymmetry due to some mismodeling effect that could enhance the observed asymmetry with respect to the prediction.
For example, a color octet with right-handed couplings that gives a good fit of all inclusive measurements could be distinguished from a SM-like effect.
In analogy with the Tevatron, at the LHC the $\pTl$ dependence of leptonic asymmetries versus $\ac$ can be used to investigate the presence of new physics~\cite{Carmona:2014gra}.

\begin{figure}[htb]
\begin{center}
\includegraphics[height=7cm,clip=]{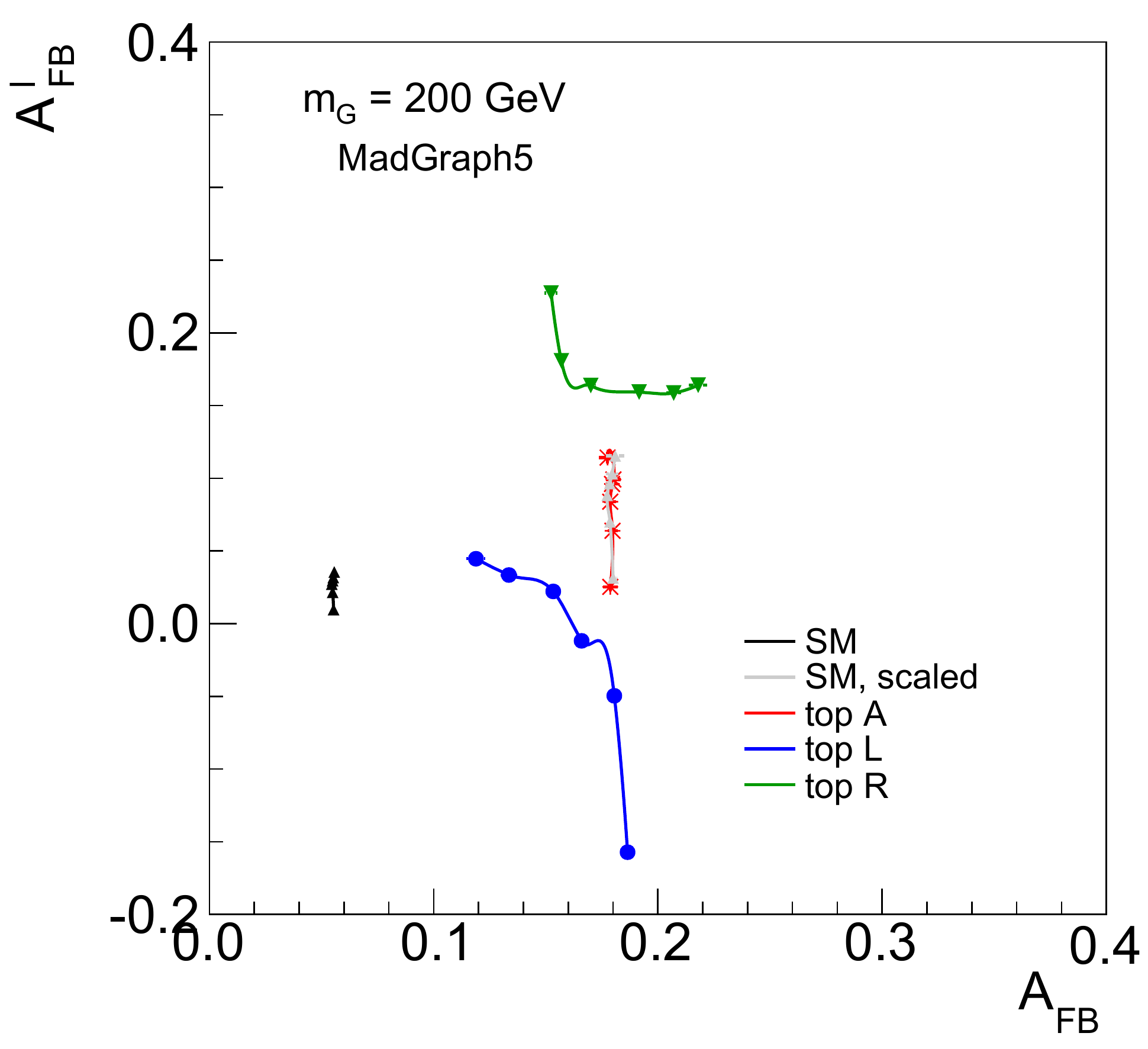}
\end{center}
\caption{Differential relation between $\afb$ and $\afbl$ (including SM and new physics contributions, if any) as a function of the lepton transverse momentum, for the SM, a `scaled SM' in which the SM predictions of $\afb$ and $\afbl$ are scaled by a common factor in order to have $\afb = 0.18$, and for a color octet with three choices of couplings to the top quark. From~\textcite{Falkowski:2012cu}.}
\label{fig:6C3}
\end{figure}


\section{Other constraints and effects}
\label{sec:7}

New-physics interpretations of the excess in $\afb$ often have other implications beyond their impact on $\ttb$ observables. They are less universal, with different kinds of effects predicted for different classes of models. In all cases, the absence of those signals of new physics puts strong constraints on the corresponding model parameter spaces. In this section we review the most important non-$\ttb$ effects associated with models with extra particles that contribute at tree-level to the charge asymmetries.


\subsection{Flavor physics}
\label{sec:7a}
Most models of new physics motivated by the anomaly in the FB asymmetry require a non-trivial flavor structure. Indeed, $t$- and $u$-channel exchanges involve intergenerational couplings of the new particles to the first and third families, while $s$-channel heavy octets must have couplings with different sign, and thus non-universal, to generate a positive $\Delta \afb$. Conciliating these features with the stringent bounds on flavor-changing neutral currents (FCNC) is not straightforward. In particular, the possible couplings to quarks are constrained.

Usually, FCNC can be avoided by aligning the couplings of the new particles with specific directions in flavor space. However, a complete alignment in both the up and the down sectors is not possible when the quark doublets $Q_L$ and $q_L$ are involved. In this case, bounds can be extracted from a combination of measurements in $B-\bar{B}$ mixing and $B$ decays, $D-\bar{D}$ mixing and $K-\bar{K}$ mixing~\cite{Blum:2009sk,Blum:2011fa,Bai:2011ed,Zhu:2011ww,Duraisamy:2011pt}. These bounds require small or near-degenerate couplings to $Q_L$ and $q_L$, or sufficiently large masses. For this reason, in many of the proposed models the extra particles are chosen to have chiral right-handed couplings to the top quark.

In the case of $s$-channel exchanges, which only involve diagonal couplings, all flavor problems would be avoided from the start if the couplings were family universal, and thus diagonal in any flavor basis. The only interesting $s$-channel multiplet for the FB asymmetry is the color octet. If heavy, it requires that the axial couplings to light and top quarks, $g_A^q$ and $g_A^t$, have opposite signs, which precludes universality. Once more, light octets with $M\lesssim 450~\mathrm{GeV}$ present an advantage here: because they need $g_A^q$ and $g_A^t$ of the same sign, all the axial and vector couplings can be chosen, in principle, to be equal for the three families~\cite{Tavares:2011zg}. In this case, the couplings to $Q_L$ do not produce FCNC, so an axigluon without vector couplings is not constrained by flavor physics. An obstacle to such a universal octet is that the usual dijet bounds require, for a sizable asymmetry, non-universal couplings with $|g_A^t|\gg |g_A^q|$ (see below). However, these bounds are relaxed when the octet has an enhanced width, which is anyway needed for agreement with other observables, see Sec.~\ref{sec:6a} and Sec.~\ref{sec:7d}.

The required large flavor-changing $tu$ (or $td$) couplings of new particles exchanged in the $t$ or $u$ channels, on the other hand, do not allow for universality. Some degree of alignment is then needed to comply with existing non-top FCNC limits. Specifically, the off-diagonal couplings mixing the first and second and the second and third families must be suppressed in the mass-eigenstate basis. While this particular pattern can be arranged in flavor models, it is unnatural and requires some tuning~\cite{Jung:2011zv,Shelton:2011hq}. This situation is improved in specific cases. Most notably, the color triplet $\omega$ has, in any basis, antisymmetric coupling matrices in flavor space~\cite{Ligeti:2011vt}. This property forbids tree-level FCNC. The most dangerous loop contributions are also absent for this field, and other flavor bounds can be avoided if the $tc$ and $cu$ couplings are small in the gauge basis \cite{Dorsner:2009mq,Giudice:2011ak}.

An interesting and natural way to avoid flavor problems and justify the unusual patterns of couplings needed in these models was proposed in~\textcite{Grinstein:2011yv}. If the interactions of the new particles respect the SM quark flavor group $G_F=U(3)_{u_R}\times U(3)_{d_R} \times U(3)_{q_L}$, or its subgroup $H_F=U(2)_{u_R}\times U(2)_{d_R} \times U(2)_{q_L} \times U(1)_3$ (with the quarks in the first two families in doublets of the corresponding $SU(2)$ factors), then the only breaking of these global symmetries comes from the SM Yukawa couplings. (More generally, a small breaking of $H_F$ from the new-physics sector can be allowed.) In this minimal-flavor-violation scenario, FCNC are under control, since they are absent before flavor breaking. Moreover, the large intergenerational couplings required in $t,u$-channel models are not only flavor-symmetric, but also a consequence of non-trivial flavor representations. Analogously, different signs of $g_A^q$ and $g_A^t$ for a color-octet are automatic if the field is also an octet under the flavor symmetry ($t$ channel exchange is also important in this case).
An additional virtue of these flavor-symmetric models is that limits from same-sign top pair production are avoided, as discussed in Sec.~\ref{sec:7b}. All the relevant flavor representations of $G_F$ and $H_F$ have been classified and analyzed in detail in~\textcite{Grinstein:2011dz}.


\subsection{Same-sign top quark pair production}
\label{sec:7b}

The production of same-sign top quark pairs would be a striking signal of physics beyond the SM. At hadron colliders, charge conservation implies that $tt$ pairs can only be produced from initial up or charm quarks. Therefore the LHC, being a $pp$ machine, is especially well suited to studying this signal.

The possible scalar and vector bosons that can produce $tt$ pairs at the tree level are a subset of the multiplets in Table~\ref{t:multiplets}: the neutral components of $Z^\prime$, $\mathcal{W}$, $G$, $H$, $\phi$ and $\Phi$, exchanged in the $t$ channel, and the charge 4/3 components of $Q^\prime$, $Y^\prime$, $\Omega$ and $\Sigma$, exchanged in the $s$ channel~\cite{AguilarSaavedra:2011zy}. The negative results so far at the LHC (see for example~\onlinecite{Aad:2012bb}) put strong constraints on particular combinations of the couplings of these fields, which we collect in Table.~\ref{t:sslimits}.

\begin{table}[htb]
\begin{center}
\caption{Limits at 95\% confidence level, from~\textcite{Aad:2012bb}, on the couplings of arbitrary heavy vector bosons and scalars that mediate the production of same-sign top-quark pairs. The fields and couplings are defined in Table~\ref{t:multiplets}. For the fields $Z^\prime$ and $G$, we have defined $|g_{13}|=(|g_{13}^q|^2+|g_{13}^u|^2)^{1/2}$.}
\label{t:sslimits}
\begin{tabular}{cc} \hline
\hline
Field & Limit  \\
$Z^\prime_\mu$ & $|g_{13}|/M < 0.57$ TeV$^{-1}$ \\[1mm]
$W^\prime_\mu$ & $|g_{13}|/M < 0.57$ TeV$^{-1}$ \\[1mm]
$G_\mu$  & $|g_{13}|/M < 0.99$ TeV$^{-1}$ \\[1mm]
$H_\mu$ &  $|g_{13}|/M < 0.99$ TeV$^{-1}$  \\[1mm]
$Q^\prime_\mu$ &  $|g_{11} g_{33}|/M^2 < 0.34$ TeV$^{-2}$ \\[1mm]
$Y^\prime_\mu$ &  $|g_{11} g_{33}|/M^2 < 0.63$ TeV$^{-2}$  \\[1mm]
$\phi$ &  $|g_{13}^u g_{31}^u|/M^2 < 0.92$ TeV$^{-2}$  \\[1mm]
$\Phi$ &  $|g_{13}^u g_{31}^u|/M^2 < 1.8$ TeV$^{-2}$ \\[1mm]
$\Omega$  &  $|g_{11} g_{33}|/M^2 <0.33$ TeV$^{-2}$  \\[1mm]
$\Sigma$ &  $|g_{11} g_{33}|/M^2 < 0.16$ TeV$^{-2}$  \\
\hline\hline
\end{tabular}
\end{center}
\end{table}

For arbitrary allowed couplings, these 10 multiplets can contribute to both $\ttb$ and $tt$ production. However, in general there is no direct relation between the observables in both processes, since they involve different combinations of couplings. In fact, a direct relation exists if and only if the following conditions are met: (i) the extra multiplet contributes to both $\ttb$ and $tt$ in the $t$ channel only and (ii) the extra multiplet is self-conjugate under CP, which is only possible for real representations of the gauge group. The reason is that in this case the new-physics amplitudes in $tt$ and $\ttb$ processes are related by a CP transformation of one of the vertices. Therefore, the stringent $tt$ bounds put in deep trouble the explanations of the FB anomaly with $t$ exchanges of $Z^\prime$~\cite{Berger:2011ua} and also of $G$, $\mathcal{W}$ and $H$~\cite{AguilarSaavedra:2011zy}. In particular, they are sufficient to exclude the simplest $Z^\prime$ models, as shown in Fig.~\ref{fig:ssZp}. 
\begin{figure}[htb]
\begin{center}
\includegraphics[width=7.5cm]{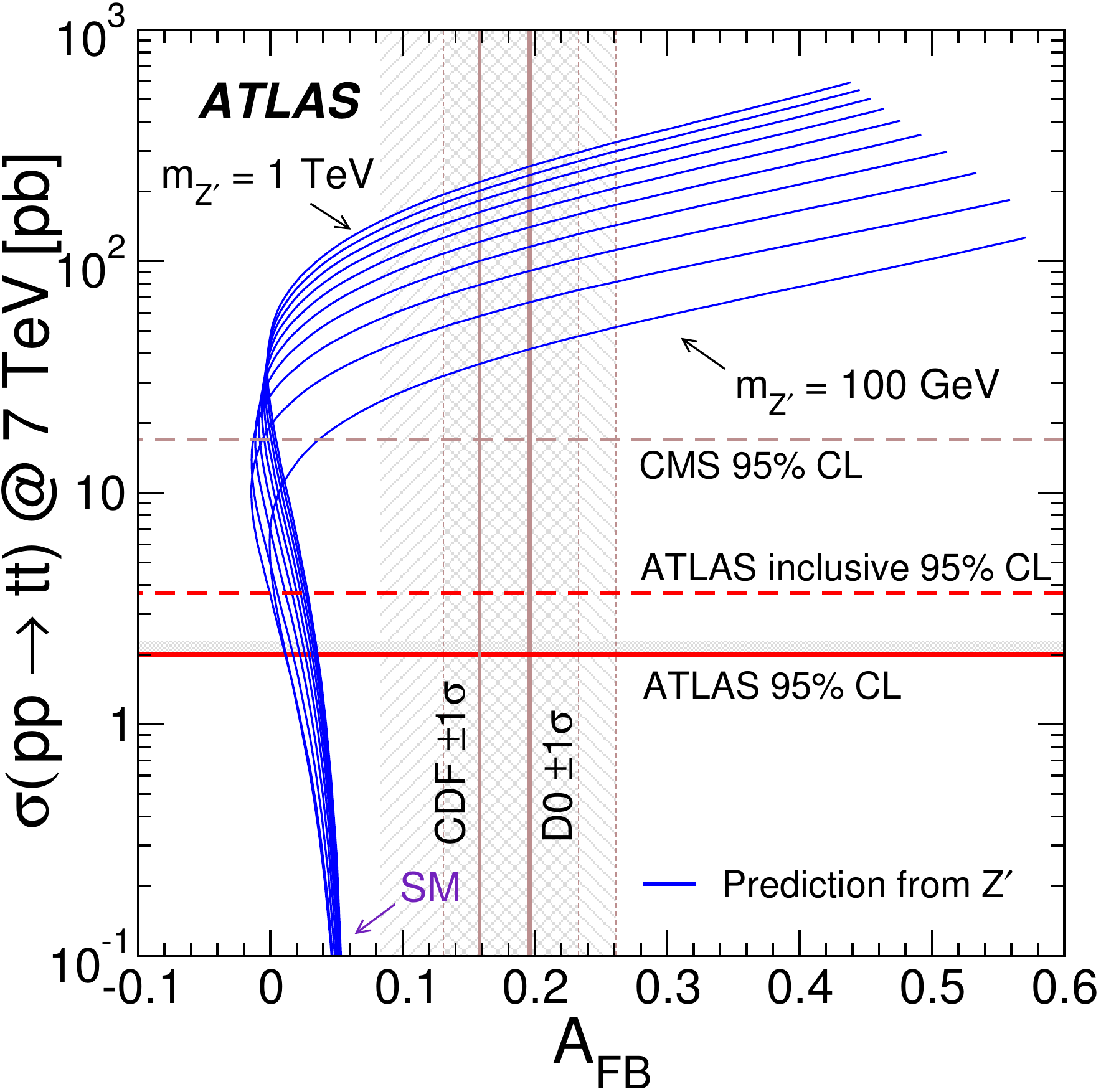} 
\caption{\label{fig:ssZp} Predictions for $\afb$ at the Tevatron and the $pp \to tt$ cross section at the LHC for a real $Z^\prime$ boson with right-handed couplings. Each curve corresponds to a different mass. The horizontal lines indicate the 95\%~C.L.\ upper limits on the cross section from measurements by the ATLAS and CMS Collaborations~\cite{Aad:2012bb,Aad:2012cg,Chatrchyan:2011dk}, while the vertical bands correspond to measurements of $\afb$, with half the final luminosity, by the CDF and D0 Collaborations~\cite{Aaltonen:2011kc,Abazov:2011rq}. From~\textcite{Aad:2012bb}.}
\end{center}
\end{figure}
A neat solution to save these models is to embed these fields in a non-trivial representation of a flavor symmetry, as discussed in the previous section. Then, the conservation of ``top number'' prevents $uu/cc \to tt$ processes~\cite{Jung:2011zv}. Equivalently, the extended symmetry ensures a cancellation of the contribution to these processes of the different irreducible components that form these reducible representations of the gauge group. Note that, even for real flavor representations, the extended fields are no longer self-conjugate when the flavor indices are fixed.

Finally, although no strong conclusions can be derived for the complex fields $\phi$ and $\Phi$, some regions of their parameter space relevant to the FB asymmetry in $\ttb$ are forbidden by the absence of $tt$ signals~\cite{AguilarSaavedra:2011zy}. 

\subsection{Top quark-jet resonances}
\label{sec:7c}
In models with $t$- or $u$-channel exchange of a new boson $R$, the same flavor-violating vertices that are necessary for $\ttb$ production will give rise to the production of $R$ through the process $qg \to R \tilde{t}$, with $\tilde{t}=t$ or~$\bar{t}$. We assume in the following that the field $R$ is not self-conjugate, to avoid the same-sign limits we have just discussed. If $M_R > m_t$, this particle can subsequently decay into $\tilde{t}q$, giving rise to $t \bar{t}$ plus jet events with a top-jet ($tj$) or anti-top-jet ($\bar{t}$) resonance~\cite{Gresham:2011dg}. At the LHC, where the initial partons are predominantly quarks, rather than anti-quarks, the resonance will be most often found in $\bar{t}j$ when $R$ is a $t$-channel mediator, and in $tj$ when it is a $u$-channel mediator. This is dictated by the couplings of the corresponding particles, and can be understood as a consequence of baryon-number conservation.

\begin{figure}[htb]
\begin{center}
\begin{tabular}{c}
\includegraphics[width=7.5cm]{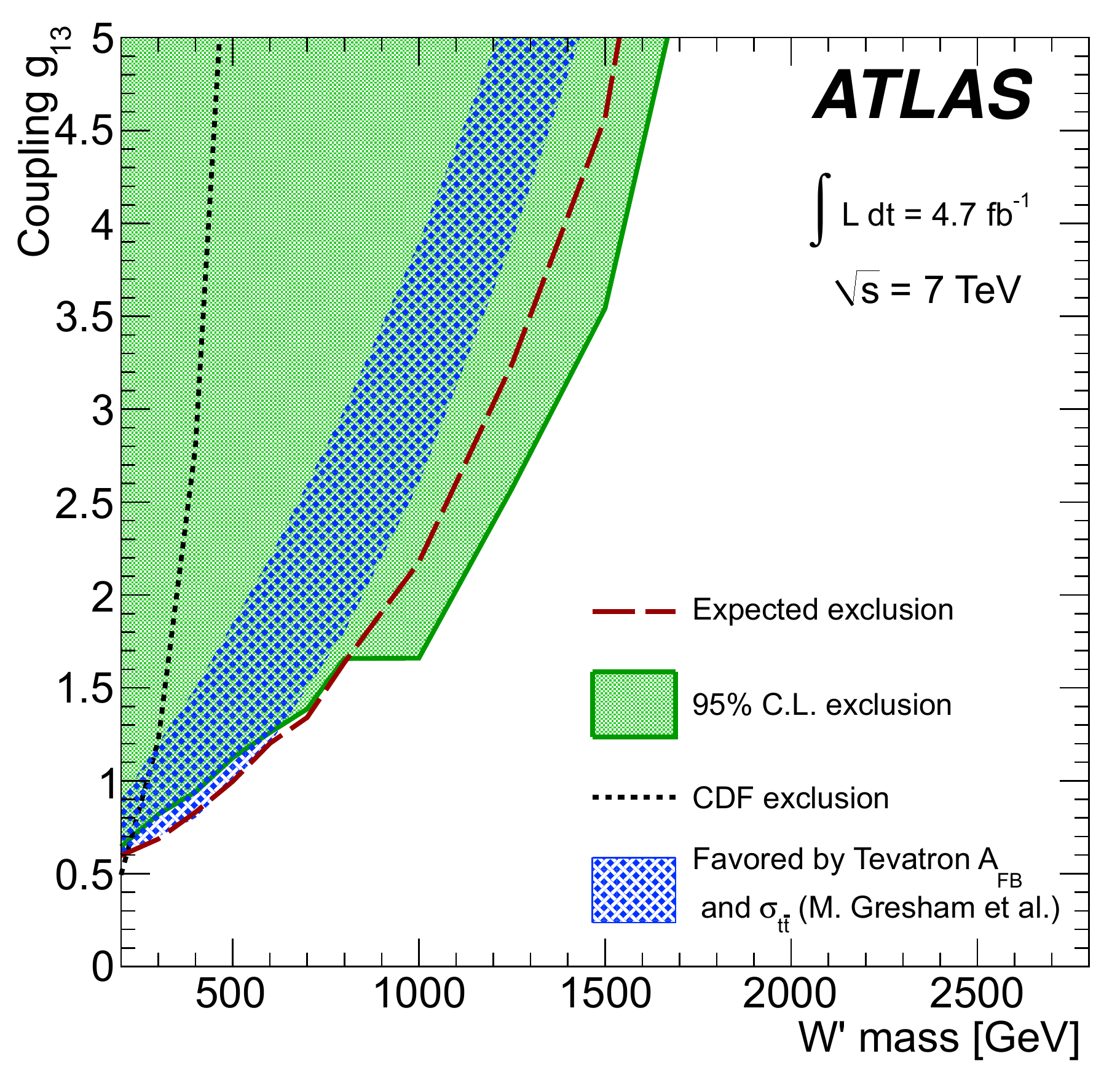} \\
\includegraphics[width=7.5cm]{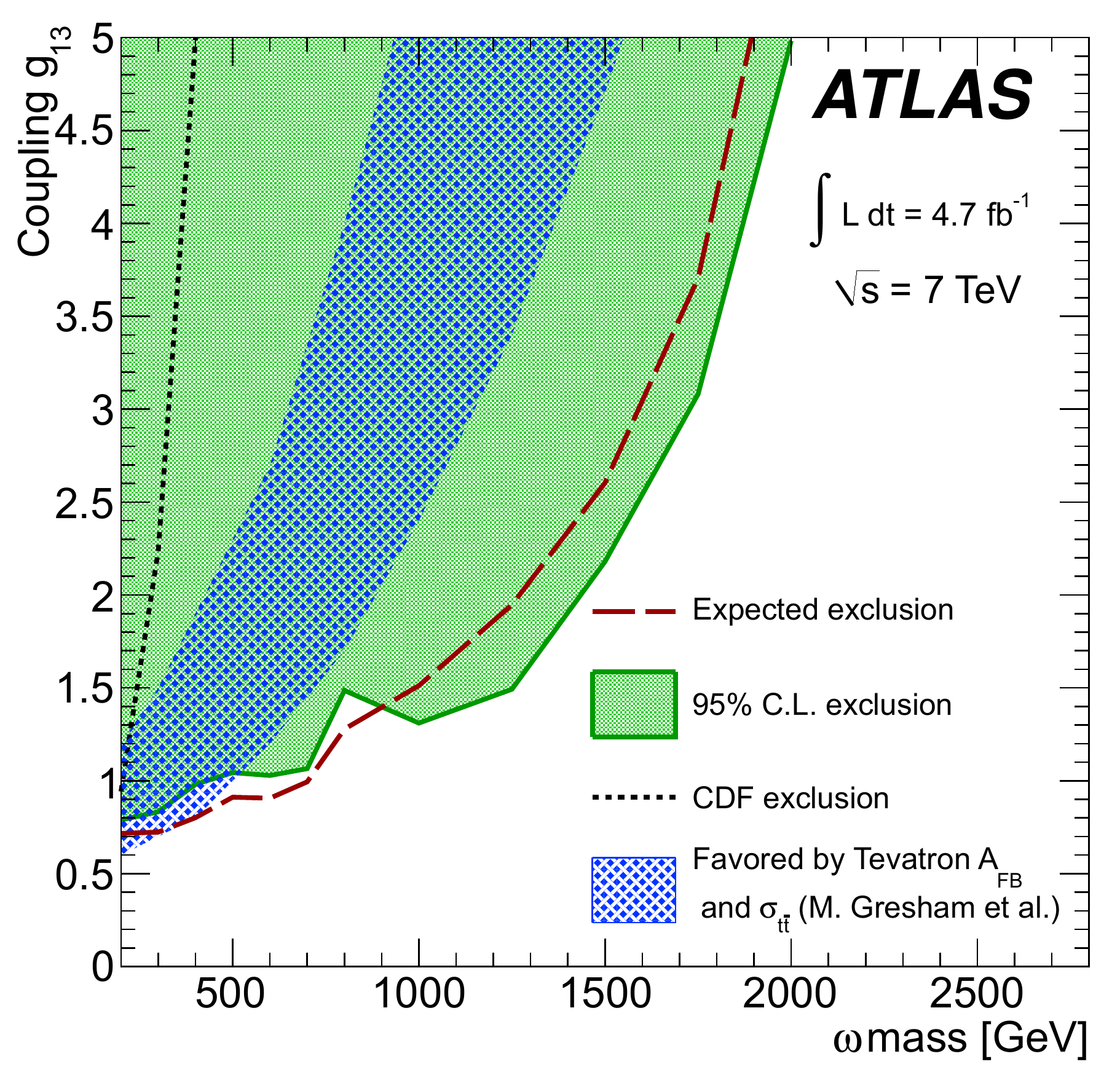} 
\end{tabular}
\caption{\label{fig:resonantbounds} Expected and observed 95\% C.L.\ upper limits on $W^\prime$ (upper plot) and $\omega$ (lower plot) in terms of their mass and the coupling. The dark blue areas are the regions favored by the $\ttb$ cross section and FB asymmetry measured by the CDF and D0 Collaborations with half the final luminosity~\cite{Aaltonen:2011kc,Abazov:2011rq}. From~\textcite{Aad:2012em}}
\end{center}
\end{figure}
These signals have been searched for by experiments at the Tevatron and the LHC. In particular, the ATLAS Collaboration has performed a search of $W^\prime$ and $\omega$ in the $\ell$+jets channel with the 2011 dataset~\cite{Aad:2012em}. The results are summarized in Fig.~\ref{fig:resonantbounds}. 
Similar bounds apply to any particle exchanged in the $t$ or $u$ channels. We see that most of the regions of parameters that could account for the large values of the FB asymmetry measured in 2011~\cite{Aaltonen:2011kc,Abazov:2011rq} are basically excluded, even without taking into account strong bounds from the $\mttb$ tail, which can independently rule out a $W^\prime$ explanation. 
This analysis is, however, not sensitive to the regions with $M_R$ below the top quark mass.

\subsection{Dijet and dijet-pair resonances}
\label{sec:7d}
All particles that mediate $\ttb$ production in the $s$ channel have $q\bar{q}$ couplings $g^q_{V,A}$, and will therefore contribute to dijet production. Searches for new phenomena in dijet final states have been performed by the UA1 and UA2 Collaborations~\cite{Albajar:1988rs,Alitti:1993pn} and also at the Tevatron~\cite{Aaltonen:2008dn,Abazov:2003tj} and the LHC~\cite{Chatrchyan:2013qha,Aad:2011fq,ATLAS:2012pu} experiments. These searches are complementary, as they cover different mass ranges. Most of them search for bumps in the dijet invariant mass distribution originating from new narrow resonances, but some analyze also angular distributions, which are useful to put bounds on broad or heavy particles, parametrized by contact interactions~\cite{Chatrchyan:2013muj}.
A convenient mass-coupling interpretation of the narrow-resonance limits from the different experiments has been given, for singlets $Z^\prime$ and octets $G$, in~\textcite{Dobrescu:2013cmh}.

Dijet bounds require a relatively weak coupling of the new particles to the light quarks, with $|g_A^q| \lesssim 0.3$ for narrow octets lighter than 2~TeV. For a given $\Delta \afb$, these limits translate into lower bounds for the axial couplings to the top quark. For instance, the limit $|g_A^q|\lesssim 0.15$ for a narrow octet with $M_G=1~\mathrm{TeV}$~\cite{Aad:2011fq,Dobrescu:2013cmh} implies that $|g_A^t| \gtrsim 6$ is required to reproduce the world average $\afb=0.13$. For lighter octets, smaller $|g_A^t|$ are allowed. At any rate, this direct interpretation of the dijet limits is actually rather conservative. Indeed, one should take into account that large values of the couplings $g_A^t$ will increase the width of the resonance and lower its branching fraction into dijets.  Open channels into other particles would further weaken these bounds.

Let us next discuss the constraints obtained from four-jet final states. The pair production of color-octets from initial-state gluons is determined by $SU(3)_C$ gauge symmetry and the unitarity of the theory~\cite{Gross:2012bz}. Pair production is proportional to $\alpha_s$ and enhanced by color and spin factors, so it is large at the LHC for light octets. When each octet decays into two jets, an event with a pair of resonant dijets is produced. The ATLAS and CMS Collaborations have studied these signatures and have basically excluded narrow octets with masses between 100 and 740~GeV~\cite{Aad:2011yh,ATLAS:2012ds,Chatrchyan:2013izb}, assuming 100\% branching ratio into dijets. This excludes the light-$G$ explanation of the FB asymmetry,\footnote{Octets $G$ generating a sizable FB asymmetry and with mass smaller than 100 GeV are excluded by electroweak precision tests, due to their loop contribution to the $\bar{q}qZ$ vertex~\cite{Gresham:2012kv}.} unless the octets are broad or have a significant branching ratio into multijets via additional intermediate resonances~\cite{Gross:2012bz}. 

\subsection{Four-top quark production}
\label{sec:7e}
Since, due to dijet bounds, large couplings $g^t_A$ are required for an explanation of the asymmetry measurements, pair-produced octets will decay dominantly into two $\ttb$ pairs, if kinematically allowed.\footnote{Four top final states also originate from diagrams with a non-resonant octet. This is especially relevant below the $\ttb$ threshold.} Such four-top quark final states have a very small background in the SM. They are difficult to reconstruct, but a simple search of this signal can be performed studying the production of same-sign dileptons~\cite{TheATLAScollaboration:2013jha} and trileptons. In this way, it is possible to exclude octets $G$ with masses between 350 and 650~GeV~\cite{AguilarSaavedra:2011ck} unless, once again, the width is enhanced by the decay to non-SM particles.


\section{Outlook}
\label{sec:8}
The large FB asymmetries observed in a succession of measurements at the Tevatron have triggered a detailed exploration, from both the experimental and the theoretical sides, of observables related to $\ttb$ production at hadron colliders. Independently of the nature of the discrepancies---which are significantly milder after the latest analyses---this effort has lead to a better understanding of the properties of the top quark and of the effects of possible new physics connected to the top-quark sector. The resulting expertise will certainly be valuable in future searches at the LHC.

The fact that the Tevatron measurements with the full dataset are closer to the SM predictions than previous measurements, with half the luminosity, strongly suggests that the former discrepancies were due  to simple statistical fluctuations in the data. However, at this point the question is not completely settled. Even if the two collaborations give average results in the $\ell+$jets channel that are statistically compatible, it is intriguing that the CDF and D0 measurements of both asymmetries, $\afb$ and $\afbl$, are actually quite similar in the 4-jet sample. It is only the inclusion of the 3-jet sample, which yields lower values of these asymmetries, that lowers the D0 averages and make them more consistent with the SM predictions. This sample is not considered in the CDF analyses. While the differences between the results from 3-jet and 4-jet samples may be purely statistical, the possibility of some mismodeling effect, either in the 4-jet or the 3-jet samples, must be investigated in more detail. On the other hand, the recent measurement of $\afb$ in the dilepton channel by the D0 Collaboration yields a large asymmetry (even larger than the CDF value in the $\ell+$jets channel but also with a larger uncertainty). A measurement by the CDF Collaboration in the dilepton channel with the full dataset might shed some light on this issue.

Because the Tevatron asymmetries cannot be directly measured at the LHC, the confirmation or rebuttal of a possible anomaly is quite difficult. One important step would be to measure the dependence of the asymmetries on the $\ttb$ velocity $\beta_z^{\ttb}$ to obtain the `collider-independent' asymmetries $A_u$, $A_d$, discussed in Sec.~\ref{sec:2c}. This measurement is quite demanding from the experimental side, since it requires a 3-dimensional unfolding in $\mttb$, $\beta$ and $|\Delta y|$. But it offers a unique possibility of testing at the LHC the same quantities that are in the origin of the Tevatron $\afb$.

The possibility of unexpectedly large higher-order QCD corrections that might significantly increase the value of the predicted FB asymmetry at the Tevatron seems now excluded by the recent NNLO calculation of $\afb$.  Indeed, the NNLO corrections turn out to be small, as expected, shifting the central value from $\afb = 0.088$ to $\afb = 0.095$. Moreover, the NNLO predictions, even when considered differentially, lie well inside the uncertainty bands of the previous NLO results, see Fig.~\ref{fig:diffafbs}. A proper combination of the CDF and D0 differential results is crucial to asess the agreement of theory with data, taking into account experimental bin-to-bin correlations and theory uncertainties.

Finally, the explanation of the asymmetry excess with new physics faces two serious problems. The first one is that almost all successful models are rather {\it ad hoc,} since they are not clearly motivated by other compelling theoretical or experimental reasons, and usually non-generic choices of parameters are required to avoid the most obvious constraints. The second problem is that, even with such parameters, most new physics models still predict a series of observable signals that have not been found. In particular, the measurements of either $\ac$, $\ttb$ differential distributions or top polarization disfavor most of the models. Searches for $tj$ resonances, on the other hand, exclude large regions of the parameter space of models with $t$-channel exchanges. Among the simple explanations in terms of just one multiplet, the model that can better account for all the $\ttb$ data, including the Tevatron and LHC asymmetries, is, arguably, an $s$-channel color-octet vector boson. It should be noted, nevertheless, that this model requires some non-trivial ingredients to comply with all the measurements. Another model that survives the different tests is a light scalar isodoublet exchanged in the $t$ channel.

The next LHC run with $13-14$ TeV and high luminosity will bring the possibility of new, independent measurements in addition to the current ones. One example is the charge asymmetry in $\ttb \gamma$ production, which has the potential of showing deviations with respect to the SM predictions even for $\ac$ in perfect agreement with the SM. Otherwise, it will further constrain the parameter space of the different models, rendering them less viable. Another example is the asymmetry in $\ttb W^\pm$. Whatever the final outcome is, it is likely that some of the questions posed by the Tevatron asymmetries will be answered in the next years. And of course, some unexpected surprises might be waiting along the road.


\section*{Acknowledgements}
This work has been supported by U.S. DOE grant  DE-SC0007859; by the Spanish MICINN Projects FPA2010-17915, FPA2012-38713 
(including ERDF funds from the European Union) and Consolider-Ingenio-2010-CSD2007-00042; MINECO Projects FPA2013-47836-C3-2-P and  Centro de Excelencia Severo Ochoa SEV-2012-0234; and by the Junta de Andaluc\'{\i}a Projects FQM 101 and FQM 6552.

\bibliography{biblio.bib}

\end{document}